\documentclass[a4paper,11pt]{article}
\usepackage{amsmath,amsthm,amsfonts,amssymb,natbib,graphicx,booktabs,float,caption,subcaption,multirow,dsfont,setspace}
\usepackage{chngcntr}
\usepackage{apptools}
\AtAppendix{\counterwithin{lemma}{section}}
\AtAppendix{\counterwithin{theorem}{section}}

\newtheorem{theorem}{Theorem}
\newtheorem{lemma}{Lemma}

\newtheorem{corollary}{Corollary}

\newtheorem{definition}{Definition}

\renewcommand{\qedsymbol}{$\blacksquare$}
\DeclareMathOperator*{\plim}{plim}
\DeclareMathOperator*{\Prb}{\text{Pr}^{b}}

\usepackage[colorinlistoftodos,textsize=tiny]{todonotes}
\newcommand{\Comments}{1}
\newcommand{\mynote}[2]{\ifnum\Comments=1\textcolor{#1}{#2}\fi}
\newcommand{\mytodo}[2]{\ifnum\Comments=1%
	\todo[linecolor=#1!80!black,backgroundcolor=#1,bordercolor=#1!80!black]{#2}\fi}

\ifnum\Comments=1               
\fi

\ifnum\Comments=1               
\fi

\ifnum\Comments=1               
\fi

\begin{document}

\title{Testing Quantile Forecast Optimality\thanks{We are grateful to  Maximo Camacho, Mike Clements, Laura Coroneo, Marc Hallin, Jos\'{e} Olmo,  Emese Lazar, Adam Lee, and Xiaohan Xue for helpful comments on the paper as well as participants of the 12th ECB Conference on Forecasting Techniques (2023, Frankfurt), the Tenth Italian Congress of Econometrics and Empirical Economics 2023 (Cagliari), the SAEe 2022 in Valencia, the  Statistical Week 2022 (M\"{u}nster) and 2021 (virtual), the International Symposium on Forecasting 2021 (virtual),  and seminar attendees at the University of Hagen, the University of Hohenheim and Goethe University Frankfurt, 2023, and at the ICMA Centre, University of Reading, 2022. Marc-Oliver Pohle is grateful for support by the Klaus Tschira Foundation, Germany.}}
\author{Jack
Fosten\thanks{%
King's Business School, King's College London, WC2B 4BG London, UK, E-mail: 
\texttt{jack.fosten@kcl.ac.uk}} \thanks{%
Data Analytics in Finance and Macro (DAFM) Research Centre, King's College
London, UK} \and Daniel Gutknecht\thanks{%
Faculty of Economics and Business, Goethe University Frankfurt, 60629
Frankfurt am Main, Germany. E-mail: \texttt{gutknecht@wiwi.uni-frankfurt.de.}%
} \and Marc-Oliver Pohle\thanks{Heidelberg Institute for Theoretical Studies, 69118 Heidelberg, Germany. E-mail: \texttt{marc-oliver.pohle@h-its.org.}%
} }
\date{\today }
\maketitle

\begin{abstract}
	\noindent Quantile forecasts made across multiple horizons have become an important output of many financial institutions, central banks and international organisations. This paper proposes misspecification tests for such quantile forecasts that assess optimality over a set of multiple forecast horizons and/or quantiles. The tests build on multiple Mincer-Zarnowitz quantile regressions cast in a moment equality framework. Our main test is for the null hypothesis of autocalibration, a concept which assesses optimality with respect to the information contained in the forecasts themselves. We provide an extension that allows to test for optimality with respect to larger information sets and a multivariate extension. Importantly, our tests do not just inform about general violations of optimality, but may also provide useful insights into specific forms of sub-optimality. A simulation study investigates the finite sample performance of our tests, and two empirical applications to financial returns and U.S. macroeconomic series illustrate that our tests can yield interesting insights into quantile forecast sub-optimality and its causes. \newline
	
	
\end{abstract}

\pagebreak
\onehalfspacing

\section{Introduction}

Economic and financial forecasters have become increasingly interested in making quantile predictions, often across different quantile levels and at multiple horizons into the future. In financial markets, for instance, such multi-step quantile predictions are produced due to the 10-day value-at-risk (VaR) requirements of the Basel Committee
on Banking Supervision.\footnote{See for instance: https://www.bis.org/publ/bcbs148.pdf [Last Accessed: 18/12/20]} In the growth-at-risk (GaR) literature on the other hand, \citet{ABG19} propose quantile models to predict downside risks to real gross domestic product (GDP) growth at horizons ranging from one quarter ahead to one year ahead. These methods are now widely implemented in academic research \citep{PRRH20,BS20} and in international institutions like the IMF \citep{PEJLAFW2019}, and are typically applied across various quantile levels. This trend for multi-horizon quantile forecasts has also developed into a growing literature in nowcasting GaR that typically uses several intra-period nowcast horizons \citep[e.g.,][]{ADP18,FMS21,CCM20}. Finally, it is common for central banks, such as the Bank of England, to produce fan charts of key economic variables such as GDP growth, unemployment or the Consumer Price Index (CPI) inflation rate across several quantile levels and horizons.

However, despite the expansion in empirical and methodological research, there is currently very little statistical guidance for assessing whether a set of  multi-step ahead, multi-quantile forecasts are consistent with respect to the outcomes observed. This consistency is often referred to as `optimality', `rationality', or `calibration' in the literature, with `full optimality' referring to optimality relative to the information set known to the forecaster, while a weaker form of optimality known as `autocalibration' is defined with respect to the information contained in the forecasts themselves \citep[see][]{gneiting2013,Tsyplakov2013}.  This paper aims to fill this gap in the literature  by proposing various (out-of-sample) optimality tests for quantile forecasts that can accommodate predictions either derived from known econometric forecasting models, or from external sources like institutional or professional forecasters. Specifically, we develop tests that assess optimality of quantile forecasts over multiple forecast horizons \textit{and} multiple quantiles simultaneously.

The main test of this paper is a joint test of autocalibration for quantile forecasts obtained across different horizons and quantile levels. The test is based on a series of quantile Mincer-Zarnowitz (MZ) regressions \citep[see][]{GLLS11} across all quantile levels and horizons, which are in turn used to construct a test statistic for the null hypothesis of autocalibration across horizons and quantiles using a set of moment equalities \citep[e.g.,][]{RS2010,AS2010}. We suggest a block bootstrap procedure to obtain critical values for the test. The bootstrap is simple to implement and avoids the need to estimate a large variance-covariance matrix that would be required in a more standard Wald-type test. We establish the first-order  asymptotic validity of these bootstrap critical values.

The test of autocalibration based on MZ regressions can provide valuable information to forecasters. In particular, failure to reject the null hypothesis of autocalibration suggests that the forecaster may proceed to use the forecasts as they are without the need to `re-calibrate' them. On the other hand, if the null hypothesis is rejected, the test hints at directions for improvement of the forecasts. That is, it informs the forecaster about the horizons, quantiles or horizon-quantile combinations that contributed strongest to the rejection of the null, and thus an improvement of the forecasts is warranted. In addition, the estimated MZ regression can be used to infer about the nature of the deviations from autocalibration, when the forecasts are plotted alongside the realisations. The estimated MZ coefficients may also be used to perform a re-calibration or the original forecasts, as has been suggested in the case of mean forecasts in the recent work of \citet{C22}. 

We provide two extensions of this test for autocalibration. The first extension allows for additional predictors in the MZ regressions, which we call the augmented quantile Mincer-Zarnowitz test. This test is operationally similar to that of the first test, but may provide richer information to the forecaster. It tests a stronger form of optimality relative to a larger information set than autocalibration. If autocalibration is not rejected, but the null hypothesis of the augmented test is rejected, it indicates that the additional variables used in the MZ regression carry additional informational content which should be used in making the forecasts themselves. The second extension allows to test optimality for multiple time series variables and not just for a single variable. Testing multiple time series variables simultaneously may be useful in cases where we are interested in testing whether one type of model delivers optimal forecasts for multiple macroeconomic variables, or across different financial asset returns, for instance different companies from the same sector. 

Finally, as a separate contribution, we also outline in the appendix a test for monotonically non-decreasing expected quantile loss as the forecast horizon increases. This extends the result of \citet{PT2012} to the quantile case whereas they focussed on the mean squared forecast error (MSFE) case for optimal multi-horizon mean forecasts. The test makes use of empirical moment inequalities using the Generalised Moment Selection (GMS) procedure of \citet{AS2010}. This test can also be seen as complementary to monotonicity tests used in the nowcasting literature for the MSFE of mean nowcasts \citep[see][and references therein]{FG17}.

We provide two empirical applications of our methodology. The first application applies the basic MZ test to classical VaR forecasts for S\&P 500 returns constructed from a GARCH(1,1) model via the GARCH bootstrap \citep{pascual2006}. We test jointly over the quantile levels 0.01, 0.025 and 0.05 and horizons from 1 to 10 trading days. Autocalibration is rejected overall and the miscalibration of the forecasts gets stronger for larger forecast horizons and more extreme quantiles. Furthermore, a clear pattern emerges over all quantiles and horizons regarding the conditional quantile bias: the VaR forecasts tend to underestimate risk in calmer times, but overestimate it in more stressful periods.

The second application applies the test in the spirit of the emerging GaR literature, where we focus on the extensions of our test using the augmented MZ test and the test with multiple time series. We expand on the work of \citet{ABG19} to formally investigate the performance of simple quantile regression models using financial conditions indicators in predicting a range of U.S. macroeconomic series. Interestingly, we find that the forecasts across four different series and a range of quantile levels and horizon are sub-optimal in that they are not autocalibrated. However, further analysis of the results shows that this sub-optimality is present only in inflation-type series and not in real series like industrial production and employment growth. We also find poorer calibration at the most extreme quantile under consideration.

In relation to the existing literature, this paper  extends the work on quantile forecast optimality or, in other words, absolute evaluation of quantile forecasts. The focus of this literature has been on single-horizon prediction at a single quantile, which mainly stems from the extensive body of research on backtesting VaR, such as \citet{christoffersen1998}, \citet{engle2004caviar}, \citet{EO10,EO11}, \citet{GLLS11} and \citet{nolde2017}. Our work also complements the literature on testing the relative forecast performance of conditional quantile models such as \citet{GK05}, \citet{M15} or more recently  \citet{CFG2022}.  Finally, as our focus lies on testing for optimality across horizons, the paper also relates to  \citet{Q17}, who emphasized the importance of multi-horizon forecast evaluation to avoid multiple testing issues in the context of relative evaluation of mean forecasts. The only work on multi-horizon optimality testing we are aware of is \citet{PT2012}, who consider the case of mean forecasts as well and discuss several implications of optimality specific to the multi-horizon context and how to construct tests for them, most notably the monotonicity of expected loss over horizons, which we extend to the quantile case in the appendix. 

The rest of the paper is organised as follows. Section \ref{sec:theory} lays out the notion of quantile forecast optimality that will provide the foundation of our tests. Section \ref{sec:testing} then introduces the test for autocalibration via MZ regression, along with the bootstrap methodology and theory. Section \ref{sec:extension} extends the test to the augmented MZ test and the test for multiple variables, while Section \ref{sec:empirical} gives the two empirical applications of our methods. Finally, Section \ref{sec:conclusion} concludes the paper. 

The appendix contains results from a Monte Carlo study (Section \ref{sec:simulations}), where we assess the finite sample properties of the MZ and augmented MZ tests across various sample sizes and bootstrap block lengths. The appendix also contains the proofs for the theoretical results (Sections \ref{sec:characoptim} and \ref{proofs}) along with the monotonicity test (Section \ref{sec:ELoss}). Sections \ref{app:Finance_Application} and  \ref{app:Macro_Application} provide additional empirical results and graphs for the VaR and the GaR application, respectively. Finally, all tests of the paper are provided as \texttt{R} functions in the \texttt{R} package \texttt{quantoptimR} available at https://github.com/MarcPohle/quantoptimR.

\section{Quantile Forecast Optimality}\label{sec:theory}

Consider a multivariate stochastic process $\{\mathbf{V}_{t}\}_{t \in \mathbb{Z}}$, where $\mathbf{V}_{t}$ is a random vector which contains a response variable of interest $y_{t}$ and other observable predictors. We denote the forecaster's information set at time $t$ by $\mathcal{F}_{t}=\sigma(\mathbf{V}_{s}; s\leq t)$, where $\sigma(.)$ denotes the $\sigma$-algebra generated by a set of random variables. Assuming a continuous outcome $y_{t}$ for the rest of the paper, our target functional is the conditional $\tau$-quantile of $y_{t}$ given $\mathcal{F}_{t-h}$:
\begin{equation*}
	q_{t} \left(\tau |\mathcal{F}_{t-h}\right)=F^{-1}_{y_{t}|\mathcal{F}_{t-h}}(\tau),
\end{equation*}%
where $F_{y_{t}|\mathcal{F}_{t-h}} \left(\cdot \right)$ is the cumulative distribution function of $y_{t}$ conditional on $\mathcal{F}_{t-h}$. We denote an $h$-step ahead forecast at time $t-h$ for this $\tau$-quantile $q_{t} \left(\tau |\mathcal{F}_{t-h} \right)$ by $\widehat{y}_{\tau,t,h}$, and assume that we observe these forecasts $\widehat{y}_{\tau,t,h}$  for each target period $t$ at multiple horizons, $h\in\mathcal{H}=\{1,\ldots,H\}$, and multiple quantile levels, $ \tau \in \mathcal{T}=\{\tau_{1},\ldots,\tau_{K}\}\subset [0+\varepsilon,1-\varepsilon]$ with $\varepsilon>0$, for some finite integers  $H$ and $K$, respectively. That is, at each time point $t$ we have a matrix of forecasts, $\left(\widehat{y}_{\tau,t,h}\right)_{\tau=\tau_1,...,\tau_K,h=1,...,H}$. In addition, throughout the paper, we will assume strict stationarity of $\{\mathbf{V}_{t}\}_{t \in \mathbb{Z}}$  and finite first moments of the forecasts $\widehat{y}_{\tau,t,h}$ and $y_t$ itself, see Assumptions A1 and A2 in Section \ref{sec:testing}.

Since our focus lies on the evaluation of quantile forecasts, the loss function used for evaluation in this context is the `tick' or `check' loss which is well-known from quantile regression. This is written as $L_{\tau} \left(y_{t+h} - \widehat{y}_{\tau,t,h} \right) = \rho_{\tau} \left( y_{t+h} - \widehat{y}_{\tau,t,h} \right)$, 
where $\rho_{\tau}(u) = u \left(\tau - 1\{u<0\} \right)$  and where $1\{.\}$ denotes the indicator function giving a value of one when the expression is true and zero otherwise.

While relative forecast evaluation deals with comparing different forecasting methods or models, mainly by ranking them via their expected loss, the subject of this paper is absolute forecast evaluation across different quantile levels and/or forecasting horizons, in other words the assessment of \textit{a particular} forecasting model or method in terms of absolute evaluation criteria for multiple quantile levels and horizons. These evaluation criteria are usually different forms of optimality (or `rationality'/`calibration'). We start by defining and discussing various forms of quantile forecast optimality before showing how to operationalise the latter for testing. 

\begin{definition}[Optimality] \label{optimality}
	An $h$-step ahead forecast	$\widehat{y}^{\ast}_{\tau,t,h|\mathcal{I}_{t-h}}$ for the $\tau$-quantile is optimal relative to an information set $\mathcal{I}_{t-h} \subset \mathcal{F}_{t-h}$ if:
	\begin{equation*} 
		\widehat{y}^{\ast}_{\tau,t,h|\mathcal{I}_{t-h}} = \arg \min_{\widehat{y}_{\tau,t,h}} \mathrm{E} \left[L_{\tau} \left( y_{t} - \widehat{y}_{\tau,t,h} \right) | \mathcal{I}_{t-h} \right].
	\end{equation*}
	We simply call it optimal and denote it by $\widehat{y}^{\ast}_{\tau,t,h}$ if $\mathcal{I}_{t-h}=\mathcal{F}_{t-h}$, i.e.\ if it is optimal relative to the full information set: $\widehat{y}^{\ast}_{\tau,t,h}\equiv \widehat{y}^{\ast}_{\tau,t,h|\mathcal{F}_{t-h}}$.
\end{definition}

Analogous to the case of mean forecasts \citep{granger1969}, an optimal quantile forecast relative to an information set can alternatively be characterised as being equal to the respective conditional quantile provided the information set is sufficiently large and includes the forecasts themselves. Specifically,   since `tick' loss is a strictly consistent scoring function for the corresponding quantile (see Definition 1 and Proposition 1 in \citealp{gneiting2011}), it holds that an $h$-step ahead forecast $\widehat{y}_{\tau,t,h}$ for the $\tau$-quantile is optimal relative to any information (sub-)set $\mathcal{I}_{t-h}$ satisfying $\sigma \left(\widehat{y}_{\tau,t,h} \right) \subset \mathcal{I}_{t-h} \subset \mathcal{F}_{t-h}$, where $\sigma \left(\widehat{y}_{\tau,t,h} \right) $ denotes the sigma algebra spanned by the forecast itself, if and only if:
\begin{equation} \label{char_optimality}
	\widehat{y}_{\tau,t,h} = q_{t} \left(\tau |\mathcal{I}_{t-h} \right).\footnote{Note that this result continues to hold for any quantile loss from the class of generalised piecewise linear loss functions, or in fact for any consistent scoring function if the $\tau$-quantile is substituted for the corresponding statistical functional for which this scoring function is consistent.}
\end{equation}

While interest often lies in testing the null hypothesis of (full) optimality relative to the information set $\mathcal{F}_{t-h}$, which amounts to testing if the forecast, $\widehat{y}_{\tau,t,h}$, is equal to its target, $q_{t}(\tau |\mathcal{F}_{t-h})$, the possibly large and generally unknown information set $\mathcal{F}_{t-h}$ usually makes direct tests of this hypothesis difficult in practice. Thus, we next discuss weaker forms of optimality that will form the basis of our test(s) in Sections \ref{sec:testing} and \ref{sec:extension} below. In fact, in Section \ref{sec:characoptim} of the appendix (see Lemma \ref{Optimality}) we show formally that these weaker forms of optimality may always be viewed as a direct implication of optimality with respect to the `full' information set $\mathcal{F}_{t-h}$. That is, any $h$-step ahead forecast  optimal with respect to the full information set $\mathcal{F}_{t-h}$, is also optimal relative to any `smaller' information (sub-)set $\mathcal{I}_{t} \subset \mathcal{F}_t$. 

A special case of this `weaker' form of optimality is optimality with respect to the information contained in the forecast itself, $\sigma \left(\widehat{y}_{\tau,t,h} \right)$, or autocalibration, a term first coined  by \cite{Tsyplakov2013} and \cite{gneiting2013} in the context of probabilistic forecasts.

\begin{definition}[Autocalibration] \label{autocalibration}
	An $h$-step ahead forecast	$\widehat{y}_{\tau,t,h}$ for the $\tau$-quantile is autocalibrated if it holds that: $\widehat{y}_{\tau,t,h} = q_{t} \left(\tau |\sigma(\widehat{y}_{\tau,t,h}) \right)$.
\end{definition}

On the one hand, autocalibration may be regarded as a direct implication of full optimality that is particularly suitable for testing as it only relies upon the forecasts themselves and does not require any assumptions on the information set $\mathcal{F}_{t-h}$ or a selection of variables from it.  On the other hand, however, autocalibration may also be viewed as a criterion for absolute forecast evaluation in its own right for several reasons. Firstly, the concept has a clear interpretation since a forecast user provided with autocalibrated forecasts should use them as they are and not transform or `recalibrate' them. Secondly, only involving forecasts and observations and no information set that depends on other quantities, it comes closest to the idea of forecast calibration as a concept of consistency between forecasts and observations \citep[see][]{gneiting2007}. Thirdly, autocalibration might often be a more reasonable criterion to demand from forecasts than full optimality, which is a often hard to fulfill in practice. Finally, the Murphy decomposition of expected loss \citep{P20} shows that autocalibration is a fundamental property of forecasts in that expected loss is driven by only two forces: deviations from autocalibration and the information content of the forecasts.

The next section will outline how to test autocalibration, viewed either as an implication of full optimality or as a forecast property its own right, across multiple quantile levels and horizons simultaneously.

\section{Quantile Mincer-Zarnowitz Test}\label{sec:testing}

\subsection{Null Hypothesis and Quantile Mincer-Zarnowitz Regressions}\label{ssec:MZTest}

While autocalibration testing has a long tradition in econometrics through the use  Mincer-Zarnowitz regressions  for mean forecasts  \citep{mincer1969}, the latter may also be used directly for the case of quantiles \cite[see][]{GLLS11}. Definition \ref{autocalibration} in fact suggests that a natural test for autocalibration of an $h$-step ahead forecast for the $\tau$-level quantile may be based on checking whether, for a given sample of outcomes and quantile forecasts at level $\tau_{k}$ and horizon $h$, it holds that: 
\[
q_{t} \left(\tau |\widehat{y}_{\tau,t,h} \right) = \alpha_{h}^{\dag}(\tau_{k})+ \widehat{y}_{\tau_{k},t,h}\beta_{h}^{\dag}(\tau_{k} )=\widehat{y}_{\tau_{k},t,h}
\]
almost surely. More generally, since our goal is to test for autocalibration over multiple forecast horizons and quantile levels jointly, we specify such a linear quantile regression model for every horizon $h\in\mathcal{H}$ and $\tau_{k}\in\mathcal{T}$ as follows:
\begin{equation}\label{POPQREG}
	y_{t}=\alpha_{h}^{\dag}(\tau_{k})+ \widehat{y}_{\tau_{k},t,h}\beta_{h}^{\dag}(\tau_{k} ) + \varepsilon_{t,h}(\tau_{k})=  \mathbf{X}_{\tau_{k},t,h}^{\prime }\boldsymbol{\beta}_{h}^{\dag} (\tau_{k} )+ \varepsilon_{t,h}(\tau_{k}) ,
\end{equation}%
where $\mathbf{X}_{\tau_{k},t,h}=(1,\widehat{y}_{\tau_{k},t,h})^{\prime}$ and $\boldsymbol{\beta}_{h}^{\dag} (\tau_{k} )=(\alpha_{h}^{\dag}(\tau_{k}),\beta_{h}^{\dag}(\tau_{k} ))^{\prime}$. Here, the population coefficient vector $\boldsymbol{\beta}^{\dag}_{h} (\tau_{k} )=(\alpha^{\dag}_{h}(\tau_{k}),\beta^{\dag}_{h}(\tau_{k} ) )^{\prime}$ of this linear quantile regression model is defined as:
\begin{equation}\label{EQPLIM}
	\boldsymbol{\beta}^{\dag}_{h} (\tau_{k} )=\arg \min_{\boldsymbol{b} \in \mathcal{B}}\mathrm{E}\left[
	\rho _{\tau_{k}  }\left( y_{t}-\mathbf{X}_{\tau_{k},t,h}^{\prime }\boldsymbol{b} \right) \right],
\end{equation}
where $\mathcal{B}$ denotes the parameter space satisfying conditions set out in Assumption A3 below. The composite null hypothesis is given by:
\begin{equation}\label{EQH0}
	H^{\text{MZ}}_{0}: \{\alpha_{h}^{\dag}(\tau_{k})= 0 \} \cap  \{ \beta_{h}^{\dag}(\tau_{k} )=1\}
\end{equation}
for all $h\in\mathcal{H}$ and $\tau_{k}\in\mathcal{T}$ versus $H^{\text{MZ}}_{1}:  \{\alpha_{h}^{\dag}(\tau_{k})\neq  0 \} \quad \text{and/or}\quad  \{ \beta_{h}^{\dag}(\tau_{k} )\neq 1\}$ for at least some $h\in\mathcal{H}$ and $\tau_{k}\in\mathcal{T}$. Testing the null hypothesis in  (\ref{EQH0}) not only yields a multi-horizon, multi-quantile test of autocalibration, but also provides us with an idea about possible deviations from the null. In particular, examining the contributions of single horizons, quantiles or horizon-quantile combinations to the overall test statistic, which will be introduced in Subsection \ref{ssec:teststat} below, may also be informative about deviations from autocalibration. Moreover, the empirical counterpart of $q_{t} \left(\tau |\widehat{y}_{\tau,t,h} \right) = \alpha_{h}^{\dag}(\tau_{k})+ \widehat{y}_{\tau_{k},t,h}\beta_{h}^{\dag}(\tau_{k} )$ may be interpreted as autocalibrated forecasts such that, for a specific value of the forecast $\widehat{y}_{\tau_{k},t,h}$, the deviations between the regression line (or recalibrated forecast) and the forecasts themselves, $\widehat{y}_{\tau_{k},t,h} - q_{t} \left(\tau |\widehat{y}_{\tau,t,h} \right)$  can be interpreted as the quantile version of a conditional bias. The direction and size of this conditional quantile bias informs us about deficiencies of forecasts in certain situations, a point that we will come back to and illustrate in the applications in Section \ref{sec:empirical}.

\subsection{Test Statistic and Bootstrap}\label{ssec:teststat}

In what follows, assume that we observe an evaluation sample of size $P$, in other words a scalar-valued time series of observations starting at some point in time $R+1 \in \mathbb{Z}$, $\{y_{t}\}_{t=R+1}^{T}$, and a matrix-valued time series of forecasts, $\left\{ \left(\widehat{y}_{\tau,t,h}\right)_{\tau=\tau_1,...,\tau_K,h=1,...,H} \right\}_{t=R+1}^{T}$. Moreover, we may also observe a vector of additional variables $\mathbf{Z}_{t-h}$ from the forecaster's information set $\mathcal{F}_{t-h}$. We will write this additional sample of vector-valued time series as $\{\mathbf{Z}_{t}\}_{t=R+1-H}^{T-1}$.

Using the taxonomy of \citet{GR2010}, forecasts $\widehat{y}_{\tau_{k},t,h}$ may stem  either from `forecasting methods'  or from `forecasting models'. In the former case, we are typically without knowledge about the underlying  model such as with forecasts from the Survey of Professional Forecasters (SPF), or may use forecasts that depend on parameters estimated in-sample using so-called limited-memory estimators based on a finite rolling estimation window. In the case of `forecasting models' on the other hand, we  need to account for the contribution of estimation uncertainty to the asymptotic distribution of the statistic. However, since the focus of this paper lies on detecting systematic forecasting bias rather than dealing with specific forms of estimation error, we consider the latter only under the expanding scheme with a `large' in-sample estimation window. Specifically, for forecasting models, we assume that we also observe $R$ additional observations of $y_{t}$ prior to $R+1$ that may be used as estimation window (note that the $R$ in-sample observations also comprise $H$ observations that are used to produce the initial out-of-sample forecast for period $R+1$), and  that $P/R \rightarrow 0$ as $P,R\rightarrow \infty$. This allows us to abstract from estimation error in the analysis and to focus on systematic features of the forecasts.\footnote{In addition, it also allows us to resample forecasts directly in the bootstrap.}

The  parametric models we consider in this paper take the form $m(\mathbf{W}_{t-h};\boldsymbol{\theta}^{\dag}_{\tau_{k},h})$, where $\mathbf{W}_{t-h}$ denotes a vector of predictor variable(s) and we assume for simplicity that $\mathbf{W}_{t-h}$ is a subset of $\mathbf{V}_{t-h}$. Moreover,  $\boldsymbol{\theta}^{\dag}_{\tau_{k},h}$ is a population parameter vector that needs to be estimated in a first step, while the function  $m(\mathbf{W}_{t-h};\cdot)$ on the other hand is assumed to be a `smooth' function of the parameter vector in the sense of Assumption A6 below. For instance, 
$m(\mathbf{W}_{t-h};\boldsymbol{\theta}^{\dag}_{\tau_{k},h})$ could itself take the form of a linear quantile regression model:
\[
q_{t,h}\left(\tau_{k}|\mathbf{W}_{t-h}\right)=m(\mathbf{W}_{t-h};\boldsymbol{\theta}^{\dag}_{\tau_{k},h})=\mathbf{W}_{t-h}^{\prime}\boldsymbol{\theta}^{\dag}_{\tau_{k},h}.
\]
Alternatively, the model could also take the form of a nonlinear location scale model:
\[
q_{t,h}\left(\tau_{k}|\mathbf{W}_{t-h}\right)=m(\mathbf{W}_{t-h};\boldsymbol{\theta}^{\dag}_{\tau_{k},h})=m_{\mu}\left(\mathbf{W}_{t-h};\boldsymbol{\theta}^{\dag}_{h,\mu}\right)+\sigma\left(\mathbf{W}_{t-h};\boldsymbol{\theta}^{\dag}_{h,\sigma}\right)q_{t,h,\epsilon}\left(\tau_{k}\right),
\]
where $\boldsymbol{\theta}^{\dag}_{\tau_{k},h}=(\boldsymbol{\theta}^{\dag\prime}_{h,\mu},\boldsymbol{\theta}^{\dag\prime}_{h,\sigma},q_{t,h,\epsilon}\left(\tau_{k}\right))^{\prime}$ and $q_{t,h,\epsilon}\left(\tau_{k}\right)$ denotes the unconditional $\tau_{k}$ quantile of the error term of the location scale model. 

To accommodate both `forecasting methods' and `forecasting models', we adopt a more generic notation  in what follows and let $\mathbf{X}_{\tau_{k},t,h}(\boldsymbol{\theta}_{\tau_{k},h}^{\dag})$  stand either for the vector of  stemming from a forecasting method or for the population vector of Mincer-Zarnowitz regressors stemming from a corresponding forecasting model. On the contrary, when forecasts have been generated from a model that has been estimated through the recursive (i.e., expanding) scheme using the first $t-h$ observations of the sample, with $t=R+1, R+2,\ldots$, we write $\mathbf{X}_{\tau_{k},t,h}(\widehat{\boldsymbol{\theta}}_{\tau_{k},t,h})$ to denote the dependence on the estimated parameter vector $\widehat{\boldsymbol{\theta}}_{\tau_{k},t,h}$. Note that even though we focus on the recursive scheme in light of the applications, the rolling estimation scheme whereby a window of the last $R$ observations is used (running from $t-h-R+1$ to $t-h$) or the fixed scheme where the parameter vector is estimated only once, i.e. $\widehat{\boldsymbol{\theta}}_{\tau_{k},t,h}=\widehat{\boldsymbol{\theta}}_{\tau_{k},R+1-h,h}$, are equally compatible with our set-up and the assumptions below could be adapted in a straightforward manner. Finally, `forecasting methods' can be accommodated by assuming $\widehat{\boldsymbol{\theta}}_{\tau_{k},t,h}=\boldsymbol{\theta}_{\tau_{k},h}^{\dag}$ almost surely.

Thus, to implement the test for the null hypothesis in (\ref{EQH0}) versus the alternative hypothesis, we first estimate the coefficient vector as:
\begin{equation}\label{EQEST}
	\widehat{\boldsymbol{\beta} }_{h}(\tau_{k} )=\left(\begin{array}{c}\widehat{\alpha}_{h}(\tau_{k})\\ \widehat{\beta}_{h}(\tau_{k} )\end{array}\right)=\arg \min_{\boldsymbol{b}\in \mathcal{B}}\frac{1}{P}%
	\sum_{t=R+1}^{T}\rho _{\tau }\left( y_{t}-\mathbf{X}_{\tau_{k},t,h}(\widehat{\boldsymbol{\theta}}_{\tau_{k},t,h})^{\prime }\boldsymbol{b}\right)
\end{equation}%
for each $h$ and $\tau_{k}$. With these estimates at hand, different possibilities to construct a suitable test statistic exist. More specifically, since the number of elements in $\mathcal{H}$ and $\mathcal{T}$ is finite, one option is to construct a Wald-type test based on the estimates in (\ref{EQEST}) together with a suitable estimator of the variance-covariance matrix. However, when interest lies in testing (\ref{EQH0}) against its complement for a larger number of quantile levels and horizons, constructing a Wald test involves estimating a large variance-covariance matrix, which can be difficult in practice and may lead to a poor finite sample performance. On the other hand, as we argue below, using a moment equality based test in combination with the nonparametric bootstrap does not suffer from this drawback. In fact, the moment equality framework can be extended easily to other set-ups which give rise to an even larger number of equalities (see Section \ref{sec:extension}).

To see the possibility of a moment equality based test, note that under $H^{\text{MZ}}_{0}$ and the Assumptions A1 to A7 outlined below it holds that:
\begin{align*}
	&\sqrt{P}\left( \widehat{\boldsymbol{\beta} }^{\dag}_{h}\left( \tau_{k} \right) - \left(\begin{array}{c}\alpha_{h}^{\dag}(\tau_{k} )\\ \beta_{h}^{\dag}(\tau_{k} )\end{array}\right) \right)\\
	&\stackrel{d}{\rightarrow}N\left(\mathbf{0},\tau_{k}(1-\tau_{k})\mathbf{J}_{h}(\tau_{k} )^{-1}\mathrm{E}\left[  \mathbf{X}_{\tau_{k},t,h}(\boldsymbol{\theta}_{\tau_{k},h}^{\dag})\mathbf{X}_{\tau_{k},t,h}(\boldsymbol{\theta}_{\tau_{k},h}^{\dag})^{\prime
	}\right]\mathbf{J}_{h}(\tau_{k} )^{-1}\right)
\end{align*}
pointwise in $h$ and $\tau_{k}$, where the matrix $\mathbf{J}_{h}(\tau_{k} )$ is given by:
\begin{equation}\label{EQJMAT}
	\mathbf{J}_{h}(\tau_{k} )\equiv \mathrm{E}\left[ f_{t,h}\left( \mathbf{X}_{\tau_{k},t,h}(\boldsymbol{\theta}_{\tau_{k},h}^{\dag})^{\prime }\boldsymbol{\beta}
	^{\dag }(\tau_{k} ) \right) \mathbf{X}_{\tau_{k},t,h}(\boldsymbol{\theta}_{\tau_{k},h}^{\dag})\mathbf{X}_{\tau_{k},t,h}(\boldsymbol{\theta}_{\tau_{k},h}^{\dag})^{\prime
	}\right] 
\end{equation}%
and $f_{t,h}(\cdot)$ is defined in Assumption A4. In fact, under $H^{\text{MZ}}_{0}$, in Section \ref{proofs} of the appendix we establish the linear Bahadur representation:
\begin{align*}
	&\sqrt{P}\left( \widehat{\boldsymbol{\beta} }_{h}\left( \tau_{k} \right) - \left(\begin{array}{c}0\\ 1\end{array}\right) \right)\\
	=&\mathbf{J}_{h}\left( \tau_{k} \right)^{-1}\left(\frac{1}{\sqrt{P}}\sum_{t=R+1}^{T}\mathbf{X}_{\tau_{k},t,h}(\boldsymbol{\theta}_{\tau_{k},h}^{\dag})%
	\left( 1\left\{ y_{t}\leq \mathbf{X}_{\tau_{k},t,h}(\boldsymbol{\theta}_{\tau_{k},h}^{\dag})^{\prime }\boldsymbol{\beta}^{\dag}_{\tau_{k},h}\left( \tau_{k} \right)
	\right\} -\tau_{k} \right)\right)+o_{\mathrm{Pr}}(1).
\end{align*}
The above representation motivates the use of a moment equality type statistic for a test of autocalibration. Thus, define the set $
\mathcal{C}^{\text{MZ}}=\left\{(h,\tau_{k},j):\ h\in \mathcal{H},\ \tau_{k}\in\mathcal{T},\ j\in\{1,2\}\right\}$, and  let $\vert \mathcal{C}^{\text{MZ}}\vert=\kappa$ denote the cardinality of $\mathcal{C}^{\text{MZ}}$, while 
$s=1,\ldots,\kappa$ is a generic element from $\mathcal{C}^{\text{MZ}}$. For the test statistic, define $\widehat{m}_{s}$ either as $\widehat{\alpha}_{h}(\tau_{k})$ or as $(\widehat{\beta}_{h}(\tau_{k} )-1)$ for a specific $\tau_{k}$ and $h$. A test statistic for the null hypothesis in (\ref{EQH0}) is then given by: 
\begin{equation}\label{EQTS1}
	\widehat{U}_{\text{MZ}}=\sum_{s=1}^{\kappa} \left(\sqrt{P}\widehat{m}_{s}\right)^{2}.
\end{equation}

Note that the non-studentised statistic in (\ref{EQTS1}) above does not require estimation of the asymptotic variance, and consequently will be non-pivotal as its asymptotic distribution does depend on the full variance-covariance matrix. Heuristically, under $H^{\text{MZ}}_{0}$, and the conditions imposed in Theorem 1 below:
\begin{equation}\label{EQLIMD}
	\sqrt{P}\left(\left( \begin{array}{c}\widehat{\alpha}_{1}(\tau_{1})\\\widehat{\beta}_{1}(\tau_{1}) \\\vdots \\\widehat{\alpha}_{H}(\tau_{K})\\\widehat{\beta}_{H}(\tau_{K})\end{array}\right)-\left(\begin{array}{c}0\\ 1\\ \vdots\\0\\ 1\end{array}\right) \right)\stackrel{d}{\rightarrow}N(\mathbf{0},\boldsymbol{\Sigma}),
\end{equation}
where $\boldsymbol{\Sigma}$ is the asymptotic variance-covariance matrix of the empirical moment equalities scaled by $\sqrt{P}$, which is unknown in practice and depends on features of the data generating process (DGP). Of course, this nuisance parameter problem can be taken into account by using a suitable bootstrap procedure \citep[e.g.,][]{HLN11}. In particular, we will generate bootstrap critical values using the moving block bootstrap (MBB) of \citet{K1989}, whose formal validity for quantile regression with time series observations has been established by \citet{GLN2018}. In doing so,  we will resample the forecasts directly as the limiting distribution will be derived under the condition that $P/R\rightarrow \pi=0$, implying that forecast estimation error does not feature into the asymptotic distribution of the test statistic \citep[cf.][]{W96}.  We do so to abstract from the dependence on a particular estimator, which would require further details about the underlying forecasting models.

We generate bootstrap samples of length $P$ consisting of $K_{b}$ blocks of length $l$ such that $P=K_{b}l$. We draw the starting index $I_{j}$ of each block $1,\ldots,K_{b}$, $\{I_{j},I_{j+1},\ldots,I_{j+l}\}$,  from a discrete random uniform distribution on $[R+1,T-l]$. These indices are used to resample from $ \left\{y_{t},\widehat{y}_{\tau,t,h}\right\}_{t=R+1}^{T}$ jointly for each $\tau=\tau_1,...,\tau_K$ and $h=1,...,H$. This way we generate $B$ bootstrap samples, each with $ \left\{y_{t}^b,\widehat{y}^b_{\tau,t,h}\right\}_{t=R+1}^{T}$ for all $\tau=\tau_1,...,\tau_K$ and $h=1,...,H$. For each bootstrap sample, we construct bootstrap equivalents of (\ref{EQEST}) and then the corresponding bootstrap statistic:
\begin{equation}\label{EQBS}
	\widehat{U}^{b}_{\text{MZ}}=\sum_{s=1}^{\kappa} \left(\sqrt{P}(\widehat{m}^{b}_{s}-\widehat{m}_{s})\right)^{2}.
\end{equation}
The critical value is then given by the $(1-\alpha)$ quantile of the empirical bootstrap distribution of $\widehat{U}^{b}_{\text{MZ}}$ over $B$ draws, say $c_{B,P,(1-\alpha)}$. 

\subsection{Assumptions and Asymptotic Validity}

For the asymptotic validity of this procedure, we make the following assumptions:\medskip

\noindent \textbf{A1}: The outcome variable $y_{t}$ is strictly stationary and $\beta$-mixing with the mixing coefficient satisfying $\beta(k)=O\left(k^{-\frac{\epsilon}{\epsilon-1}}\right)<\infty$ for $\epsilon>1$.\medskip

\noindent \textbf{A2}: For all $h\in \mathcal{H}$, $\tau_{k}\in \mathcal{T}$, and $\boldsymbol{\theta}\in \boldsymbol{\Theta}$, $\mathbf{X}_{\tau_{k},t,h}(\boldsymbol{\theta})$ is strictly stationary, and satisfies the mixing condition from A1 as well as $\mathrm{E}\left[\left\Vert \mathbf{X}_{\tau_{k},t,h}(\boldsymbol{\theta})\right\Vert^{2\epsilon+2}\right]<\infty$, where $\Vert\cdot\Vert$ denotes the Euclidean norm and  $\boldsymbol{\Theta}$ is defined in A6 below. The distribution of $\mathbf{X}_{\tau_{k},t,h}(\boldsymbol{\theta})$ is absolutely continuous with Lebesgue density.\medskip

\noindent \textbf{A3}: For every $\tau_{k}\in\mathcal{T}$ and $h\in\mathcal{H}$, assume that the parameter space of $\boldsymbol{\beta}_{h} (\tau_{k} )$, $\mathcal{B}$, is a compact and convex set. For every $\tau_{k}\in\mathcal{T}$ and $h\in\mathcal{H}$, the coefficient vector $\boldsymbol{\beta}_{h}^{\dag} (\tau_{k} )$ from (\ref{EQPLIM}) lies in the interior of $\mathcal{B}$.\medskip

\noindent \textbf{A4}: For all $h\in\mathcal{H}$ and $\tau_{k}\in \mathcal{T}$, the conditional distribution function of $y_{t}$ (given $\mathbf{X}_{\tau_{k},t,h}(\boldsymbol{\theta}^{\dag}_{\tau_{k},h})$), $F(\cdot|\mathbf{X}_{\tau_{k},t,h}(\boldsymbol{\theta}^{\dag}_{\tau_{k},h}))\equiv F_{t,h}(\cdot) $, admits a continuous Lebesgue density, $ f(\cdot|\mathbf{X}_{\tau_{k},t,h}(\boldsymbol{\theta}))\equiv f_{t,h}(\cdot)$, which is bounded away from zero and infinity for all $u$ in $\mathcal{U}=\{u:0<F_{t,h}(u)<1\}$  almost surely. For all $h$, the density $f_{t,h}(\cdot)$ is integrable uniformly over $\mathcal{U}$.\medskip

\noindent \textbf{A5}: For every $h\in\mathcal{H}$ and $\tau_{k}\in\mathcal{T}$, assume that the matrix $\boldsymbol{J}_{h}(\tau_{k})$ defined in (\ref{EQJMAT}) is positive definite and that:
\[
\mathrm{E}\left[\mathbf{X}_{\tau_{k},t,h}(\boldsymbol{\theta}_{\tau_{k},h}^{\dag})%
\left( 1\left\{ y_{t}\leq \mathbf{X}_{\tau_{k},t,h}(\boldsymbol{\theta}_{\tau_{k},h}^{\dag})^{\prime }\boldsymbol{\beta}^{\dag}_{\tau_{k},h}\left( \tau_{k} \right)
\right\} -\tau_{k} \right)\right]=\mathbf{0}.
\] 
\medskip

\noindent \textbf{A6}: Assume that $\boldsymbol{\Theta}$ is compact and that, for each $\tau_{k}\in\mathcal{T}$ and $h\in\mathcal{H}$, $\boldsymbol{\theta}_{\tau_{k},h}^{\dag}$ lies in its interior. For all $\boldsymbol{\theta}_{1},\boldsymbol{\theta}_{2}\in\boldsymbol{\Theta}$ and $t$, it holds that:
\[
\Vert\mathbf{X}_{\tau_{k},t,h}(\boldsymbol{\theta}_{1})-\mathbf{X}_{\tau_{k},t,h}(\boldsymbol{\theta}_{2}) \Vert \leq B(\mathbf{X}_{\tau_{k},t,h})\Vert \boldsymbol{\theta}_{1}-\boldsymbol{\theta}_{2}\Vert 
\]
for some non-negative, real-valued function $B(\mathbf{X}_{\tau_{k},t,h})$ satisfying $\mathrm{E}[ B(\mathbf{X}_{\tau_{k},t,h})^2 ]<\infty$. In addition, assume that for every $h\in\mathcal{H}$ and $\tau_{k}\in\mathcal{T}$, the estimator $\widehat{\boldsymbol{\theta}}_{\tau_{k},t,h}$ satisfies:
\[
\sup_{t\geq R+1}\Vert \widehat{\boldsymbol{\theta}}_{\tau_{k},t,h}-\boldsymbol{\theta}_{\tau_{k},h}^{\dag} \Vert =O_{\mathrm{Pr}}\left(\frac{1}{\sqrt{R}}\right) .
\]

\medskip
\noindent \textbf{A7}: Assume that $R,P,l\rightarrow \infty$ as $T\rightarrow \infty$ with $P/R\rightarrow \pi= 0$ and $l/P\rightarrow 0$. 

\medskip

Assumption A1 imposes some mild restrictions on the time dependence of the data that are in turn linked to the existence and finiteness of corresponding moments in A2. On the other hand, the continuity of $\mathbf{X}_{\tau_{k},t,h}(\boldsymbol{\theta})$ for any given $\tau_{k}$, $h$, and $\boldsymbol{\theta}$ only serves the purpose to simplify some of the arguments in the proofs of Section \ref{proofs} in the appendix, and could be relaxed at the expense of more cumbersome notation.  Assumptions A3-A5 are required to derive the limiting distribution of the linear quantile regression estimator \citep[e.g., ][]{KX06,GMO2011}. In fact, A3 and A4 represent standard assumptions on the parameter space and the smoothness of the (conditional) distribution of $y_{t}$. A5 ensures asymptotic normality of the quantile regression estimator, with the moment condition representing the quantile equivalent of the well known orthogonality condition from ordinary least squares \citep{KW2003}.   Assumption A6 on the other hand is only needed when the focus lies on forecasting models. Specifically, it places restrictions on the underlying parametric models, but is in fact compatible with commonly used location scale or linear quantile regression models that satisfy the Lipschitz condition in A6 and that can be estimated at rate $\sqrt{R}$. In particular, as \citet{KX2009} propose a two-step estimation procedure for linear GARCH models based on linear quantile regression, our set-up also comprises the latter type of models that are frequently used in finance applications. Finally, Assumption A7 governs the rates at which $P$ and $R$ as well as the block length $l$ may grow to infinity. In particular, and in analogy to \citet{W96}, we require $\pi=0$ for estimation error to be ignorable asymptotically and to be able to resample directly from the forecasts (rather than to resample from the realised predictors). In turn, this allows us to focus on miscalibration as a structural feature of the models. 

We are now ready to derive the asymptotic properties of the statistic under the null hypothesis:

\begin{theorem}\label{MZTest} Assume that A1 to A7 hold, and that $\boldsymbol{\Sigma}\in\mathbb{R}^{\kappa\times\kappa}$ is positive definite. Then under $H^{\text{MZ}}_{0}$: 
	\[
	\lim_{T,B\rightarrow \infty}\Pr\left(\widehat{U}_{\text{MZ}}>c_{B,P,(1-\alpha)}\right)=\alpha.
	\]
	
\end{theorem}

Theorem \ref{MZTest} establishes the asymptotic size control of the moment equality test. It is easy to implement using the moving block bootstrap and performs very well in terms of finite sample size and power, which we assess through several simulations. These can be found in Section \ref{sec:simulations} of the appendix. There we provide two contrasting simulation set-ups in subsections \ref{sec:sim-MZ} and \ref{sec:sim-Fin} to match both the macroeconomic and financial applications below, which differ in sample sizes, target quantile levels and time series characteristics (we choose AR and GARCH processes as DGPs). In terms of block length choice, we find that values like $l=10$ work well in financial applications with thousands of daily observations and a lower length like $l=4$ is adequate in macroeconomic situations with only hundreds of monthly observations. However, the results do not change significantly when tweaking the block length. We also provide additional simulations in Subsection \ref{sim-aug} for augmented quantile Mincer-Zarnowitz test that we discuss next.

\section{Extensions}\label{sec:extension}

In the following two subsections, we will describe two extensions of the test from Section \ref{sec:testing}, firstly to accommodate additional predictors $\mathbf{Z}_{t-h}$ in the Mincer-Zarnowitz regression to test different forms of optimality, and secondly  to test for autocalibration of quantile forecasts for multiple time series simultaneously.

\subsection{Augmented Quantile Mincer-Zarnowitz Test}

Recalling the characterisation of optimality relative to any information set $\mathcal{I}_{t-h}\subset \mathcal{F}_{t-h}$ from \eqref{char_optimality}, it becomes clear that the Mincer-Zarnowitz set-up from the previous section may also be used to test stronger forms of optimality with respect to larger information sets than $\sigma(\widehat{y}_{\tau_{k},t,h})$. More precisely, while $H_{0}^{MZ}$ vs. $H_{1}^{MZ}$ is a test of autocalibration, it does not check if all available valuable information from $\mathcal{F}_{t-h}$ was incorporated into the forecasting model or taken into account by the forecaster. We therefore suggest the idea of augmented quantile Mincer-Zarnowitz regressions, where a vector of additional regressors $\mathbf{Z}_{t-h} \in \mathcal{F}_{t-h}$ is added to the regression model in (\ref{POPQREG}) to test for optimality relative to $\sigma(\widehat{y}_{\tau_{k},t,h},\mathbf{Z}_{t-h})$, see also \citet[chapter 15]{ET2016} for a discussion of augmented Mincer-Zarnowitz regressions in the context of mean forecasts.  

That is, in analogy to the previous section, we again specify a linear quantile regression model for every horizon $h\in\mathcal{H}$ and $\tau_{k}\in\mathcal{T}$ as follows:
\begin{equation}\label{AUGPOPQREG}
	y_{t}=\alpha_{h}^{\dag}(\tau_{k})+ \widehat{y}_{\tau_{k},t,h}\beta_{h}^{\dag}(\tau_{k} ) + \mathbf{Z}_{t-h}^{\prime }\boldsymbol{\gamma}_{h}^{\dag} (\tau_{k} ) + \varepsilon_{t,h}(\tau_{k}),
\end{equation}%
where with slight abuse of notation we use the same symbols as in the previous section for the first two regression coefficients as well as the error term, and suppress the possible dependence of the forecasts on some parameter vector $\boldsymbol{\theta}^{\dag}$.  If the coefficients of $\mathbf{Z}_{t-h}$ in the population augmented Mincer-Zarnowitz regression are non-zero, i.e. $\boldsymbol{\gamma}_{h}^{\dag} (\tau_{k})\neq \boldsymbol{0}$, there is valuable information in $\mathbf{Z}_{t-h}$ that has not been incorporated into the forecasts yet. As a result, those variables or a subset thereof should be included into the model to improve forecast accuracy.  

Formally, the null hypothesis we test is given by:
\begin{equation}\label{AUGEQH0}
	H^{\text{AMZ}}_{0}: \{\alpha_{h}^{\dag}(\tau_{k})= 0 \} \cap  \{ \beta_{h}^{\dag}(\tau_{k} )=1\} \cap  \{\boldsymbol{\gamma}_{h}^{\dag} (\tau_{k} )= \boldsymbol{0} \}
\end{equation}
for all $h\in\mathcal{H}$ and $\tau_{k}\in\mathcal{T}$ versus $H^{\text{AMZ}}_{1}:  \{\alpha_{h}^{\dag}(\tau_{k})\neq  0 \}$ and/or $\{ \beta_{h}^{\dag}(\tau_{k} )\neq 1\}$ and/or $\{\boldsymbol{\gamma}_{h}^{\dag} (\tau_{k} ) \neq \boldsymbol{0} \}$ for at least some $h\in\mathcal{H}$ and $\tau_{k}\in\mathcal{T}$.

In contrast to a standard Mincer-Zarnowitz quantile regression, the augmented version requires a choice of variables from $\mathcal{F}_{t-h}$. These variables included in $\mathbf{Z}_{t-h}$ have to be chosen \textit{a priori} and may, in some situations,  suggest themselves naturally as  illustrated in our macroeconomic application. In other situations, however, it might be hard to pick those variables from a potentially very large information set. In those cases, regularised or factor-augmented Mincer-Zarnowitz regressions could be used instead of (\ref{AUGPOPQREG}). We leave this extension to future research and state the asymptotic size result of a moment equality based test of $H^{\text{AMZ}}_{0}$ vs. $H^{\text{AMZ}}_{1}$. Specifically, let $\widehat{\widetilde{m}}_{s}$ denote the quantile estimators $\widehat{\alpha}_{h}(\tau_{k})$, $(\widehat{\beta}_{h}(\tau_{k} )-1)$, or $\widehat{\boldsymbol{\gamma}}_{h}(\tau_{k})$ for a specific quantile level $\tau_{k}$ and horizon $h$. Similarly, in analogy to before, let $\widetilde{\kappa}$ denote the total (finite) number of moment equalities to be tested. The corresponding test statistic, $\widehat{U}_{\text{AMZ}}$, is given by:
\begin{equation}\label{EQATS1}
	\widehat{U}_{\text{AMZ}}=\sum_{s=1}^{\kappa} \left(\sqrt{P}\widehat{\widetilde{m}}_{s}\right)^{2}.
\end{equation}

Finally, define the `augmented' covariate   and coefficient vectors:
\[
\widetilde{\mathbf{X}}_{\tau_{k},t,h}(\boldsymbol{\theta}_{\tau_{k},h}^{\dag} )=(\mathbf{X}_{\tau_{k},t,h}(\boldsymbol{\theta}_{\tau_{k},h}^{\dag} )^{\prime},\mathbf{Z}_{t-h}^{\prime })^{\prime}
\] 
and $\widetilde{\boldsymbol{\beta}}_{h}^{\dag}(\tau_{k})=(\alpha^{\dag}_{h}(\tau_{k}),\beta^{\dag}_{h}(\tau_{k}),\boldsymbol{\gamma}^{\dag}_{h}(\tau_{k})^{\prime})^{\prime}$, respectively, the latter with parameter space $\widetilde{\mathcal{B}}$. We obtain the following result for the  test defined by (\ref{EQATS1}):

\begin{corollary} Assume that A1 to A7 hold with $\mathbf{X}_{\tau_{k},t,h}(\cdot)$, $\boldsymbol{\beta}_{h}^{\dag}(\tau_{k})$, and $\mathcal{B}$ replaced by $\widetilde{\mathbf{X}}_{\tau_{k},t,h}(\cdot)$, $\widetilde{\boldsymbol{\beta}}_{h}^{\dag}(\tau_{k})$, and $\widetilde{\mathcal{B}}$, respectively.  Moreover, assume that:
	\[
	\plim_{P\rightarrow \infty} \mathrm{Var}\left[\sqrt{P}\left(\begin{array}{c} \widehat{\widetilde{m}}_{1}\\\vdots \\\widehat{\widetilde{m}}_{\overline{\kappa}}\end{array}\right)\right]\equiv \widetilde{\mathbf{\Sigma}}\in\mathbb{R}^{\widetilde{\kappa}\times\widetilde{\kappa}}
	\]
	is positive definite,	where $\mathrm{Var}\left[\cdot\right]$ denotes the variance operator. Then under $H^{\text{AMZ}}_{0}$: 
	\[
	\lim_{T,B\rightarrow \infty}\Pr\left(\widehat{U}_{\text{AMZ}}>c_{B,P,(1-\alpha)}\right)=\alpha.
	\]
	
\end{corollary}

\subsection{Multivariate Quantile Mincer-Zarnowitz Test}\label{ssec:AugmentedMZTest}

While Section \ref{sec:testing} establishes a test for autocalibration of quantile forecasts of a single time series $y_{t}$, researchers may sometimes be interested in testing this property  across several time series $\mathbf{y}_{t}=(y_{1,t},\ldots,y_{G,t})^{\prime}$ using an array of $h$ step ahead $\tau_{k}$-level forecasts $\left(\widehat{\mathbf{y}}_{\tau,t,h}\right)_{\tau=\tau_1,...,\tau_K,h=1,...,H}$, where $\widehat{\mathbf{y}}_{\tau_{k},t,h}=(\widehat{y}_{1,\tau_{k},t,h},\ldots,\widehat{y}_{G,\tau_{k},t,h})^{\prime}$, where $h\in\mathcal{H}$, $\tau_{k}\in\mathcal{T}$, and finite $G \in \mathbb{N}$. For instance, we may be interested in testing for forecast autocalibration jointly across different industries or sectors; components of GDP growth like consumption or export growth; or different macro series like in our application in Section \ref{sec:empirical_macro}. 

We now sketch the extension of the autocalibration test to such a multivariate set-up. To this end,  consider again the following linear quantile regression model:
\begin{equation}\label{EQCLUSTER}
	y_{i,t}=\alpha_{i,h}^{\dag}(\tau_{k})+ \widehat{y}_{i,\tau_{k},t,h}\beta_{i,h}^{\dag}(\tau_{k} ) + \varepsilon_{i,t,h}(\tau_{k}),\quad h\in\mathcal{H}, \quad \tau_{k}\in\mathcal{T}, \quad i=1,\ldots,G.
\end{equation}%
Here, for each group $i\in\{1,\ldots,G\}$, the coefficient vector $\boldsymbol{\beta}^{\dag}_{i,h} (\tau_{k} )=(\alpha^{\dag}_{i,h}(\tau_{k}),\beta^{\dag}_{i,h}(\tau_{k} ) )^{\prime}$ is defined as:
\begin{equation}\label{EQPLIMPANEL}
	\boldsymbol{\beta}^{\dag}_{i,h} (\tau_{k} )=\arg \min_{\boldsymbol{b}_{i} \in \mathcal{B}}\mathrm{E}\left[
	\left(\rho _{\tau_{k}  }\left( y_{i,t}-\mathbf{X}_{i,\tau_{k},t,h}(\boldsymbol{\theta}^{\dag}_{i,\tau_{k},h})^{\prime }\boldsymbol{b}_{i} \right)\right) \right]
\end{equation}
with $\mathbf{X}_{i,\tau_{k},t,h}(\boldsymbol{\theta}^{\dag}_{i,\tau_{k},h})^{\prime}=(1, \widehat{y}_{i,\tau_{k},t,h})^{\prime}$.\footnote{Note that we use $\boldsymbol{\theta}_{i}^{\dag}$ to denote the possible dependence of the forecast for series $i$ on a forecasting model.} Therefore, the set-up in (\ref{EQCLUSTER}) allows for miscalibration also at the individual time series level (e.g., industry or sector) if for some $i$,$h$, and $\tau_{k}$ it holds that $\alpha_{i,h}(\tau_{k})\neq 0$ and/or $\beta_{i,h}(\tau_{k})\neq 1$, respectively. 

The sample analogue of (\ref{EQPLIMPANEL}), using the evaluation sample of observations $\{y_{t}\}_{t=R+1}^{T}$ and array-valued forecasts $\widehat{y}_{i,\tau,t,h}$ for each $i$ is given by:
\begin{equation}\label{EQESTPANEL}
	\widehat{\boldsymbol{\beta} }_{i,h}(\tau_{k} )=\arg \min_{\boldsymbol{b}_{i}\in \mathcal{B}}\frac{1}{P}%
	\sum_{s=R+1}^{T}
	\left(\rho _{\tau }\left( y_{i,s}-\mathbf{X}_{i,\tau_{k},s,h}(\widehat{\boldsymbol{\theta}}_{i,\tau_{k},s,h})^{\prime }\boldsymbol{b}_{i}\right)\right).
\end{equation}%
As in Section \ref{sec:testing}, we are interested in testing the composite null hypothesis:
\begin{equation}\label{EQH0P}
	H^{\text{MMZ}}_{0}: \{\alpha_{i,h}(\tau_{k})= 0 \} \cap  \{ \beta_{i,h}(\tau_{k} )=1\}
\end{equation}
for all $h\in\mathcal{H}$, $\tau_{k}\in\mathcal{T}$, $i=1,\ldots,G$, versus $H^{\text{MMZ}}_{1}:  \{\alpha_{i,h}(\tau_{k})\neq  0 \}$ and/or $\{ \beta_{i,h}(\tau_{k} )\neq 1\}$ for at least some $h$, $\tau_{k}$, and $i$. Note that, since $G$ is finite,  for a given horizon $h$ and quantile level $\tau_{k}$ (with A1-A7 adapted to hold for $y_{i,t}$ and $\mathbf{X}_{i,\tau_{k},t,h}(\boldsymbol{\theta}_{i,\tau_{k},h}^{\dag})$, $i=1,\ldots,G$), the estimator in (\ref{EQESTPANEL}) is consistent for $\boldsymbol{\beta}^{\dag}_{i,h} (\tau_{k} )$ for each $i$. The test statistic for the null hypothesis in (\ref{EQH0P}) versus its complement is therefore given by $\widehat{U}_{\text{MMZ}}=\sum_{s=1}^{\overline{\kappa}} \left(\sqrt{P}\widehat{\overline{m}}_{s}\right)^{2}$,
where either $\widehat{\overline{m}}_{s}=\widehat{\alpha}_{i,h}(\tau_{k})$ or $\widehat{\overline{m}}_{s}=\widehat{\beta}_{i,h}(\tau_{k})-1$, respectively, and in a similar way as above, $\overline{\kappa}$ denotes the total number of moment conditions which in this case accounts for the $G$ different series being used in the test.

In order to construct a suitable bootstrap statistic in analogue to Subsection  \ref{ssec:teststat}, we construct bootstrap analogues $\widehat{\boldsymbol{\beta}}^{b}_{i,h}(\tau_{k})$ of (\ref{EQESTPANEL}) from bootstrap samples of length $P=K_{b}l$ from $K_{b}$ blocks of length $l$ by resampling again from the series of forecast-observation pairs, where the forecasts in this case are array-valued. The bootstrap procedure does not just take the horizon, but also the group structure as given, which ensures that the dependence of the original data across horizons $h$ as well as across series $i=1,\ldots,G$ is maintained. More precisely, we again draw the starting index $I_{j}$ of each block of forecasts and observations $1,\ldots,K_{b}$ from a discrete random uniform distribution on $[R+1,T-l]$. These indices are used to resample from $ \left\{y_{t},\left(\widehat{\mathbf{y}}_{\tau,t,h}\right)_{\tau=\tau_1,...,\tau_K,h=1,...,H}\right\}_{t=R+1}^{T}$. This way we generate $B$ bootstrap samples, each with $ \left\{y_{t}^b,\left(\widehat{\mathbf{y}}^b_{\tau,t,h}\right)_{\tau=\tau_1,...,\tau_K,h=1,...,H}\right\}_{t=R+1}^{T}$.

For each $i=1,\ldots,G$, we then construct a corresponding bootstrap estimator given by:
\[
\widehat{\boldsymbol{\beta} }^{b}_{i,h}(\tau_{k} )=\arg \min_{\boldsymbol{b}_{i}\in \mathcal{B}}\frac{1}{P}%
\sum_{s=R+1}^{T}\left(\rho _{\tau }\left( y_{i,s}^{b}-\mathbf{X}_{i,\tau_{k},s,h}(\widehat{\boldsymbol{\theta}}_{i,\tau_{k},s,h}^{b})^{\prime}\boldsymbol{b}_{i}\right)\right).
\]
The final bootstrap statistic becomes $\widehat{U}^{b}_{\text{MMZ}}=\sum_{s=1}^{\overline{\kappa}} \left(\sqrt{T}(\widehat{\overline{m}}^{b}_{s}-\widehat{\overline{m}}_{s})\right)^{2}$, where $\widehat{\overline{m}}^{b}_{s}$ is equal to  $\widehat{\alpha}^{b}_{i,h}(\tau_{k})$ or $\widehat{\beta}^{b}_{i,h}(\tau_{k})-1$, respectively. Constructing critical values on the basis of $\widehat{U}^{b}_{\text{MMZ}}$, $b=1,\ldots,B$, as in Subsection \ref{ssec:teststat}, the following corollary holds:

\begin{corollary}\label{testMMZ}
	
	Assume that Assumptions A1 to A7 hold  with $y_{t}$, $\mathbf{X}_{\tau_{k},t,h}(\boldsymbol{\theta}_{\tau_{k},h}^{\dag})$, and $\widehat{\boldsymbol{\theta}}_{\tau_{k},t,h}$ replaced by $y_{i,t}$, $\mathbf{X}_{i,\tau_{k},t,h}(\boldsymbol{\theta}_{i,\tau_{k},h}^{\dag})$, and $\widehat{\boldsymbol{\theta}}_{i,\tau_{k},t,h}$, respectively, for every $i=1,\ldots,G$. Moreover, assume that:
	\[
	\plim_{P\rightarrow \infty} \mathrm{Var}\left[\sqrt{P}\left(\begin{array}{c} \widehat{\overline{m}}_{1}\\\vdots \\\widehat{\overline{m}}_{\overline{\kappa}}\end{array}\right)\right]\equiv \overline{\mathbf{\Sigma}}\in\mathbb{R}^{\overline{\kappa}\times\overline{\kappa}},
	\]
	is positive definite, where $\mathrm{Var}\left[\cdot\right]$ denotes the variance operator. Then,  under $H^{\text{MMZ}}_{0}$: 
	\[
	\lim_{T,B\rightarrow \infty}\Pr\left(\widehat{U}_{\text{MMZ}}>c_{B,P,(1-\alpha)}\right)=\alpha.
	\]

\end{corollary}

\section{Empirical Applications}\label{sec:empirical}

In this section, we provide two empirical applications. The first is a finance application applying the MZ test to test the optimality of GARCH predictions of the tail quantiles of financial returns. The second application uses the MZ test extensions to assess the optimality of GaR forecasts made across a range of U.S. macroeconomic variables.

\subsection{Empirical Application 1: Financial Returns}\label{sec:empirical_finance}

Forecasts of lower tail quantiles of returns distributions (or upper tail quantiles of loss distributions) play an important role in financial risk management as the most prominent risk measures are either themselves tail quantiles or defined in terms of tail quantiles (see \cite{he2022} for a recent overview). In particular, the VaR at level $\tau$ is just the $\tau$-quantile of the returns $y_{t}$, $VaR(\tau)=q(\tau)$, where usually $\tau$ is chosen to be either 0.05 or 0.01. Expected shortfall as well as median shortfall are also defined in terms of quantiles of the return distribution and our multi-quantile evaluation framework can be useful in the evaluation of those risk measures as well (see Subsection \ref{subsec:expected_shortfall} of the appendix for a discussion). Producing and backtesting VaR forecasts is thus a central task in financial risk management. Therefore, as discussed in the introduction, the majority of contributions to quantile forecast optimality testing was motivated by this problem. However, the literature focused on single-quantile forecasts, while it may be of interest to check VaR at both the 0.01 and the 0.05 level, or even a grid of several VaR levels to approximate the whole tail distribution. In addition, note that risk management requires forecasts of risk measures over multiple horizons, e.g.\ for one day ahead and cumulative losses over the next ten trading days. Nevertheless, extant evaluation methods focus on a single horizon (with the noteworthy exception of \cite{barendse2023} who also consider multi-horizon evaluation of VaR and expected shortfall) and consequently one-day-ahead forecasts are usually evaluated. Our tests solve those problems as they enable joint evaluation of VaR forecasts over multiple levels and horizons.

To illustrate the use of the Mincer-Zarnowitz test for the evaluation of forecasts of financial risk measures, we apply it to multi-horizon, multi-quantile forecasts for daily S\&P 500 returns. We consider horizons from $h=1$ through $h=10$ in terms of trading days and three quantile levels $\tau\in\{0.01,0.025,0.05\}$. The  classic model for return volatility and VaR forecasting is the GARCH(1,1) model \citep{bollerslev1986}. As no closed-form formula for multi-period-ahead GARCH quantile forecasts is available, except for the case of Gaussian innovations, we use the GARCH bootstrap of \cite{pascual2006}. It draws standardised residuals from the estimated one-period-ahead model to simulate draws multiple periods in the future, from which quantiles can be obtained.\footnote{We use the implementation of the GARCH bootstrap from the \texttt{rugarch} package in R \citep{ghalanos2014}.} We choose student-$t$ errors for the estimation of the model. 

Our sample consists of daily S\&P 500 returns from January 3rd 2000 to June 27th 2022, amounting to 5634 observations.\footnote{Data taken from the Oxford-Man Realized Library: https://realized.oxford-man.ox.ac.uk/data/download [Last accessed: 05/07/22]} We use recursive pseudo-out-of-sample forecasting with an initial estimation window of size 3000 for $h=10$ or, in other words, $R=3009$, leading to an evaluation sample of size $P=2625$. Figure \ref{fig:forecastploth1} in the appendix displays the one-day ahead forecasts for the three quantiles and the realisations. The forecasts for the other horizons look very similar, but are expectedly a bit wider. 

We first use our Mincer-Zarnowitz test over the three quantiles, $\mathcal{T}=\{0.01,0.025,0.05\}$, and ten horizons, $\mathcal{H}=\{1,...,10\}$. We use $B=1000$ bootstrap draws and a block length of $l=10$. Table \ref{tab:MZ_Finance_Application} presents the results. With a $p$-value of 0.01 there is clear evidence against the null of autocalibration. Alternative block length choices of $l=5$ and $l=20$ lead to very similar $p$-values (0.011 and 0.015) which is promising in that the results are insensitive to block length. 

\begin{table}[!htb]
	\centering
	\caption{Mincer-Zarnowitz Test Results, Finance Application}
	\begin{tabular}{lccccc}
		\toprule
		& Stat     & 90\%      & 95\%      & 99\%      & p-value \\
		\midrule
		& 9834.131 & 5009.153 & 6569.754 & 9821.539 & 0.01  \\
		\bottomrule
	\end{tabular}
	\label{tab:MZ_Finance_Application}
\end{table}

As the test statistic from \eqref{EQTS1} can directly be interpreted as an empirical distance from the null, consisting of scaled (by $\sqrt{P}$), squared deviations of all the Mincer-Zarnowitz regression coefficients from their values under the null, we can also look at the individual contributions to this statistic from single quantiles and single horizons or single quantile-horizon combinations, displayed in Table \ref{tab:MZ_Finance_Application_Ind_Cont}.   From this table a clear picture emerges. The outer quantiles and the longer forecast horizons contribute more to the test statistic, and thus show stronger evidence for miscalibration. Since risk management is typically concerned about the performance of a certain risk model such as the GARCH(1,1) across a range of quantile levels or horizons, this also demonstrates that a common practice to evaluate those models only for a specific choice of the latter may lead to incorrect conclusions about the overall performance of the prediction model.\footnote{In fact, Table \ref{tab:MZ_Finance_Application_pValues} in Section \ref{app:Finance_Application} of the appendix, which contains the $p$-values for individual autocalibration tests at given values $\tau$ and $h$, illustrates that such `telescoping' practice may indeed be misleading as there is no strong evidence against autocalibration from some individual level quantile horizon combinations.}

\begin{table}[!htb]
	\centering
	\caption{Individual Contributions to Test Statistic, Mincer-Zarnowitz Test, Finance Application}
	\begin{tabular}{lcccc}
		\toprule
		& $\tau=0.01$ & $\tau=0.025$ & $\tau=0.05$ & Sum   \\
		\midrule
		$h=1 $ & 427.463  & 81.455    & 39.217   & 548.135  \\
		$h=2 $ & 439.467  & 195.77    & 50.126   & 685.363  \\
		$h=3 $ & 672.67   & 266.256   & 127.524  & 1066.45  \\
		$h=4 $ & 591.907  & 265.84    & 99.559   & 957.306  \\
		$h=5 $ & 549.574  & 431.091   & 141.886  & 1122.551 \\
		$h=6 $ & 553.68   & 431.926   & 114.26   & 1099.866 \\
		$h=7 $ & 149.554  & 291.555   & 230.722  & 671.831  \\
		$h=8  $& 258.922  & 298.486   & 223.656  & 781.063  \\
		$h=9 $ & 560.313  & 405.563   & 402.132  & 1368.008 \\
		$h=10$ & 497.402  & 562.498   & 473.658  & 1533.558 \\
		Sum  & 4700.952 & 3230.439  & 1902.74  & 9834.131 \\
		\bottomrule
	\end{tabular}
	\label{tab:MZ_Finance_Application_Ind_Cont}
\end{table}

The tests may also convey information about how models could be improved by a closer look at the Mincer-Zarnowitz regression lines themselves.  For instance, using $h=1$ and $\tau=0.01$ as an illustrative example since all estimated intercepts are negative and all slopes of the regression lines less than  one (see Subsection \ref{ssec:AdditionalMaterial} of the appendix), Figure \ref{fig:mzplot} shows the scatter plot of forecast-observation pairs alongside the estimated Mincer-Zarnowitz regression line and the diagonal. The latter represents the population regression line under $H_{0}^{MZ}$, in other words when $\alpha_{1}^{\dag}(0.01)=0$ and $\beta_{1}^{\dag}(0.01)=1$, respectively. The discrepancy between the Mincer-Zarnowitz regression line and the diagonal thus suggests that the forecasts are in fact mis-calibrated. Contrasting the two, it becomes clear that in calmer times (when the forecasts and realizations are less extreme, i.e. closer to 0) the GARCH(1,1) forecasts tend to under-predict the actual risk, in other words the quantile forecasts are not extreme enough, while in more volatile times the forecasts tend to overestimate risk. 

\begin{figure}
	\centering
	\caption{Scatter Plot of Forecast-Realisation Pairs with Mincer-Zarnowitz Regression Line (red) and Diagonal (orange) for $h=1$ and $\tau=0.01$}
	\label{fig:mzplot}
	\includegraphics[width=0.7\linewidth]{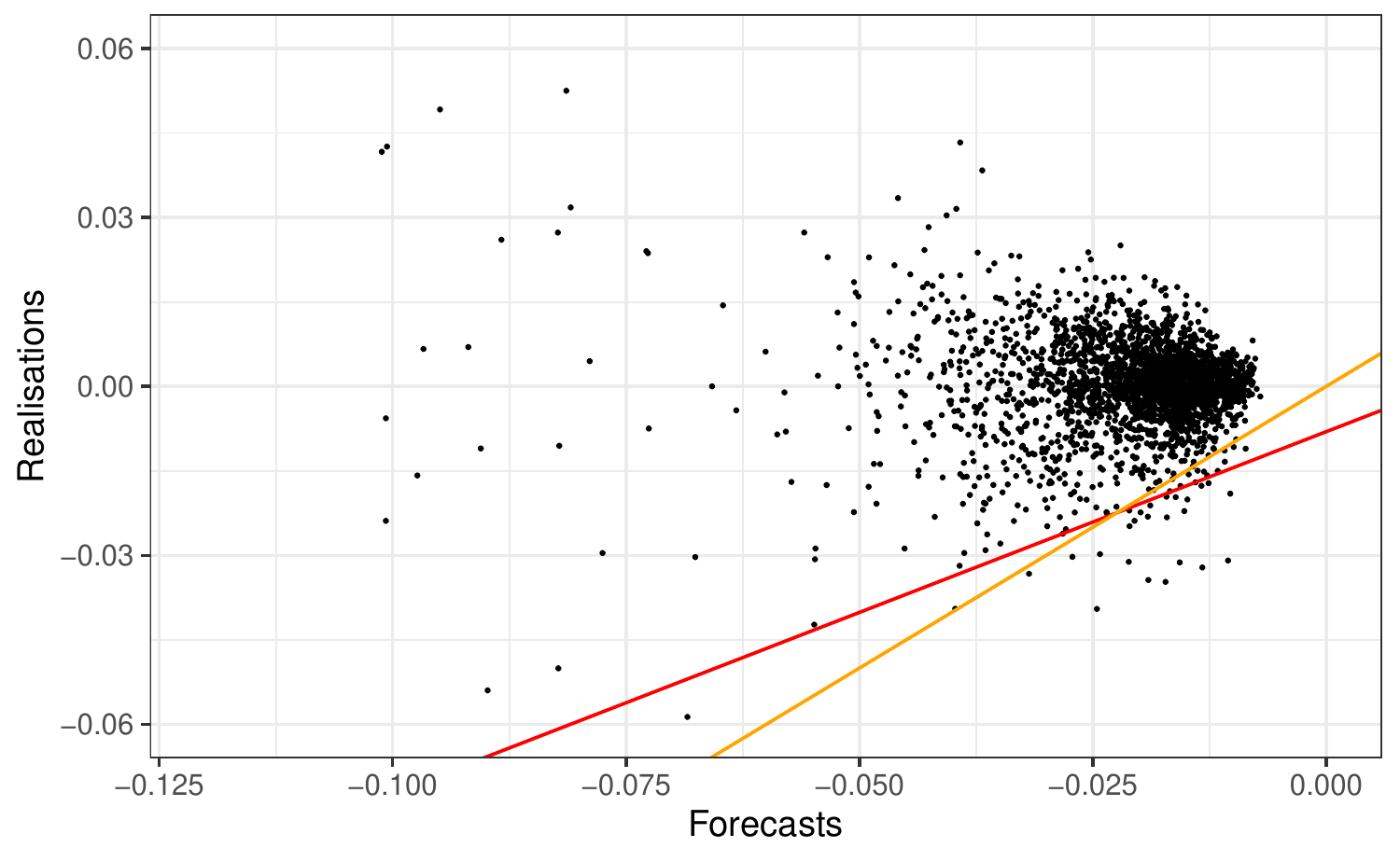}
\end{figure}

Finally, in Subsection \ref{subsec:robustness_rolling} of the appendix, we examine the robustness of our results against various specification changes and examine other popular calibration tests designed for single horizon and quantile pairs. Specifically, we find that the above results do not change qualitatively when we alter the specification to a different estimation scheme (rolling) and to smaller estimation window sizes. Similarly, results remained also unchanged when omitting the COVID-19 period, or when experimenting with the GJR-GARCH model \citep{glosten1993}. Finally, when looking at the autocalibration tests of \citet{engle2004caviar} with the respective quantile forecast as regressor and the test of \citet{christoffersen1998}, we find that the former provides $p$-values that are very close to the individual level $p$-values of the Mincer-Zarnowitz test, while this is not the case for the test of \citet{christoffersen1998} which tests  different implications of full optimality. 

\subsection{Empirical Application 2: U.S. Macro Series}\label{sec:empirical_macro}

In this section we use our tests to explore the optimality of model-based forecasts of various U.S. macroeconomic series. The analysis of quantile forecasts for macroeconomic series has become widespread since studies like \citet{M15}. More recently, the GaR literature has emerged to provide a tool to monitor downside risk to economic growth using quantile predictions. This approach typically analyses quarterly real GDP growth using financial conditions indicators (see \citealp{ABG19}), and has been subsequently applied to other quarterly macro series like employment and inflation by \citet{AABG21}.

However, in spite of the increasing interest in quantile forecasting in macroeconomics, none of these papers subject their models to the type of forecast optimality test we develop in this paper. We aim to fill this gap in the empirical literature, applying our tests to shed light on the optimality of commonly-used models in predicting various macro series. 

Instead of using quarterly data we propose the use of monthly variables (also used recently in similar contexts by  \citealp{CFG2022}) and we will focus on the same four target variables analysed in \citet{M15}. These series, all transformed to stationarity using the growth rate, are the Consumer Price Index for All Urban Consumers (CPIAUCSL), Industrial Production: Total Index (INDPRO), All Employees, Total Nonfarm (PAYEMS) and Personal Consumption Expenditures Excluding Food and Energy (Chain-Type Price Index) (PCEPILFE).\footnote{All series in the study are taken from the Federal Reserve Economic Data (FRED). Url: https://fred.stlouisfed.org/ [Last accessed: 08/03/22]} These series are very close in nature to the quarterly series analysed in \citet{AABG21} and will be regressed on an autoregressive term and the Chicago Fed National Financial Conditions Index (NFCI) as in \citet{ABG19}. 

Specifically, we use the direct forecasting scheme to generate quantile forecasts at quantile levels $\tau_k$ for $k=1,...,K$ and horizons $h=1,...,H$ as follows:
\begin{equation}\label{eq:EmpQF}
	\widehat{y}_{\tau_k,t,h}=\widehat{\gamma}_{0,h,t}(\tau_k)+\widehat{\gamma}_{1,h,t}(\tau_k)y_{t-h}+\widehat{\gamma}_{2,h,t}(\tau_k)x_{t-h}
\end{equation}
where $y_{t-h}$ is the autoregressive term corresponding to one of the four target variables mentioned above and $x_{t-h}$ is the NFCI. The parameter estimates are obtained by the standard quantile regression estimator and are indexed both by $\tau_k$ and $h$ to denote that a separate quantile regression is run at each quantile and horizon as in the direct scheme, as well as by $t$ as the forecasts are generate in a pseudo out-of-sample fashion as mentioned below. In essence, (\ref{eq:EmpQF}) boils down to a forecast made by a quantile autoregressive distributed lag (QADL) model \citep{galvao2013} using the direct forecasting scheme.

The data series span the period 1984M1 to 2019M12, giving a total number of $T=432$ monthly observations. We use the recursive out-of-sample scheme and split the sample into equal portions for the initial estimation sample and the evaluation sample, $R=P=216$. This gives an evaluation sample size, $P$, around the middle of the range of Monte Carlo simulations in Section \ref{sec:simulations} of the appendix. In making forecasts using (\ref{eq:EmpQF}) we will use horizons $h=1,...,12$ and quantile levels $\tau_k\in\{0.1,0.25,0.5\}$. The use of these quantile levels allows us to focus on the left part of the distribution, as is common in GaR studies such as \citet{AABG21}, but also includes the median as an important case of predicting the centre of the distribution. For the bootstrap implementation we use $B=1000$ bootstrap draws and employ a block length of $l=4$ as this is seen to work well in the simulation study in the appendix. 

The results in Table \ref{tab:EmpResults} display the results of the Mincer-Zarnowitz tests for autocalibration, with further graphical insight into the behaviour of the out-of-sample predictions given in Subsection \ref{secappoos} of the appendix. We first analyse the joint Mincer-Zarnowitz test (`Joint')  which works on multiple time series, as described above, where in this context we have $G=4$ target variables and we jointly test for autocalibration across all series to avoid the multiple testing problem. The results in the first row of Table \ref{tab:EmpResults} show that there is indeed some evidence against autocalibration when looking across all four macro series. The $p$-value of 0.07 indicates that there is evidence at the 10\% significance level that the QADL-type model does not produce well-calibrated forecasts jointly across these four series, for forecast horizons $h=1,...,12$ and quantile levels $\tau_k\in\{0.1,0.25,0.5\}$.

\begin{table}[!htb]
	\centering
	\caption{Mincer-Zarnowitz Test Results}
	\begin{tabular}{lccccc} \toprule
		& Stat  & 90\%  & 95\%  & 99\%  & $p$-value \\ \midrule
		Joint & 38264.280 & 28908.454 & 45259.085 & 86531.304 & 0.067 \\ \midrule
		CPIAUCSL & 18269.966 & 18033.852 & 32452.813 & 66594.353 & 0.099 \\ 
		INDPRO & 4258.078 & 7578.204 & 11224.918 & 24413.160 & 0.222 \\ 
		PAYEMS & 871.704 & 1574.085 & 2060.305 & 4994.712 & 0.308 \\ 
		PCEPILFE & 14864.532 & 2316.907 & 2792.387 & 3678.394 & 0.000 \\ \bottomrule
	\end{tabular}%
	\label{tab:EmpResults}%
\end{table}%

With this in mind, it is useful to dig further into the individual series to see which of them are likely to be causing the rejection of the joint null of autocalibration. The remainder of Table \ref{tab:EmpResults} displays the Mincer-Zarnowitz test when performed individually for each series. For the two real series, industrial production and employment, we see little evidence against the null. On the other hand, for the two price-type series we see somewhat different results with a clear rejection in the case of PCEPILFE and a $p$-value just under 10\% for CPIAUCSL. This suggests that the QADL-type approach suggested by \citet{ABG19} does indeed appear appropriate for real macroeconomic series but less-so for price series.\footnote{In Section \ref{app:Macro_Application} of the appendix, we also try to isolate the specific horizons and/or quantile levels that contribute the most to the rejection. Our findings suggest that for PCEPILFE and CPIAUCSL the smallest contribution to the statistic comes from quantile level $\tau_k=0.5$, while the $\tau_k=0.1$ quantile level contributes more substantially, even though no systematic conclusion can be drawn beyond $h=1$. This seems to indicate that further study should consider investigating the types of series which might deliver better-calibrated predictions in the far-left tail of the distribution of inflation-type series. }

One final exercise we perform is to apply the augmented MZ test where we use additional predictors in the MZ regression. Table \ref{tab:EmpResults2} displays the results for each of the four series above, where in each case the remaining three variables were used as the augmenting regressors. This serves as a simple check to see if any of these other variables would have been able to improve the forecasts if they were added to the forecasting model, especially for the real variables for which the weaker null of autocalibration was not rejected. However, the results in Table \ref{tab:EmpResults2} are similar to the non-augmented version of the test. As expected, the stronger null is rejected as well for the inflation type series, which already showed rejections for the weaker null of autocalibration. More interestingly, for the real variables, we still get no rejections. This suggests that we are not able to improve these forecasts by the addition of inflation type variables to the forecasting model.

\begin{table}[!htb]
	\centering
	\caption{Augmented Mincer-Zarnowitz Test Results}
	\begin{tabular}{lccccc} \toprule
		& Stat  & 90\%  & 95\%  & 99\%  & $p$-value \\ \midrule
		CPIAUCSL & 21984.030 & 19794.203 & 29896.138 & 57657.304 & 0.085 \\ 
		INDPRO & 5194.690 & 8722.551 & 12596.841 & 27604.813 & 0.224 \\ 
		PAYEMS & 723.354 & 1494.399 & 2011.985 & 4470.360 & 0.350 \\ 
		PCEPILFE & 15648.207 & 2455.174 & 2938.071 & 3801.048 & 0.000 \\  \bottomrule
	\end{tabular}%
	\label{tab:EmpResults2}%
\end{table}%

\section{Conclusion}\label{sec:conclusion}

This paper deals with the absolute evaluation of quantile forecasts in situations where predictions are made over multiple horizons and possibly multiple quantile levels. We propose multi-horizon, multi-quantile tests for  optimality by employing quantile Mincer-Zarnowitz regressions and a moment equality framework with a bootstrap methodology which avoids the estimation of a large covariance matrix. The main quantile Mincer-Zarnowitz test is of the null hypothesis of autocalibration, which is a fundamental property of forecast consistency. We also provide two extensions. The first extension tests a stronger null hypothesis, which allows us to add further important variables to the information set with respect to which optimality is tested. This augmented quantile Mincer-Zarnowitz test thus makes it possible to examine if the information contained in those variables was used optimally by the forecaster. The second extension is a multivariate quantile Mincer-Zarnowitz test and allows us to check autocalibration of forecasts for multiple time series at possibly multiple horizons and quantiles.

Our tests allow for an overall decision about the quality of a forecasting approach, whether it is a single model used over multiple horizons and quantiles or a mix of different models and expert judgement employed by an institution. Crucially, it avoids the multiple testing problem inherent to most practical situations, where many forecasts are made over horizons, quantiles or multiple variables. Importantly, our testing framework is constructive in that it does not only provide a formal procedure to reach this overall decision, but may also provide valuable feedback about possible weaknesses of the forecasting approach under consideration and how it could be improved.

There are many possible future avenues arising from our work, for instance the evaluation of distributional or probabilistic forecasts \citep{gneiting2014}. Since these distributional forecasts are considered quantile calibrated  when the corresponding quantile forecasts for all quantiles are autocalibrated, one future extension of our work may look into optimality testing across many quantiles.

\newpage\clearpage
\bibliographystyle{chicago}
\bibliography{DecompositionMonotonicity}

\begin{thebibliography}{}

\bibitem[\protect\citeauthoryear{Adams, Adrian, Boyarchenko, and
  Giannone}{Adams et~al.}{2021}]{AABG21}
Adams, P.~A., T.~Adrian, N.~Boyarchenko, and D.~Giannone (2021).
\newblock {Forecasting macroeconomic risks}.
\newblock {\em International Journal of Forecasting\/}~{\em 37\/}(3),
  1173--1191.

\bibitem[\protect\citeauthoryear{Adrian, Boyarchenko, and Giannone}{Adrian
  et~al.}{2019}]{ABG19}
Adrian, T., N.~Boyarchenko, and D.~Giannone (2019).
\newblock {Vulnerable growth}.
\newblock {\em American Economic Review\/}~{\em 109\/}(4), 1263--1289.

\bibitem[\protect\citeauthoryear{Andrews and Guggenberger}{Andrews and
  Guggenberger}{2009}]{AG2009}
Andrews, D. and P.~Guggenberger (2009).
\newblock Validity of subsampling and ` plug-in asymptotic' inference for
  parameters defined by moment inequalities.
\newblock {\em Econometric Theory\/}~{\em 25}, 669--709.

\bibitem[\protect\citeauthoryear{Andrews and Pollard}{Andrews and
  Pollard}{1994}]{AP1994}
Andrews, D. and D.~Pollard (1994).
\newblock An introduction to functional central limit theorems for dependent
  stochastic processes.
\newblock {\em International Statistical Review\/}~{\em 62\/}(1), 119--132.

\bibitem[\protect\citeauthoryear{Andrews and Soares}{Andrews and
  Soares}{2010}]{AS2010}
Andrews, D. and G.~Soares (2010).
\newblock Inference for parameters defined by moment inequalities using
  generalized moment selection.
\newblock {\em Econometrica\/}~{\em 78}, 119--157.

\bibitem[\protect\citeauthoryear{{Antolin Diaz}, Drechsel, and
  Petrella}{{Antolin Diaz} et~al.}{2021}]{ADP18}
{Antolin Diaz}, J., T.~Drechsel, and I.~Petrella (2021).
\newblock Advances in nowcasting economic activity.
\newblock {\em Mimeo\/}.

\bibitem[\protect\citeauthoryear{Ardia, Boudt, and Catania}{Ardia
  et~al.}{2019}]{Ardia2019}
Ardia, D., K.~Boudt, and L.~Catania (2019).
\newblock Generalized autoregressive score models in {R}: The {GAS} package.
\newblock {\em Journal of Statistical Software\/}~{\em 88\/}(6), 1–28.

\bibitem[\protect\citeauthoryear{Barendse, Kole, and van Dijk}{Barendse
  et~al.}{2023}]{barendse2023}
Barendse, S., E.~Kole, and D.~van Dijk (2023).
\newblock Backtesting value-at-risk and expected shortfall in the presence of
  estimation error.
\newblock {\em Journal of Financial Econometrics\/}~{\em 21\/}(2), 528--568.

\bibitem[\protect\citeauthoryear{Bayer and Dimitriadis}{Bayer and
  Dimitriadis}{2022}]{bayer2022}
Bayer, S. and T.~Dimitriadis (2022).
\newblock Regression-based expected shortfall backtesting.
\newblock {\em Journal of Financial Econometrics\/}~{\em 20\/}(3), 437--471.

\bibitem[\protect\citeauthoryear{Bollerslev}{Bollerslev}{1986}]{bollerslev1986}
Bollerslev, T. (1986).
\newblock Generalized autoregressive conditional heteroskedasticity.
\newblock {\em Journal of Econometrics\/}~{\em 31\/}(3), 307--327.

\bibitem[\protect\citeauthoryear{Brownlees and Souza}{Brownlees and
  Souza}{2021}]{BS20}
Brownlees, C.~T. and A.~Souza (2021).
\newblock {Backtesting Global Growth-at-Risk}.
\newblock {\em Journal of Monetary Economics\/}~{\em 118}, 312--330.

\bibitem[\protect\citeauthoryear{Carriero, Clark, and Marcellino}{Carriero
  et~al.}{2020}]{CCM20}
Carriero, A., T.~E. Clark, and M.~Marcellino (2020).
\newblock {Nowcasting Tail Risks to Economic Activity with Many Indicators}.
\newblock {\em Federal Reserve Bank of Cleveland, Working Paper No. 20-13\/}.

\bibitem[\protect\citeauthoryear{Christoffersen}{Christoffersen}{1998}]{christoffersen1998}
Christoffersen, P.~F. (1998).
\newblock Evaluating interval forecasts.
\newblock {\em International Economic Review\/}, 841--862.

\bibitem[\protect\citeauthoryear{Clements}{Clements}{2022}]{C22}
Clements, M.~P. (2022).
\newblock {Forecaster Efficiency, Accuracy, and Disagreement: Evidence Using
  Individual-Level Survey Data}.
\newblock {\em Journal of Money, Credit and Banking\/}~{\em 54\/}(2-3),
  537--568.

\bibitem[\protect\citeauthoryear{Corradi, Fosten, and Gutknecht}{Corradi
  et~al.}{2023}]{CFG2022}
Corradi, V., J.~Fosten, and D.~Gutknecht (2023).
\newblock {Conditional Quantile Coverage: an Application to Growth-at-Risk}.
\newblock {\em Journal of Econometrics\/}~{\em 236\/}(2).

\bibitem[\protect\citeauthoryear{Doukhan, Massart, and Rio}{Doukhan
  et~al.}{1995}]{DMR1995}
Doukhan, P., P.~Massart, and E.~Rio (1995).
\newblock Invariance principles for absolutely regular empirical processes.
\newblock {\em Annales de l'Inst. Henri Poincar\'{e}\/}~{\em 31\/}(2),
  393--427.

\bibitem[\protect\citeauthoryear{Elliott and Timmermann}{Elliott and
  Timmermann}{2016}]{ET2016}
Elliott, G. and A.~Timmermann (2016).
\newblock {\em Economic Forecasting}.
\newblock Princeton University Press.

\bibitem[\protect\citeauthoryear{Engle and Manganelli}{Engle and
  Manganelli}{2004}]{engle2004caviar}
Engle, R.~F. and S.~Manganelli (2004).
\newblock {CAV}ia{R}: Conditional autoregressive value at risk by regression
  quantiles.
\newblock {\em Journal of Business \& Economic Statistics\/}~{\em 22\/}(4),
  367--381.

\bibitem[\protect\citeauthoryear{Escanciano and Olmo}{Escanciano and
  Olmo}{2010}]{EO10}
Escanciano, J.~C. and J.~Olmo (2010).
\newblock {Backtesting Parametric Value-at-Risk With Estimation Risk}.
\newblock {\em Journal of Business {\&} Economic Statistics\/}~{\em 28\/}(1),
  36--51.

\bibitem[\protect\citeauthoryear{Escanciano and Olmo}{Escanciano and
  Olmo}{2011}]{EO11}
Escanciano, J.~C. and J.~Olmo (2011, jun).
\newblock {Robust Backtesting Tests for Value-at-risk Models}.
\newblock {\em Journal of Financial Econometrics\/}~{\em 9\/}(1), 132--161.

\bibitem[\protect\citeauthoryear{Ferrara, Mogliani, and Sahuc}{Ferrara
  et~al.}{2021}]{FMS21}
Ferrara, L., M.~Mogliani, and J.-G. Sahuc (2021).
\newblock {High-frequency monitoring of growth at risk}.
\newblock {\em International Journal of Forecasting\/}.

\bibitem[\protect\citeauthoryear{Fissler, Ziegel, and Gneiting}{Fissler
  et~al.}{2015}]{fissler2015}
Fissler, T., J.~F. Ziegel, and T.~Gneiting (2015).
\newblock Expected shortfall is jointly elicitable with value at
  risk-implications for backtesting.
\newblock {\em arXiv preprint arXiv:1507.00244\/}.

\bibitem[\protect\citeauthoryear{Fitzenberger}{Fitzenberger}{1997}]{F1997}
Fitzenberger, B. (1997).
\newblock The moving block bootstrap and robust inference for linear least
  squares and quantile regressions.
\newblock {\em Journal of Econometrics\/}~{\em 82}, 235--287.

\bibitem[\protect\citeauthoryear{Fosten and Gutknecht}{Fosten and
  Gutknecht}{2020}]{FG17}
Fosten, J. and D.~Gutknecht (2020).
\newblock {Testing Nowcast Monotonicity with Estimated Factors}.
\newblock {\em Journal of Business {\&} Economic Statistics\/}~{\em 38\/}(1),
  107--123.

\bibitem[\protect\citeauthoryear{Gaglianone, Lima, Linton, and
  Smith}{Gaglianone et~al.}{2011}]{GLLS11}
Gaglianone, W.~P., L.~R. Lima, O.~Linton, and D.~R. Smith (2011).
\newblock {Evaluating Value-at-Risk Models via Quantile Regression}.
\newblock {\em Journal of Business {\&} Economic Statistics\/}~{\em 29\/}(1),
  150--160.

\bibitem[\protect\citeauthoryear{Galvao, Montes-Rojas, and Olmo}{Galvao
  et~al.}{2011}]{GMO2011}
Galvao, A., G.~Montes-Rojas, and J.~Olmo (2011).
\newblock Threshold quantile autoregressive models.
\newblock {\em Journal of Time Series Analysis\/}~{\em 32}, 253--267.

\bibitem[\protect\citeauthoryear{Galvao~Jr., Montes-Rojas, and Park}{Galvao~Jr.
  et~al.}{2013}]{galvao2013}
Galvao~Jr., A.~F., G.~Montes-Rojas, and S.~Y. Park (2013).
\newblock Quantile autoregressive distributed lag model with an application to
  house price returns.
\newblock {\em Oxford Bulletin of Economics and Statistics\/}~{\em 75\/}(2),
  307--321.

\bibitem[\protect\citeauthoryear{Ghalanos}{Ghalanos}{2022}]{ghalanos2014}
Ghalanos, A. (2022).
\newblock rugarch: Univariate {GARCH} models.
\newblock {\em R package version\/}~{\em 1.4-9}.

\bibitem[\protect\citeauthoryear{Giacomini and Komunjer}{Giacomini and
  Komunjer}{2005}]{GK05}
Giacomini, R. and I.~Komunjer (2005).
\newblock {Evaluation and Combination of Conditional Quantile Forecasts}.
\newblock {\em Journal of Business {\&} Economic Statistics\/}~{\em 23\/}(4),
  416--431.

\bibitem[\protect\citeauthoryear{Giacomini, Politis, and White}{Giacomini
  et~al.}{2013}]{GPW13}
Giacomini, R., D.~N. Politis, and H.~White (2013).
\newblock {A Warp-Speed Method for Conducting Monte Carlo Experiments Involving
  Bootstrap Estimators}.
\newblock {\em Econometric Theory\/}~{\em 29\/}(3), 567--589.

\bibitem[\protect\citeauthoryear{Giacomini and Rossi}{Giacomini and
  Rossi}{2010}]{GR2010}
Giacomini, R. and B.~Rossi (2010).
\newblock Forecast comparisons in unstable environments.
\newblock {\em Journal of Applied Econometrics\/}~{\em 25\/}(4), 595--620.

\bibitem[\protect\citeauthoryear{Glosten, Jagannathan, and Runkle}{Glosten
  et~al.}{1993}]{glosten1993}
Glosten, L.~R., R.~Jagannathan, and D.~E. Runkle (1993).
\newblock On the relation between the expected value and the volatility of the
  nominal excess return on stocks.
\newblock {\em The Journal of Finance\/}~{\em 48\/}(5), 1779--1801.

\bibitem[\protect\citeauthoryear{Gneiting}{Gneiting}{2011}]{gneiting2011}
Gneiting, T. (2011).
\newblock Making and evaluating point forecasts.
\newblock {\em Journal of the American Statistical Association\/}~{\em
  106\/}(494), 746--762.

\bibitem[\protect\citeauthoryear{Gneiting, Balabdaoui, and Raftery}{Gneiting
  et~al.}{2007}]{gneiting2007}
Gneiting, T., F.~Balabdaoui, and A.~E. Raftery (2007).
\newblock Probabilistic forecasts, calibration and sharpness.
\newblock {\em Journal of the Royal Statistical Society: Series B (Statistical
  Methodology)\/}~{\em 69\/}(2), 243--268.

\bibitem[\protect\citeauthoryear{Gneiting and Katzfuss}{Gneiting and
  Katzfuss}{2014}]{gneiting2014}
Gneiting, T. and M.~Katzfuss (2014).
\newblock Probabilistic forecasting.
\newblock {\em Annual Review of Statistics and Its Application\/}~{\em 1},
  125--151.

\bibitem[\protect\citeauthoryear{Gneiting and Ranjan}{Gneiting and
  Ranjan}{2013}]{gneiting2013}
Gneiting, T. and R.~Ranjan (2013).
\newblock Combining predictive distributions.
\newblock {\em Electronic Journal of Statistics\/}~{\em 7}, 1747--1782.

\bibitem[\protect\citeauthoryear{Goncalves and White}{Goncalves and
  White}{2002}]{GW2002}
Goncalves, S. and H.~White (2002).
\newblock The bootstrap of the mean for dependent and heterogeneous arrays.
\newblock {\em Econometric Theory\/}~{\em 18}, 1367--1384.

\bibitem[\protect\citeauthoryear{Goncalves and White}{Goncalves and
  White}{2004}]{GW2004}
Goncalves, S. and H.~White (2004).
\newblock Maximum likelihood and the bootstrap for nonlinear dynamic models.
\newblock {\em Journal of Econometrics\/}~{\em 119}, 199--219.

\bibitem[\protect\citeauthoryear{G\"{o}tze and K\"{u}nsch}{G\"{o}tze and
  K\"{u}nsch}{1996}]{GK1996}
G\"{o}tze, F. and H.~K\"{u}nsch (1996).
\newblock Second-order correctness of the blockwise bootstrap for stationary
  observations.
\newblock {\em The Annals of Statistics\/}~{\em 24\/}(5), 1914--1933.

\bibitem[\protect\citeauthoryear{Granger}{Granger}{1969}]{granger1969}
Granger, C.~W. (1969).
\newblock Prediction with a generalized cost of error function.
\newblock {\em Journal of the Operational Research Society\/}~{\em 20\/}(2),
  199--207.

\bibitem[\protect\citeauthoryear{Gregory, Lahiri, and Nordman}{Gregory
  et~al.}{2018}]{GLN2018}
Gregory, K., S.~Lahiri, and D.~Nordman (2018).
\newblock A smooth block bootstrap for quantile regression with time series.
\newblock {\em Annals of Statistics\/}~{\em 46\/}(3), 1138--1166.

\bibitem[\protect\citeauthoryear{Hansen, Lunde, and Nason}{Hansen
  et~al.}{2011}]{HLN11}
Hansen, P.~R., A.~Lunde, and J.~M. Nason (2011).
\newblock {The Model Confidence Set}.
\newblock {\em Econometrica\/}~{\em 79\/}(2), 453--497.

\bibitem[\protect\citeauthoryear{Hassler and Pohle}{Hassler and
  Pohle}{2023}]{Hassler2023}
Hassler, U. and M.-O. Pohle (2023).
\newblock Forecasting under long memory.
\newblock {\em Journal of Financial Econometrics\/}~{\em 21\/}(3), 742--778.

\bibitem[\protect\citeauthoryear{He, Kou, and Peng}{He et~al.}{2022}]{he2022}
He, X.~D., S.~Kou, and X.~Peng (2022).
\newblock Risk measures: robustness, elicitability, and backtesting.
\newblock {\em Annual Review of Statistics and Its Application\/}~{\em 9},
  141--166.

\bibitem[\protect\citeauthoryear{Kim and White}{Kim and White}{2003}]{KW2003}
Kim, T.-H. and H.~White (2003).
\newblock Estimation, inference, and specification testing for possibly
  misspecified quantile regression.
\newblock In T.~Fomby and R.~Carter~Hill (Eds.), {\em Maximum Likelihood
  Estimation of Misspecified Models: Twenty Years Later}, Volume~17 of {\em
  Advances in Econometrics}, pp.\  107--132. Emerald Group Publishing Limited,
  Bingley.

\bibitem[\protect\citeauthoryear{Koenker and Xiao}{Koenker and
  Xiao}{2006}]{KX06}
Koenker, R. and Z.~Xiao (2006).
\newblock {Quantile autoregression}.
\newblock {\em Journal of the American Statistical Association\/}~{\em
  101\/}(475), 980--990.

\bibitem[\protect\citeauthoryear{Koenker and Xiao}{Koenker and
  Xiao}{2009}]{KX2009}
Koenker, R. and Z.~Xiao (2009).
\newblock Conditional quantile estimation for generalized autoregressive
  conditional heteroscedasticity models.
\newblock {\em Journal of the American Statistical Association\/}~{\em
  104\/}(488), 1696--1712.

\bibitem[\protect\citeauthoryear{Kratz, Lok, and McNeil}{Kratz
  et~al.}{2018}]{kratz2018}
Kratz, M., Y.~H. Lok, and A.~J. McNeil (2018).
\newblock Multinomial {VaR} backtests: A simple implicit approach to
  backtesting expected shortfall.
\newblock {\em Journal of Banking \& Finance\/}~{\em 88}, 393--407.

\bibitem[\protect\citeauthoryear{K\"{u}nsch}{K\"{u}nsch}{1989}]{K1989}
K\"{u}nsch, H. (1989).
\newblock The jackknife and the bootstrap for general stationary observations.
\newblock {\em Annals of Statistics\/}~{\em 17\/}(3), 1217--1241.

\bibitem[\protect\citeauthoryear{Manzan}{Manzan}{2015}]{M15}
Manzan, S. (2015).
\newblock {Forecasting the distribution of economic variables in a data-rich
  environment}.
\newblock {\em Journal of Business {\&} Economic Statistics\/}~{\em 33\/}(1),
  144--164.

\bibitem[\protect\citeauthoryear{Mincer and Zarnowitz}{Mincer and
  Zarnowitz}{1969}]{mincer1969}
Mincer, J.~A. and V.~Zarnowitz (1969).
\newblock The evaluation of economic forecasts.
\newblock In {\em Economic forecasts and expectations: Analysis of forecasting
  behavior and performance}, pp.\  3--46. NBER.

\bibitem[\protect\citeauthoryear{Nolde and Ziegel}{Nolde and
  Ziegel}{2017}]{nolde2017}
Nolde, N. and J.~F. Ziegel (2017).
\newblock Elicitability and backtesting: Perspectives for banking regulation.
\newblock {\em The Annals of Applied Statistics\/}~{\em 11\/}(4), 1833--1874.

\bibitem[\protect\citeauthoryear{Oodaira and Yoshihara}{Oodaira and
  Yoshihara}{1971}]{OY1971}
Oodaira, H. and K.~Yoshihara (1971).
\newblock The law of the iterated logarithm for stationary processes satisfying
  mixing conditions.
\newblock {\em Kodai Mathematical Seminar Reports\/}~{\em 23\/}(3), 311--334.

\bibitem[\protect\citeauthoryear{Pascual, Romo, and Ruiz}{Pascual
  et~al.}{2006}]{pascual2006}
Pascual, L., J.~Romo, and E.~Ruiz (2006).
\newblock Bootstrap prediction for returns and volatilities in {GARCH} models.
\newblock {\em Computational Statistics \& Data Analysis\/}~{\em 50\/}(9),
  2293--2312.

\bibitem[\protect\citeauthoryear{Patton and Timmermann}{Patton and
  Timmermann}{2012}]{PT2012}
Patton, A. and A.~Timmermann (2012).
\newblock Forecast rationality tests based on multi-horizon bounds.
\newblock {\em Journal of Business and Economic Statistics\/}~{\em 30\/}(1).

\bibitem[\protect\citeauthoryear{Plagborg-M{\o}ller, Reichlin, Ricco, and
  Hasenzagl}{Plagborg-M{\o}ller et~al.}{2020}]{PRRH20}
Plagborg-M{\o}ller, M., L.~Reichlin, G.~Ricco, and T.~Hasenzagl (2020).
\newblock {When is growth at risk?}
\newblock {\em BPEA Conference Draft, Spring\/}.

\bibitem[\protect\citeauthoryear{Pohle}{Pohle}{2020}]{P20}
Pohle, M.-O. (2020).
\newblock {The Murphy Decomposition and the Calibration-Resolution Principle: A
  New Perspective on Forecast Evaluation}.
\newblock {\em arXiv preprint arXiv:2005.01835\/}.

\bibitem[\protect\citeauthoryear{Prasad, Elekdag, Jeasakul, Lafarguette, Alter,
  Feng, and Wang}{Prasad et~al.}{2019}]{PEJLAFW2019}
Prasad, A., S.~Elekdag, P.~Jeasakul, R.~Lafarguette, A.~Alter, A.~X. Feng, and
  C.~Wang (2019).
\newblock {Growth at Risk: Concept and Application in IMF Country
  Surveillance}.
\newblock IMF Working Paper 19/36, International Monetary Fund.

\bibitem[\protect\citeauthoryear{Quaedvlieg}{Quaedvlieg}{2021}]{Q17}
Quaedvlieg, R. (2021).
\newblock {Multi-Horizon Forecast Comparison}.
\newblock {\em Journal of Business {\&} Economic Statistics\/}~{\em 39\/}(1),
  40--53.

\bibitem[\protect\citeauthoryear{Romano and Shaikh}{Romano and
  Shaikh}{2010}]{RS2010}
Romano, J. and A.~Shaikh (2010).
\newblock Inference for the identified set in partially identified econometric
  models.
\newblock {\em Econometrica\/}~{\em 78}, 169--211.

\bibitem[\protect\citeauthoryear{Tsyplakov}{Tsyplakov}{2013}]{Tsyplakov2013}
Tsyplakov, A. (2013).
\newblock Evaluation of probabilistic forecasts: proper scoring rules and
  moments.
\newblock {\em Available at SSRN 2236605\/}.

\bibitem[\protect\citeauthoryear{West}{West}{1996}]{W96}
West, K.~D. (1996).
\newblock {Asymptotic Inference about Predictive Ability}.
\newblock {\em Econometrica\/}~{\em 64\/}(5), 1067--1084.

\bibitem[\protect\citeauthoryear{White}{White}{2000}]{W00}
White, H. (2000).
\newblock {A Reality Check for Data Snooping}.
\newblock {\em Econometrica\/}~{\em 68\/}(5), 1097--1126.

\end{thebibliography}

\newpage

\section*{Appendix}
\appendix

This appendix contains additional material for this paper. Section \ref{sec:simulations} illustrates results from a Monte Carlo study, where the finite sample properties of the MZ and augmented MZ tests across different sample sizes and bootstrap block lengths are assessed. Sections \ref{sec:characoptim} and \ref{proofs} contain the proofs for the theoretical results of the paper, while Section \ref{sec:ELoss}  outlines a separate monotonicity test. Finally, Sections \ref{app:Finance_Application} and  \ref{app:Macro_Application} contain additional empirical results and graphs for the VaR and GaR forecast application, respectively.

\section{Monte Carlo Simulations}\label{sec:simulations}

In this section we will present results to explore the finite sample properties of the Mincer-Zarnowitz and augmented Mincer-Zarnowitz tests described above using a variety of set-ups, DGPs and sample sizes.

\subsection{Mincer-Zarnowitz Simulations}\label{sec:sim-MZ}

We first develop a set-up which is designed to assess the size and power of the Mincer-Zarnowitz test in a simulation environment designed to be similar to our macroeconomic application. In a later section we will also provide a brief example in a financial context. In this setting we explore a case where the target variable is generated according to a simple AR(1) process. The forecasts, on the other hand, will be computed using a potentially incorrect value for the AR(1) parameter which will drive cases of mis-calibration. Specifically, the DGP for $y_t$ is:
\begin{equation}\label{MCDGP}
	y_t=b y_{t-1}+\varepsilon_t
\end{equation} 
where we generate \(\varepsilon_t\) as an i.i.d.\ normal variable with mean zero and
variance \(1-b^2\) which fixes the variance of \(y_t\) to unity. The
optimal \(h\)-step ahead quantile
forecast is therefore
\(q_{t+h}(\tau|\mathcal{F}_t)=b^hy_t+\sqrt{1-b^{2h}}\Phi^{-1}(\tau)\).

The actual quantile forecasts for quantile level $\tau$ are generated with the same functional form as the optimal forecast but using a (potentially) incorrect value $\tilde{b}$ for the AR(1) parameter
instead of \(b\). In other words:
\begin{equation}\label{Eq:incquant}
	\widehat{q}^{inc}_{t+h}(\tau)=\tilde{b}^h y_t +\sqrt{1-\tilde{b}^{2h}}\Phi^{-1}(\tau).
\end{equation} 

In this specification, the forecasts are autocalibrated when we have $\tilde{b}=b$ whereas we have miscalibration when
\(\tilde{b}\neq b\). This gives us a simple way to assess the size and power of the Mincer-Zarnowitz type test: in the former case we expect the slope coefficient in the Mincer-Zarnowitz regression (equation (\ref{POPQREG}) in the main text) to be equal to unity and we are under the null hypothesis, whereas in the latter case it is not equal to unity and we are under the alternative hypothesis.\footnote{Note that an analytical expression for the Mincer-Zarnowitz coefficients is simple to derive in the case of making mean predictions. In that case, with MSFE being the criterion used to assess the forecasts, it can be shown that the slope of the Mincer-Zarnowitz regression is
	\((\beta/b)^h\) which is different from unity when \(b\neq\beta\). An analogous result will hold in the quantile case.}  We will use values of
\(b=0.6\) and \(\tilde{b}=\{0.6,0.8\}\), where the case with \(\tilde{b}=0.6\)
allows us to assess the size performance and \(\tilde{b}=0.8\) will assess
power.

We perform the simulations for a range of evaluation sample sizes
$P\in\{120,240,480\}$ and we let there be a total of $H=4$ horizons. We will
generate a sample of size $P+H+1$ for this AR(1) case which includes the single initial condition (drawn from the stationary distribution of the AR(1) process) and allows for up to
$H$-step ahead forecasts to be made for all $P$ periods. We will consider multiple quantile levels
$\mathcal{T}=\{0.25,0.5,0.75\}$. For the bootstrap we will use various block
lengths $l\in\{4,8,12\}$ and we will perform \(B=1999\) Monte Carlo
replications, saving a single bootstrap draw each time as in the Warp
Speed bootstrap of \citet{GPW13}. We compute rejection rates for
a nominal size of \(\alpha=0.05\).

We now present the size and power results for the moment equality Mincer-Zarnowitz test from (\ref{EQTS1}) in the main text. Table \ref{tab:MC-MZ-1} presents the rejection rates of the test with $\tilde{b}$ set to 0.6 which assesses the size of the test. The results are displayed for the different bootstrap block lengths and sample sizes discussed above. Then the power of the test is displayed in Table \ref{tab:MC-MZ-3} for the case when $\tilde{b}=0.8$ which results in mis-calibration.

\begin{table}[!htb]
	\centering
	\caption{MZ Test - Size - $\tilde{b}=0.6$ }
	\begin{tabular}{lrrr} \toprule
		
		& $P=120$ & $P=240$ & $P=480$ \\ \midrule
		$l=4$ & 0.037 & 0.051 & 0.055 \\ 
		$l=8$ & 0.053 & 0.038 & 0.034 \\ 
		$l=12$ & 0.039 & 0.044 & 0.045 \\  \bottomrule
	\end{tabular}%
	\label{tab:MC-MZ-1}
	
	\bigskip
	
	\caption{MZ Test - Power - $\tilde{b}=0.8$ }
	\begin{tabular}{lrrr} \toprule
		& \multicolumn{1}{l}{$P=120$} & \multicolumn{1}{l}{$P=240$} & \multicolumn{1}{l}{$P=480$} \\ \midrule
		$l=4$ & 0.792 & 0.970 & 1.000\\ 
		$l=8$ & 0.747 & 0.959 & 1.000\\ 
		$l=12$ & 0.740 & 0.961 & 1.000\\   \bottomrule
	\end{tabular}%
	\label{tab:MC-MZ-3}%
\end{table}%

The results in Table \ref{tab:MC-MZ-1} show that the test has fairly good size properties for a reasonable block length selection. For instance, we see rejection rates very close to nominal size (5.1\% and 5.5\%) for block length $l=4$ for the two larger samples size $P=240$ and $P=480$. Even in smaller samples the size is fairly close to nominal size. Regarding the power of the test, Table \ref{tab:MC-MZ-3} shows very good power properties with rejection rates roughly 75\% or above, even at the lowest sample size $P=120$. These rejection rates rise to above 95\% as the sample size increases to $P=240$ and are equal to unity when $P=480$.

\subsection{Augmented Mincer-Zarnowitz Simulations}\label{sim-aug}

In order to display the properties of the augmented Mincer-Zarnowitz test, we take the simplest possible case where $\mathbf{Z}_t$ comprises a single variable $z_{t}$ which is to be included in the regression described in (\ref{AUGPOPQREG}) in the main text. We also generate this variable according to an AR(1) process:
\begin{equation}\label{MCDGPz}
	z_t=d z_{t-1}+\nu_t
\end{equation}
where $\nu_t$ is similarly drawn from an i.i.d.\ normal distribution with mean zero and variance $1-d^2$, and for simplicity we set the AR parameter $d$ equal to that in (\ref{MCDGP}), in other words $d=b$. 

We can easily assess the size of the augmented MZ test if we let (\ref{MCDGP}) be the true DGP and we use the augmented MZ regression (\ref{AUGPOPQREG}) from the main text with the additional regressor $z_t$ which will enter the MZ equation with a coefficient of zero. If we use the forecasts described in (\ref{Eq:incquant}) with $\tilde{b}=b=0.6$ we are therefore under the null hypothesis. On the other hand, to display the power of the augmented MZ test, we need to use an additional DGP for $y_t$ which involves $z_t$ and a set of forecasts which omit this additional variable. As such we propose an ADL(1,1) process for $y_t$:
\begin{equation}\label{MCDGP2}
	y_t=b y_{t-1}+c z_{t-1}+\varepsilon_t
\end{equation} 
where we use the same parameter values and random draws $b$ and $\varepsilon_t$ as in (\ref{MCDGP}) with the addition of $z_t$ which is drawn according to (\ref{MCDGPz}) and we set $c=0.5$. If we make the forecasts in the same way as (\ref{Eq:incquant}) above, which only makes use of the variable $y_t$ then the coefficient on $z_t$ in the augmented MZ regression is non-zero which violates the condition in the null in (\ref{AUGEQH0}) from the main text. For the $\tilde{b}$ coefficient used in making the forecast, we use the projection coefficient of $y_t$ on $y_{t-1}$, which leads to autocalibrated forecasts, that is, the null of the basic MZ test is not violated. The null of the augmented MZ test is, however, violated, and both coefficients ($\beta$ and $\gamma$) differ from their values under the null in (\ref{AUGEQH0}) in the main text. For the parameter values described above, the projection coefficient of $y_t$ on $y_{t-1}$ under (\ref{MCDGP2}) is roughly $\tilde{b}=0.70$. 

The results for size and power are displayed in Tables \ref{tab:MC-Aug-MZ-1} and \ref{tab:MC-Aug-MZ-2} below. The size results show that the augmented MZ test has good size properties with rejection rates fairly close to the nominal size of 5\% across sample sizes and block lengths. This shows that the augmented version of the test, indeed, has similar size properties to the standard MZ test. For the power, we also see results in line with what is expected. In Table \ref{tab:MC-Aug-MZ-2} we see that power is above 60\% in the small sample size $P=120$. This rapidly improves to above 95\% as the sample size rises to $P=240$ and to unity when $P=480$. 

\begin{table}[!htb]
	\centering
	\caption{Augmented MZ Test - Size - $\tilde{b}=0.6$ }
	\begin{tabular}{lrrr} \toprule
		& \multicolumn{1}{l}{$P=120$} & \multicolumn{1}{l}{$P=240$} & \multicolumn{1}{l}{$P=480$} \\ \midrule
		$l=4$ & 0.053 & 0.058 & 0.052 \\ 
		$l=8$ & 0.034 & 0.055 & 0.058 \\ 
		$l=12$ & 0.039 & 0.038 & 0.052 \\  \bottomrule
	\end{tabular}%
	
	\parbox[t]{6.5cm}{\vspace{1mm} \textbf{Notes:} DGP for size is equation (\ref{MCDGP}).}
	\label{tab:MC-Aug-MZ-1}
	
	\bigskip
	
	\caption{Augmented MZ Test - Power - $\tilde{b}=0.70$ }
	\begin{tabular}{lrrr} \toprule
		& \multicolumn{1}{l}{$P=120$} & \multicolumn{1}{l}{$P=240$} & \multicolumn{1}{l}{$P=480$} \\ \midrule
		$l=4$ & 0.667 & 0.979 & 1.000\\ 
		$l=8$ & 0.646 & 0.983 & 1.000\\ 
		$l=12$ & 0.638 & 0.976 & 1.000\\  \bottomrule
	\end{tabular}%
	
	\parbox[t]{6.7cm}{\vspace{1mm} \textbf{Notes:} DGP for power is equation (\ref{MCDGP2}).}
	\label{tab:MC-Aug-MZ-2}%

\end{table}%

\subsection{Simulations for Financial Applications}\label{sec:sim-Fin}

In this section we will revisit the size and power simulations for the main Mincer-Zarnowitz test but with a target variable designed to be closer to those found in financial applications. To that end we generate the target variable $y_t$ as a GARCH(1,1) process and, for simplicity, we will focus only on the 1-step ahead forecasting case across multiple quantiles. The variable evolves according to $y_t=b_0+\sigma_t\varepsilon_t$ where we assume $\varepsilon_t$ to follow an i.i.d. Student's $t$ distribution with $\nu=30$ degrees of freedom as in \citet{EO10}. We denote this distribution by $F(.)$ for simplicity. The optimal 1-step ahead quantile
forecast is therefore
\(q_{t+1}(\tau|\mathcal{F}_t)=b_0+\sqrt{b_1+b_2y_{t-1}^2+b_3\sigma_{t-1}^2}F^{-1}(\tau)\) where $\sigma^2_t=b_1+b_2y_{t-1}^2+b_3\sigma_{t-1}^2$ is the modelled conditional variance of $y_t$. We set the parameters of this GARCH(1,1) process to be $\mathbf{b}=(b_0,b_1,b_2,b_3)^{\prime}=(0,0.05,0.1,0.85)^{\prime}$ so the process has a mean of zero and the conditional variance, following \citet{EO10}, shows realistic levels of volatility clustering.

As above, the actual quantile forecasts are generated as a mis-specified version of the correct optimal forecast. In this GARCH(1,1) case we do this by specifying potentially incorrect values for the GARCH coefficients $\tilde{\mathbf{b}}=(\tilde{b}_0,\tilde{b}_1,\tilde{b}_2,\tilde{b}_3)^{\prime}$, in other words:
\begin{equation}\label{Eq:incquant2}
	\widehat{q}^{inc}_{t+1}(\tau)=\tilde{b}_0+\sqrt{\tilde{b}_1+\tilde{b}_2y_{t-1}^2+\tilde{b}_3\sigma_{t-1}^2}F^{-1}(\tau)
\end{equation}
so that the forecasts are based on a potentially incorrect conditional mean and/or variance estimate. In this set-up, the case where $\tilde{\mathbf{b}}=\mathbf{b}=(0,0.05,0.1,0.85)^{\prime}$ clearly corresponds to the autocalibrated case under the null which will allow us to assess the size of the test. On the other hand, cases with $\tilde{\mathbf{b}}\neq\mathbf{b}$ generate mis-calibration and allow us to assess the power of the test. We will try out two cases: $\tilde{\mathbf{b}}=(1,0.05,0.1,0.85)^{\prime}$ is the case where the conditional mean is misspecified, and $\tilde{\mathbf{b}}=(0,0.05,0.45,0.4)^{\prime}$ is where the conditional variance is misspecified where the ARCH and GARCH parameters are incorrect but still sum up to the same value as in the true DGP.

We will use samples sizes in line with empirical VaR studies $P\in\{1000,3000,5000\}$. As mentioned before, we only consider one forecast horizon $H=1$, but will look at multiple quantile levels
$\mathcal{T}=\{0.01,0.025,0.05\}$ as in our empirical application. For the bootstrap we will use larger block lengths than in the previous simulation which had smaller samples sizes with the macroeconomic context in mind. Here we use $l\in\{10,20,30\}$. The number of bootstrap draws and nominal size are \(B=1999\) and \(\alpha=0.05\) as above. 

The results in Tables \ref{tab:MC-MZ-Fin1}, \ref{tab:MC-MZ-Fin2} and \ref{tab:MC-MZ-Fin3} display the rejection rates in both the null and two alternative cases. Table \ref{tab:MC-MZ-Fin1} shows that the test has good size control. With a block length choice of $l=10$ the size is around 3\%, slightly below the nominal level of 5\%, whereas for a larger block length like $l=30$ the rejection rate gets close to the nominal size as the sample size reaches $P=5000$. Moving to the power results, Table \ref{tab:MC-MZ-Fin2} also shows the expected result that the power rises to unity as we increase the sample size towards $P=5000$ in the misspecified conditional mean case. In the more realistic case where the conditional variance is misspecified, we also observe similar findings. Table \ref{tab:MC-MZ-Fin3} shows that the power starts around 80\% at the lowest sample size $P=1000$ and then increases to over 98\% for the sample size $P=3000$ which is similar to the sample size used in our empirical application. It is worth noting here that we use the quantile level 1\% which is far lower than in the previous simulations, so it is encouraging to see the test works well in this scenario.

\begin{table}[!htb]
	\centering
	\caption{MZ Test - Size - $\tilde{\mathbf{b}}=(0,0.05,0.1,0.85)^{\prime}$ }
	\begin{tabular}{lrrr} \toprule
		
		& \multicolumn{1}{l}{$P=1000$} & \multicolumn{1}{l}{$P=3000$} & \multicolumn{1}{l}{$P=5000$} \\ \midrule
		$l=10$ & 0.029 & 0.028 & 0.027 \\
		$l=20$ & 0.026 & 0.029 & 0.034 \\
		$l=30$ & 0.030 & 0.038 & 0.045 \\\bottomrule
	\end{tabular}%
	\label{tab:MC-MZ-Fin1}
	
	\bigskip
	
	\caption{MZ Test - Power 1 - $\tilde{\mathbf{b}}=(1,0.05,0.1,0.85)^{\prime}$ }
	\begin{tabular}{lrrr} \toprule
		& \multicolumn{1}{l}{$P=1000$} & \multicolumn{1}{l}{$P=3000$} & \multicolumn{1}{l}{$P=5000$} \\ \midrule
		$l=10$ & 0.897 & 1.000 & 1.000 \\
		$l=20$ & 0.869 & 1.000 & 1.000 \\
		$l=30$ & 0.906 & 1.000 & 1.000 \\  \bottomrule
	\end{tabular}%
	\label{tab:MC-MZ-Fin2}%

	\bigskip
	
	\caption{MZ Test - Power 2 - $\tilde{\mathbf{b}}=(0,0.05,0.45,0.4)^{\prime}$ }
	\begin{tabular}{lrrr} \toprule
		& \multicolumn{1}{l}{$P=1000$} & \multicolumn{1}{l}{$P=3000$} & \multicolumn{1}{l}{$P=5000$} \\ \midrule
		$l=10$ & 0.792 & 0.988 & 0.999 \\
		$l=20$ & 0.797 & 0.988 & 0.999 \\
		$l=30$ & 0.802 & 0.993 & 0.999 \\  \bottomrule
	\end{tabular}%
	\label{tab:MC-MZ-Fin3}%
	
\end{table}%

\newpage

\section{Characterisations of Optimality}\label{sec:characoptim}

\begin{lemma} \label{Optimality}
	Let $\widehat{y}^{\ast}_{\tau,t,h}$ be an optimal $h$-step ahead forecast for the $\tau$-quantile (cf. Definition \ref{optimality}). Then $\widehat{y}^{\ast}_{\tau,t,h}$ is optimal relative to $\mathcal{I}_{t} \subset \mathcal{F}_t$ as well.
\end{lemma}

\noindent\textbf{Proof of Lemma \ref{Optimality}}: 	By optimality it holds for $\widehat{y}^{\ast}_{\tau,t,h}$ that
$$\mathrm{E} \left[L_{\tau} \left( y_{t} - \widehat{y}^{\ast}_{\tau,t,h} \right) | \mathcal{F}_{t-h} \right] \leq \mathrm{E} \left[L_{\tau} \left( y_{t} - \widehat{y}_{\tau,t,h} )\right | \mathcal{F}_{t-h} \right]$$
for all possible forecasts $\widehat{y}_{\tau,t,h}$ based on $\mathcal{F}_{t-h}$. By monotonicity of (conditional) expectation and the law of iterated expectation, this implies
$$\mathrm{E} \left[L_{\tau} \left( y_{t} - \widehat{y}^{\ast}_{\tau,t,h} \right) | \mathcal{I}_{t-h} \right] \leq \mathrm{E} \left[L_{\tau} \left( y_{t} - \widehat{y}_{\tau,t,h} \right) | \mathcal{I}_{t-h} \right],$$
which proves the claim. \qedsymbol

\section{Proofs of Lemmas and Theorems}\label{proofs}

This section provides an asymptotic linear Bahadur representation for the quantile regression estimator and a stochastic equicontinuity result in Lemma \ref{Bahadur}, which is subsequently used in the proof of Theorem \ref{MZTest}. The proof of the auxiliary Lemma \ref{Bahadur} follows at the end of this section.

\begin{lemma}\label{Bahadur}  Under
	Assumptions A1 and A7, it holds that: \medskip
	
	\noindent \textbf{(i)} For each $\tau\in\mathcal{T}$ and $h\in\mathcal{H}$: 
	\begin{equation*}
		\left\Vert\widehat{\boldsymbol{\beta} }_{h}(\tau )-\boldsymbol{\beta}_{h}
		^{\dag }(\tau )\right\Vert =o_{\Pr}(1),
	\end{equation*}%
	where $\boldsymbol{\beta}_{h}^{\dag }(\tau )$ is defined in Equation (\ref{EQPLIM}) and $\widehat{\boldsymbol{\beta} }_{h}(\tau )$ in Equation (\ref{EQEST}). \medskip
	
	\noindent \textbf{(ii)} The empirical process: 
	\begin{equation*}
		\frac{1}{\sqrt{P}}\sum_{t=R+1}^{T}\left(\mathbf{X}_{\tau,t,h}(\boldsymbol{\theta})\left( 1\left\{ y_{t}\leq
		\mathbf{X}_{\tau,t,h}(\boldsymbol{\theta})^{\prime }\boldsymbol{\beta} \right\} -\tau \right) -\mathrm{E}%
		\left(\mathbf{X}_{\tau,t,h}(\boldsymbol{\theta})\left( 1\left\{ y_{t}\leq \mathbf{X}_{\tau,t,h}(\boldsymbol{\theta})^{\prime }\boldsymbol{\beta}
		\right\} -\tau \right) \right)\right)
	\end{equation*}
	is stochastically equicontinuous in $\boldsymbol{\theta}\in\Theta$, $\boldsymbol{\beta} \in \mathcal{B}$, and 
	$\tau \in \mathcal{T}$ w.r.t. the $L_{2}$ pseudo-metric: 
	\begin{align*}
		&\rho_{\Theta\times \mathcal{B}\times\mathcal{T}}((\boldsymbol{\theta}_{1},\boldsymbol{\beta}_{1},\tau_{1}),(\boldsymbol{\theta}_{2},%
		\boldsymbol{\beta}_{2},\tau_{2}))^2\\
		=& \max_{l\in d}\mathrm{E}%
		\left(\left(X_{l,t,h}(\boldsymbol{\theta}_{1})\left( 1\left\{ y_{t}\leq \mathbf{X}_{\tau,t,h}(\boldsymbol{\theta}_{1})^{\prime }\boldsymbol{%
			\beta} \right\} -\tau \right)-X_{l,t,h}(\boldsymbol{\theta}_{2})\left( 1\left\{ y_{t}\leq
		\mathbf{X}_{\tau,t,h}(\boldsymbol{\theta}_{2})^{\prime }\boldsymbol{\beta}_{2} \right\} -\tau_{2}\right)\right)^2\right)
	\end{align*}
	where $X_{l,t,h}(\boldsymbol{\theta})$ denotes the $l$-th element of $\mathbf{X}_{\tau,t,h}(\boldsymbol{\theta})$ and $d$ is the
	dimension of $\mathbf{X}_{\tau,t,h}(\boldsymbol{\theta})$. \medskip
	
	\noindent \textbf{(iii)} For each $\tau \in \mathcal{T}$ and $h\in\mathcal{H}$: 
	\begin{eqnarray*}
		&&\sqrt{P}\left( \widehat{\boldsymbol{\beta }}_{h}(\tau )-\boldsymbol{%
			\beta }_{h}^{\dag }(\tau )\right)  \\
		&=&\mathbf{J}_{h}\left( \tau \right) ^{-1}\left( \frac{1}{\sqrt{P}}%
		\sum_{t=R+1}^{T}\mathbf{X}_{\tau,t,h}(\boldsymbol{\theta}_{\tau,h}^{\dag})\left( 1\left\{ y_{t}\leq \mathbf{X}_{\tau,t,h}(\boldsymbol{\theta}_{\tau,h}^{\dag})^{\prime }%
		\boldsymbol{\beta }^{\dag }\left( \tau \right) \right\} -\tau \right)
		\right) +o_{\Pr}(1),
	\end{eqnarray*}%
	where $\mathbf{J}_{h}\left( \tau \right) $ is defined in Assumption A3.
\end{lemma}

\noindent\textbf{Proof of Theorem \ref{MZTest}}: Under A1-A7 and Lemma \ref{Bahadur}, for each given $s\in\{1,\ldots,\kappa\}$, it follows under the null hypothesis that:
\[
\sqrt{P}\widehat{m}_{s}\stackrel{d}{\rightarrow}N(0,\Omega_{s}),
\]
where, for a given $h$ and  $\tau_{k}$, $\Omega_{s}$ is a diagonal element of the $2\times 2$ positive definite covariance matrix. Moreover, since for the asymptotic covariance matrix of the empirical moment equalities (scaled by $\sqrt{P}$) we have that $\boldsymbol{\Sigma}\in\mathbb{R}^{\kappa\times\kappa}$ is positive definite, and $H$ and $K$ are finite dimensional integers,  it holds under $H^{\text{MZ}}_{0}$ that:
\[
\sqrt{P}\left( \begin{array}{c}\widehat{m}_{1}(\tau_{1})\\\vdots \\\widehat{m}_{\kappa}\end{array}\right)\stackrel{d}{\rightarrow}N(\mathbf{0},\boldsymbol{\Sigma}).
\]
by an application of the Cram\'{e}r-Wold device. Thus, as a consequence of continuous mapping:
\[
\sum_{s=1}^{\kappa} \left(\sqrt{P}\widehat{m}_{s}\right)^{2}\stackrel{d}{\rightarrow}\sum_{s=1}^{\kappa} 
Z_{s}^{2},
\]
where $Z_{s}$ is an element of:
\[
\mathbf{Z}=\left(\begin{array}{c} Z_{1}\\\vdots\\ Z_{\kappa}\end{array}\right)\sim N\left(\mathbf{0},\mathbf{\Sigma}\right).
\] 
Note also that the first stage estimation error does not feature into this asymptotic variance when $\pi=0$ as shown in the proof of Lemma \ref{Bahadur}. As a result, we may directly resample from the generated forecasts.  That is, under A1 to A7 with $\epsilon>2$ and the block length condition, the first order validity of the MBB follows from Theorem 2 of \citet{GLN2018} as a special case of the SETBB using untapered blocks for each $\tau_{k}$ and $h$. That is, pointwise in $\tau_{k}\in\mathcal{T}$, $h\in\mathcal{H}$, and for any $\nu>0$: 
\[
\Pr\left(\sup_{x\in\mathbb{R}^2}\left\vert \Prb \left(\sqrt{P}(\widehat{\boldsymbol{\beta}}^{b}_{h}(\tau_{k})-\widehat{\boldsymbol{\beta}}_{h}(\tau_{k}))\leq x\right)-\Pr\left(\sqrt{P}(\widehat{\boldsymbol{\beta}}_{h}(\tau_{k})-\boldsymbol{\beta}^{\dag}_{h}(\tau_{k}))\leq x\right)\right\vert >\nu \right)\rightarrow 0,
\]
where $\Prb$ denotes the bootstrap probability measure, conditional on the original sample (defined on a given probability space $(\Omega, \mathcal{F}, \Pr)$). Now, to establish uniform convergence of the bootstrap for a finite number of quantile levels $\tau_{k}$ and horizons $h$, we need to show in addition that the bootstrap covariances adequately mimic the actual covariances, so that again by an application of the Cram\'{e}r-Wold device uniform convergence follows. To this end, denote by $(y_{k}^{b},\mathbf{X}_{\tau_{k},k,h}^{b \prime})$ the bootstrapped observations, where we suppress the dependence on the estimated parameters as we resample from the forecasts directly. Moreover, in what follows, we write $o^{b}_{\Pr}(1)$, in probability when for any $\delta>0$, $\Prb\left(|T^{b}|>\delta\right)=o_{\Pr}(1)$ for any bootstrap statistic $T^{b}$. Finally, let $\mathrm{E}^{b}\left[\cdot\right]$ and $\text{Cov}^{b}\left[\cdot\right]$ denote the expectation and covariance under the bootstrap probability measure $\Prb$, respectively. 

Thus, first note that by \citet[][p.1163]{GLN2018}, we have that:
\[
\sqrt{P}\left(\widehat{\boldsymbol{\beta}}^{b}_{h}(\tau_{k})-\widetilde{\boldsymbol{\beta}}_{h}(\tau_{k})\right)=\left(-\boldsymbol{J}_{h}^{-1}(\tau_{k})\right) \frac{1}{\sqrt{P}}\sum_{t=R+1}^{T} \mathbf{X}_{\tau_{k},t,h}^{b}\left(1\{y_{t}^{b}\leq  \mathbf{X}_{\tau_{k},t,h}^{b \prime}\widetilde{\boldsymbol{\beta}}_{h}(\tau_{k})\} -  \tau_{k}\right) + o_{\Pr}^{b}(1)
\]
in probability, where:
\[
\widetilde{\boldsymbol{\beta}}_{h}(\tau_{k})=\arg \min_{\boldsymbol{b}\in \mathcal{B}}\mathrm{E}^{b}\left[\rho _{\tau }\left( y_{t}^{b} -\mathbf{X}_{\tau_{k},t,h}^{b \prime }\boldsymbol{b}\right)\right]
\] 
is the bootstrap centering analog of $\boldsymbol{\beta}_{h}^{\dag}(\tau_{k})$.  Moreover, while $\widetilde{\boldsymbol{\beta}}_{h}(\tau_{k})$ is not the same as $\widehat{\boldsymbol{\beta}}_{h}(\tau_{k})$, for the standard MBB it holds that $\sqrt{P}\left(\widetilde{\boldsymbol{\beta}}_{h}(\tau_{k})-\widehat{\boldsymbol{\beta}}_{h}(\tau_{k})\right)=o_{\Pr}(1)$ \citep[][p.1147]{GLN2018}, so that as a result:
\begin{align}
	&\sqrt{P}\left(\widehat{\boldsymbol{\beta}}^{b}_{h}(\tau_{k})-\widehat{\boldsymbol{\beta}}_{h}(\tau_{k})\right)\notag\\
	=&\left(-\boldsymbol{J}_{h}^{-1}(\tau_{k})\right) \frac{1}{\sqrt{P}}\sum_{t=R+1}^{T} \mathbf{X}_{\tau_{k},t,h}^{b}\left(1\{y_{t}^{b}\leq  \mathbf{X}_{\tau_{k},t,h}^{b \prime}\widetilde{\boldsymbol{\beta}}_{h}(\tau_{k})\} -  \tau_{k}\right) + o_{\Pr}^{b}(1)\label{EQBOOT1}
\end{align}
in probability. Next, for some $\tau_{1},\tau_{2}\in\mathcal{T}$ and $h_{1},h_{2}\in\mathcal{H}$, we examine:
\footnotesize\begin{align*}
	&\text{Cov}^{b}\left[ \sqrt{P}\left(\widehat{\boldsymbol{\beta}}^{b}_{h_{1}}(\tau_{1})-\widehat{\boldsymbol{\beta}}_{h_{1}}(\tau_{1})\right),\sqrt{P}\left(\widehat{\boldsymbol{\beta}}^{b}_{h_{2}}(\tau_{2})-\widehat{\boldsymbol{\beta}}_{h_{2}}(\tau_{2})\right)\right]\\
	=&\mathrm{E}^{b}\left[ P\left(\widehat{\boldsymbol{\beta}}^{b}_{h_{1}}(\tau_{1})-\widehat{\boldsymbol{\beta}}_{h_{1}}(\tau_{1})\right)\times\left(\widehat{\boldsymbol{\beta}}^{b}_{h_{2}}(\tau_{2})-\widehat{\boldsymbol{\beta}}_{h_{2}}(\tau_{2})\right)\right]-\mathrm{E}^{b}\left[\sqrt{P}\left(\widehat{\boldsymbol{\beta}}^{b}_{h_{1}}(\tau_{1})-\widehat{\boldsymbol{\beta}}_{h_{1}}(\tau_{1})\right)\right]\mathrm{E}^{b}\left[\sqrt{P}\left(\widehat{\boldsymbol{\beta}}^{b}_{h_{2}}(\tau_{2})-\widehat{\boldsymbol{\beta}}_{h_{2}}(\tau_{2})\right)\right].
\end{align*}\normalsize
where, using  (\ref{EQBOOT1}) we have that:
\footnotesize\begin{align*}
	&\mathrm{E}^{b}\left[ P\left(\widehat{\boldsymbol{\beta}}^{b}_{h_{1}}(\tau_{1})-\widehat{\boldsymbol{\beta}}_{h_{1}}(\tau_{1})\right)\times\left(\widehat{\boldsymbol{\beta}}^{b}_{h_{2}}(\tau_{2})-\widehat{\boldsymbol{\beta}}_{h_{2}}(\tau_{2})\right)\right]\\
	=&\mathrm{E}^{b}\left[ \left(\left(-\boldsymbol{J}_{h_{1}}^{-1}(\tau_{1})\right) \frac{1}{\sqrt{P}}\sum_{t=R+1}^{T} \mathbf{X}_{\tau_{1},t,h_{1}}^{b}\left(1\{y_{t}^{b}\leq  \mathbf{X}_{\tau_{1},t,h_{1}}^{b \prime}\widetilde{\boldsymbol{\beta}}_{h_{1}}(\tau_{1})\} -  \tau_{1}\right)\right)\right.\\
	&\left.\times \left( \frac{1}{\sqrt{P}}\sum_{t=R+1}^{T} \mathbf{X}_{\tau_{2},t,h_{2}}^{b \prime}\left(1\{y_{t}^{b}\leq  \mathbf{X}_{\tau_{2},t,h_{2}}^{b \prime}\widetilde{\boldsymbol{\beta}}_{h_{2}}(\tau_{2})\} -  \tau_{2}\right)\left(-\boldsymbol{J}_{h_{2}}^{-1}(\tau_{2})\right)\right)\right] + o_{\Pr}(1)\equiv I_{P}^{b} + o_{\Pr}(1),
\end{align*}\normalsize
and  similarly:
\footnotesize\begin{align*}
	&\mathrm{E}^{b}\left[ P\left(\widehat{\boldsymbol{\beta}}^{b}_{h_{1}}(\tau_{1})-\widehat{\boldsymbol{\beta}}_{h_{1}}(\tau_{1})\right)\right]\times\mathrm{E}^{b}\left[\left(\widehat{\boldsymbol{\beta}}^{b}_{h_{2}}(\tau_{2})-\widehat{\boldsymbol{\beta}}_{h_{2}}(\tau_{2})\right)\right]\\
	=&\mathrm{E}^{b}\left[ \left(\left(-\boldsymbol{J}_{h_{1}}^{-1}(\tau_{1})\right) \frac{1}{\sqrt{P}}\sum_{t=R+1}^{T} \mathbf{X}_{\tau_{1},t,h_{1}}^{b}\left(1\{y_{t}^{b}\leq  \mathbf{X}_{\tau_{1},t,h_{1}}^{b \prime}\widetilde{\boldsymbol{\beta}}_{h_{1}}(\tau_{1})\} -  \tau_{1}\right)\right)\right]\\
	&\times \mathrm{E}^{b}\left[\left( \frac{1}{\sqrt{P}}\sum_{t=R+1}^{T} \mathbf{X}_{\tau_{2},t,h_{2}}^{b \prime}\left(1\{y_{t}^{b}\leq  \mathbf{X}_{\tau_{2},t,h_{2}}^{b \prime}\widetilde{\boldsymbol{\beta}}_{h_{2}}(\tau_{2})\} -  \tau_{2}\right)\left(-\boldsymbol{J}_{h_{2}}^{-1}(\tau_{2})\right)\right)\right] + o_{\Pr}(1)\equiv  II_{P}^{b} + o_{\Pr}(1).
\end{align*}\normalsize
Next, observe that by the properties of the MBB, it holds that:
\footnotesize\begin{align*}
	&I_{P}^{b} - II_{P}^{b}\\
	=& \mathrm{E}^{b}\left[\left(\left(-\boldsymbol{J}_{h_{1}}^{-1}(\tau_{1})\right) \frac{1}{\sqrt{P}}\sum_{k=1}^{K_{b}}\sum_{i=1}^{l} \mathbf{X}_{\tau_{1},I_{k}+i,h_{1}}\left(1\{y_{I_{k}+i}\leq  \mathbf{X}_{\tau_{1},I_{k}+i,h_{1}}^{\prime}\widetilde{\boldsymbol{\beta}}_{h_{1}}(\tau_{1})\} -  \tau_{1}\right)\right)\right.\\
	&\left.\times \left( \frac{1}{\sqrt{P}}\sum_{k=1}^{K_{b}}\sum_{i=1}^{l}  \mathbf{X}_{\tau_{2},I_{k}+i,h_{2}}^{ \prime}\left(1\{y_{I_{k}+i}\leq  \mathbf{X}_{\tau_{2},I_{k}+i,h_{2}}^{\prime}\widetilde{\boldsymbol{\beta}}_{h_{2}}(\tau_{2})\} -  \tau_{2}\right)\left(-\boldsymbol{J}_{h_{2}}^{-1}(\tau_{2})\right)\right)\right]\\
	&-\mathrm{E}^{b}\left[\left(-\boldsymbol{J}_{h_{1}}^{-1}(\tau_{1})\right) \frac{1}{\sqrt{P}}\sum_{t=1}^{K_{b}}\sum_{i=1}^{l} \mathbf{X}_{\tau_{1},I_{t}+i,h_{1}}\left(1\{y_{I_{t}+i}\leq  \mathbf{X}_{\tau_{1},I_{t}+i,h_{1}}^{\prime}\widetilde{\boldsymbol{\beta}}_{h_{1}}(\tau_{1})\} -  \tau_{1}\right)\right]\\
	&\times \mathrm{E}^{b}\left[ \frac{1}{\sqrt{P}}\sum_{s=1}^{K_{b}}\sum_{j=1}^{l}  \mathbf{X}_{\tau_{2},I_{s}+j,h_{2}}^{ \prime}\left(1\{y_{I_{s}+j}\leq  \mathbf{X}_{\tau_{2},I_{s}+j,h_{2}}^{\prime}\widetilde{\boldsymbol{\beta}}_{h_{2}}(\tau_{2})\} -  \tau_{2}\right)\left(-\boldsymbol{J}_{h_{2}}^{-1}(\tau_{2})\right)\right]\\
	=&\left(-\boldsymbol{J}_{h_{1}}^{-1}(\tau_{1})\right)\mathrm{E}^{b}\left[\frac{1}{l(T-l+1)} \sum_{k=1}^{K_{b}}\sum_{j=1}^{l}\sum_{i=1}^{l}\mathbf{X}_{\tau_{1},I_{k}+i,h_{1}}\mathbf{X}_{\tau_{2},I_{k}+j,h_{2}}^{ \prime}\left(1\{y_{I_{k}+i}\leq  \mathbf{X}_{\tau_{1},I_{k}+i,h_{1}}^{\prime}\widetilde{\boldsymbol{\beta}}_{h_{1}}(\tau_{1})\} -  \tau_{1}\right)\right.\\
	&\left.\times\left(1\{y_{I_{k}+j}\leq  \mathbf{X}_{\tau_{2},I_{k}+j,h_{2}}^{\prime}\widetilde{\boldsymbol{\beta}}_{h_{2}}(\tau_{2})\} -  \tau_{2}\right)\right]\left(-\boldsymbol{J}_{h_{2}}^{-1}(\tau_{2})\right) + o_{\Pr}(1),
\end{align*}\normalsize
where the last line follows from the fact that the blocks are, conditional on the sample, independent, and because $\mathrm{E}\left[\mathbf{X}_{\tau,t,h}\left(1\{y_{t}\leq  \mathbf{X}_{\tau,t,h}^{\prime}\boldsymbol{\beta}^{\dag}_{h}(\tau)\} -  \tau\right)\right]=0$
by A5. Now, from MBB-Lemma A.2 in \citet{F1997} (which also applies to vectors), we have that the last expression is equal to:
\footnotesize\begin{align*}
	&\left(-\boldsymbol{J}_{h_{1}}^{-1}(\tau_{1})\right) \left(\sum_{j=-l+1}^{l-1}\left(1 - \frac{\vert j\vert }{l}\right)\left(\frac{1}{P}\sum_{t=R+1}^{T-\vert j\vert}\mathbf{X}_{\tau_{1},t,h_{1}}\mathbf{X}_{\tau_{2},t+\vert j \vert,h_{2}}^{ \prime}\left(1\{y_{t}\leq  \mathbf{X}_{\tau_{1},t,h_{1}}^{\prime}\widetilde{\boldsymbol{\beta}}_{h_{1}}(\tau_{1})\} -  \tau_{1}\right)\right.\right.\\
	&\left.\left.\times\left(1\{y_{t+\vert j\vert }\leq  \mathbf{X}_{\tau_{2},t+\vert j\vert ,h_{2}}^{\prime}\widetilde{\boldsymbol{\beta}}_{h_{2}}(\tau_{2})\} -  \tau_{2}\right)\right)\right)\left(-\boldsymbol{J}_{h_{2}}^{-1}(\tau_{2})\right) + O_{\Pr}\left(\frac{l^{2}}{P}\right)\\
	=&\left(-\boldsymbol{J}_{h_{1}}^{-1}(\tau_{1})\right)\left(\frac{1}{P}\sum_{t=R+1}^{T-l+1} \sum_{j=-l+1}^{l-1} \left(1 - \frac{\vert j\vert }{l}\right) \mathbf{X}_{\tau_{1},t,h_{1}}\mathbf{X}_{\tau_{2},t+j,h_{2}}^{ \prime}\left(1\{y_{t}\leq  \mathbf{X}_{\tau_{1},t,h_{1}}^{\prime}\widetilde{\boldsymbol{\beta}}_{h_{1}}(\tau_{1})\} -  \tau_{1}\right)\right.\\
	&\left.\times\left(1\{y_{t+j }\leq  \mathbf{X}_{\tau_{2},t+j ,h_{2}}^{\prime}\widetilde{\boldsymbol{\beta}}_{h_{2}}(\tau_{2})\} -  \tau_{2}\right)\right)\left(-\boldsymbol{J}_{h_{2}}^{-1}(\tau_{2})\right) + O_{\Pr}\left(\frac{l^{2}}{P}\right).
\end{align*}\normalsize
Now adding and subtracting:
\[
\left(-\boldsymbol{J}_{h_{1}}^{-1}(\tau_{1})\right)\sum_{k=-l}^{+l}  \boldsymbol{\Gamma}(k)\left(-\boldsymbol{J}_{h_{2}}^{-1}(\tau_{2})\right)
\]
with
\footnotesize\[
\boldsymbol{\Gamma}(k)= \mathrm{E}\left[\mathbf{X}_{\tau_{1},t,h_{1}}\mathbf{X}_{\tau_{2},t+k,h_{2}}^{ \prime}\left(1\{y_{t}\leq  \mathbf{X}_{\tau_{1},t,h_{1}}^{\prime}\boldsymbol{\beta}^{\dag}_{h_{1}}(\tau_{1})\} -  \tau_{1}\right)\left(1\{y_{t+k }\leq  \mathbf{X}_{\tau_{2},t+k ,h_{2}}^{\prime}\boldsymbol{\beta}^{\dag}_{h_{2}}(\tau_{2})\} -  \tau_{2}\right)\right]
\]\normalsize
we obtain:
\footnotesize\begin{align*}
	&\left(-\boldsymbol{J}_{h_{1}}^{-1}(\tau_{1})\right)\left(\frac{1}{P}\sum_{t=R+1}^{T-l+1} \sum_{j=-l+1}^{l-1} \left(1 - \frac{\vert j\vert }{l}\right) \mathbf{X}_{\tau_{1},t,h_{1}}\mathbf{X}_{\tau_{2},t+j,h_{2}}^{ \prime}\left(1\{y_{t}\leq  \mathbf{X}_{\tau_{1},t,h_{1}}^{\prime}\widetilde{\boldsymbol{\beta}}_{h_{1}}(\tau_{1})\} -  \tau_{1}\right)\right.\\
	&\left.\times\left(1\{y_{t+j }\leq  \mathbf{X}_{\tau_{2},t+j ,h_{2}}^{\prime}\widetilde{\boldsymbol{\beta}}_{h_{2}}(\tau_{2})\} -  \tau_{2}\right) -   \boldsymbol{\Gamma}(j)\right)\left(-\boldsymbol{J}_{h_{2}}^{-1}(\tau_{2})\right) +\left(-\boldsymbol{J}_{h_{1}}^{-1}(\tau_{1})\right)\sum_{k=-l}^{+l}  \boldsymbol{\Gamma}(k)\left(-\boldsymbol{J}_{h_{2}}^{-1}(\tau_{2})\right) + O_{\Pr}\left(\frac{l^{2}}{P}\right).
\end{align*}\normalsize
Since $\widetilde{\boldsymbol{\beta}}_{h}(\tau_{k}) \stackrel{\Pr}{\rightarrow} \boldsymbol{\beta}_{h}^{\dag}(\tau_{k})$ for any $\tau_{k}\in\mathcal{T}$ and $h\in\mathcal{H}$, the first term converges in probability to zero, while the second term converges to the asymptotic long run covariance as $T,l\rightarrow \infty$. Since the number of quantile levels and horizons is finite and $\boldsymbol{\Sigma}$ is positive definite, uniform convergence now follows by an application of the Cram\'{e}r-Wold device, see for instance \citet{W00}. Since $\widehat{m}^{b}_{s}$ either $\widehat{m}^{b}_{s}=\widehat{\alpha}_{h}^{b}(\tau_{k})$ or $\widehat{m}^{b}_{s}=\widehat{\beta}_{h}^{b}(\tau_{k})-1$, the result now follows by continuous mapping. \qedsymbol

\bigskip

\noindent \textbf{Proof of Lemma \ref{Bahadur}:}\footnote{For notational simplicity, we will not make the dependence of $\mathbf{X}_{\tau_{k},t,h}$ on the forecast $\widehat{y}_{\tau_{k},t,h}$ explicit, and just write $\mathbf{X}_{t}$ instead.} \medskip

\noindent \textbf{(i)} First, observe that for any value $\boldsymbol{\beta}\in\mathcal{B}$, $\tau\in\mathcal{T}$, and $h\in\mathcal{H}$:
\begin{equation}\label{EQConsistency}
	Q_{P}(\widehat{\boldsymbol{\theta}}_{\tau,j,h};\tau ,\boldsymbol{\beta} )= Q_{P}(\boldsymbol{\theta}_{\tau,h}^{\dag};\tau ,\boldsymbol{\beta} )- \left(Q_{P}(\widehat{\boldsymbol{\theta}}_{\tau,j,h};\tau ,\boldsymbol{\beta} )- Q_{P}(\boldsymbol{\theta}_{\tau,h}^{\dag};\tau ,\boldsymbol{\beta} )\right)
\end{equation}
where:
\begin{equation*}
	\widehat{Q}_{P}(\widehat{\boldsymbol{\theta}}_{\tau,j,h};\tau ,\boldsymbol{\beta} )\equiv \frac{1}{P}\sum_{j=R+1}^{T}\rho _{\tau
	}(y_{j}-\mathbf{X}_{j}(\widehat{\boldsymbol{\theta}}_{\tau,j,h})^{\prime }\boldsymbol{\beta} )
\end{equation*}
and
\begin{equation*}
	Q_{P}(\boldsymbol{\theta}_{\tau,h}^{\dag};\tau ,\boldsymbol{\beta} )\equiv \frac{1}{P}\sum_{j=R+1}^{P}\rho _{\tau
	}(y_{j}-\mathbf{X}_{j}(\boldsymbol{\theta}_{\tau,h}^{\dag})^{\prime }\boldsymbol{\beta} ).
\end{equation*}
Next, note that:
\begin{align*}
	&	\sup_{\widehat{\boldsymbol{\theta}}_{\tau,j,h};\;  j\geq R+1}\sup_{\boldsymbol{\beta}\in\boldsymbol{\mathcal{B}}}\left\vert\left(Q_{P}(\widehat{\boldsymbol{\theta}}_{\tau,j,h};\tau ,\boldsymbol{\beta} )- Q_{P}(\boldsymbol{\theta}_{\tau,h}^{\dag};	\tau ,\boldsymbol{\beta} )\right)\right\vert\\
	\leq&\sup_{\widehat{\boldsymbol{\theta}}_{\tau,j,h};\;  j\geq R+1}\sup_{\boldsymbol{\beta}\in\boldsymbol{\mathcal{B}}}\left( \frac{1}{P}\sum_{j=R+1}^{T}\left\vert \rho_{\tau}\left( y_{j}-\mathbf{X}_{j}(\widehat{\boldsymbol{\theta}}_{\tau,j,h})^{\prime }\boldsymbol{\beta}\right)-\rho_{\tau}\left( y_{j}-\mathbf{X}_{j}(\boldsymbol{\theta}_{\tau,h}^{\dag})^{\prime }\boldsymbol{\beta}\right)\right\vert\right)\\
	\leq& \sup_{\widehat{\boldsymbol{\theta}}_{\tau,j,h};\;  j\geq R+1}\left(\frac{1}{P}\sum_{j=R+1}^{T}\left\Vert \mathbf{X}_{j}(\widehat{\boldsymbol{\theta}}_{\tau,j,h})-\mathbf{X}_{j}(\boldsymbol{\theta}_{\tau,h}^{\dag})\right\Vert\right)\times  \sup_{\boldsymbol{\beta}\in\boldsymbol{\mathcal{B}}}\Vert \boldsymbol{\beta}\Vert \times\max\{\tau,1-\tau\} \\
	\leq &\frac{1}{P}\sum_{j=R+1}^{T} \sup_{\widehat{\boldsymbol{\theta}}_{\tau,j,h};\;  j\geq R+1}\left\Vert \mathbf{X}_{j}(\widehat{\boldsymbol{\theta}}_{\tau,j,h})-\mathbf{X}_{j}(\boldsymbol{\theta}_{\tau,h}^{\dag})\right\Vert\times  \sup_{\boldsymbol{\beta}\in\boldsymbol{\mathcal{B}}}\Vert \boldsymbol{\beta}\Vert \\
	\leq & C\cdot O_{\Pr}\left(R^{-\frac{1}{2}}\right) 
\end{align*}
for some generic constant $C>0$, where the second inequality follows from the properties of the `check function', the third one from Jensen's inequality, and the last one from Markov's inequality and Assumptions A3 and A6.  The rest of the proof can now proceed by standard arguments from the quantile regression literature. In particular, observe that the definition of $Q_{P}(\boldsymbol{\theta}_{\tau,h}^{\dag}; \tau ,\boldsymbol{\beta} )$ (and $Q_{P}(\widehat{\boldsymbol{\theta}}_{\tau,j,h}; \tau ,\boldsymbol{\beta} )$) may be changed to:
\[
Q_{P}(\boldsymbol{\theta}_{\tau,h}^{\dag}; \tau ,\boldsymbol{\beta} )\equiv \frac{1}{P}\sum_{j=R+1}^{T}\left( \rho _{\tau
}(y_{j}-\mathbf{X}_{j}(\boldsymbol{\theta}_{\tau,h}^{\dag})^{\prime }\boldsymbol{\beta} )-\rho _{\tau
}(y_{j}-\mathbf{X}_{j}(\boldsymbol{\theta}_{\tau,h}^{\dag})^{\prime }\boldsymbol{\beta}_{h}^{\dag}(\tau) )\right)
\]
without affecting minimisation or any of the preceeding arguments. In addition, note that by A2, A3, and A4, it follows that: 
\begin{equation*}
	Q_{\infty}(\tau ,\boldsymbol{\beta} )\equiv \mathrm{E}\left[ \left( \rho _{\tau
	}(y_{j}-\mathbf{X}_{j}(\boldsymbol{\theta}_{\tau,h}^{\dag})^{\prime }\boldsymbol{\beta} )-\rho _{\tau
	}(y_{j}-\mathbf{X}_{j}(\boldsymbol{\theta}_{\tau,h}^{\dag})^{\prime }\boldsymbol{\beta}_{h}^{\dag}(\tau) )\right)\right]=O(1) .
\end{equation*}%
As a result, pointwise in $\boldsymbol{\beta} $, $\tau$, $h$, $Q_{P}(\boldsymbol{\theta}_{\tau,h}^{\dag}; \tau ,\boldsymbol{\beta} )\overset{\Pr}{%
	\rightarrow}Q_{\infty ,\boldsymbol{\theta}^{\dag}}(\tau ,\boldsymbol{\beta} )$ by A1, A3, and McLeish's law of
large numbers for strong mixing processes.  Consistency of $\widehat{\boldsymbol{\beta} }$ for $\boldsymbol{\beta}_{h}^{\dag}(\tau)$ pointwise in $\tau$ and $h$ then follows since $Q_{\infty}(\tau ,\boldsymbol{\beta} )$ is uniquely minimised at $\boldsymbol{\beta}_{h}^{\dag}(\tau)$ by the stated assumptions. 
\medskip

\noindent \textbf{(ii)} We prove part (ii) by appealing to Theorem 1, application 4, case (2.15) of \citet{DMR1995}.  First, define the function class:
\[
\mathcal{F}=\left\{\mathbf{X}_{t}(\boldsymbol{\theta}) \left(1\{y_{t} \leq \mathbf{X}_{t}(\boldsymbol{\theta})^{\prime}\boldsymbol{\beta}\} - \tau\right):\; \boldsymbol{\theta}\in \boldsymbol{\Theta},\;\boldsymbol{\beta}\in\mathcal{B},\;\tau\in \mathcal{T}\right\}.
\]  
Moreover, note that, for some $\epsilon>1$, by assumption A2:
\begin{align*}
	&\mathrm{E}\left[\sup_{(\boldsymbol{\theta},\boldsymbol{\beta},\tau)\in \boldsymbol{\Theta}\times \mathcal{B}\times \mathcal{T}}\left\Vert\mathbf{X}_{t}(\boldsymbol{\theta}) \left(1\{y_{t} \leq \mathbf{X}_{t}(\boldsymbol{\theta})^{\prime}\boldsymbol{\beta}\} - \tau\right)\right\Vert^{2\epsilon} \right]\\
	\leq &C \mathrm{E}\left[\sup_{\boldsymbol{\theta}\in \boldsymbol{\Theta}}\left\Vert\mathbf{X}_{t}(\boldsymbol{\theta}) \right\Vert^{2\epsilon} \right]<\infty,
\end{align*}
the envelope condition is satisfied. Now, in what follows, let: 
\[
\sup_{\boldsymbol{\theta}^{\ast},\boldsymbol{\beta}^{\ast},\tau^{\ast}}\equiv \sup_{\substack{(\boldsymbol{\theta}^{\ast},\boldsymbol{\beta}^{\ast},\tau^{\ast})\in \boldsymbol{\Theta}\times \mathcal{B}\times \mathcal{T},\\
		\Vert \boldsymbol{\theta}^{\ast} - \boldsymbol{\theta}\Vert \leq r_{1}, \Vert \boldsymbol{\beta}^{\ast} - \boldsymbol{\beta}\Vert \leq r_{2},\vert \tau^{\ast} - \tau\vert \leq r_{3},\\
		\sqrt{r_{1}^2 +r_{2}^2}\leq \overline{r}, (r_{1}^{2}  +r_{3}^2 )\leq \widetilde{r},\\ \sqrt{\overline{r}+\widetilde{r}^2 }\leq r}}.
\]
Note that by the $c_{r}$ inequality:
\begin{align}
	&\mathrm{E}\left[ \sup_{\boldsymbol{\theta}^{\ast},\boldsymbol{\beta}^{\ast},\tau^{\ast}}\left\Vert \mathbf{X}_{t}(\boldsymbol{\theta}^{\ast}) \left(1\{y_{t} \leq \mathbf{X}_{t}(\boldsymbol{\theta}^{\ast})^{\prime}\boldsymbol{\beta}^{\ast}\} - \tau^{\ast}\right)-\mathbf{X}_{t}(\boldsymbol{\theta}) \left(1\{y_{t} \leq \mathbf{X}_{t}(\boldsymbol{\theta})^{\prime}\boldsymbol{\beta}\} - \tau\right) \right\Vert^2 \right]\notag\\
	\leq &2 \left( \mathrm{E}\left[ \sup_{\boldsymbol{\theta}^{\ast},\boldsymbol{\beta}^{\ast}}\left\Vert \mathbf{X}_{t}(\boldsymbol{\theta}^{\ast}) 1\{y_{t} \leq \mathbf{X}_{t}(\boldsymbol{\theta}^{\ast})^{\prime}\boldsymbol{\beta}^{\ast}\} - \mathbf{X}_{t}(\boldsymbol{\theta}) 1\{y_{t} \leq \mathbf{X}_{t}(\boldsymbol{\theta})^{\prime}\boldsymbol{\beta}\}  \right\Vert^2 \right]\right.\notag\\
	&\left. + \mathrm{E}\left[ \sup_{\boldsymbol{\theta}^{\ast},\tau^{\ast}}\left\Vert \mathbf{X}_{t}(\boldsymbol{\theta}) \tau-\mathbf{X}_{t}(\boldsymbol{\theta}^{\ast})  \tau^{\ast} \right\Vert^2 \right]\right)\label{EQL2CONT}
\end{align}
We focus on the first term on the Right Hand Side (RHS) of the last line, the second term will follow by similar arguments. To start with, note that:
\begin{align*}
	& \mathrm{E}\left[ \sup_{\boldsymbol{\theta}^{\ast},\boldsymbol{\beta}^{\ast}}\left\Vert \mathbf{X}_{t}(\boldsymbol{\theta}^{\ast}) 1\{y_{t} \leq \mathbf{X}_{t}(\boldsymbol{\theta}^{\ast})^{\prime}\boldsymbol{\beta}^{\ast}\} - \mathbf{X}_{t}(\boldsymbol{\theta}) 1\{y_{t} \leq \mathbf{X}_{t}(\boldsymbol{\theta})^{\prime}\boldsymbol{\beta}\}  \right\Vert^2 \right]\\
	= & \mathrm{E}\left[ \sup_{\boldsymbol{\theta}^{\ast},\boldsymbol{\beta}^{\ast}}\left\Vert \mathbf{X}_{t}(\boldsymbol{\theta}^{\ast}) 1\{y_{t} \leq \mathbf{X}_{t}(\boldsymbol{\theta}^{\ast})^{\prime}\boldsymbol{\beta}^{\ast}\} - \mathbf{X}_{t}(\boldsymbol{\theta}) 1\{y_{t} \leq \mathbf{X}_{t}(\boldsymbol{\theta}^{\ast})^{\prime}\boldsymbol{\beta}^{\ast}\}  \right.\right.\\
	&\left.\left.+ \mathbf{X}_{t}(\boldsymbol{\theta}) 1\{y_{t} \leq \mathbf{X}_{t}(\boldsymbol{\theta}^{\ast})^{\prime}\boldsymbol{\beta}^{\ast}\} - \mathbf{X}_{t}(\boldsymbol{\theta}) 1\{y_{t} \leq \mathbf{X}_{t}(\boldsymbol{\theta})^{\prime}\boldsymbol{\beta}\}  \right\Vert^2 \right]\\
	\leq &2\left(\mathrm{E}\left[ \sup_{\boldsymbol{\theta}^{\ast},\boldsymbol{\beta}^{\ast}}\left\Vert \mathbf{X}_{t}(\boldsymbol{\theta}^{\ast}) 1\{y_{t} \leq \mathbf{X}_{t}(\boldsymbol{\theta}^{\ast})^{\prime}\boldsymbol{\beta}^{\ast}\} - \mathbf{X}_{t}(\boldsymbol{\theta}) 1\{y_{t} \leq \mathbf{X}_{t}(\boldsymbol{\theta}^{\ast})^{\prime}\boldsymbol{\beta}^{\ast}\}  \right\Vert^2 \right] \right.\\
	&\left. +\mathrm{E}\left[ \sup_{\boldsymbol{\theta}^{\ast},\boldsymbol{\beta}^{\ast}}\left\Vert \mathbf{X}_{t}(\boldsymbol{\theta}) 1\{y_{t} \leq \mathbf{X}_{t}(\boldsymbol{\theta}^{\ast})^{\prime}\boldsymbol{\beta}^{\ast}\} - \mathbf{X}_{t}(\boldsymbol{\theta}) 1\{y_{t} \leq \mathbf{X}_{t}(\boldsymbol{\theta})^{\prime}\boldsymbol{\beta}\}  \right\Vert^2 \right]\right)
\end{align*}
For the first term, we have by A6:
\begin{align*}
	&2\mathrm{E}\left[ \sup_{\boldsymbol{\theta}^{\ast},\boldsymbol{\beta}^{\ast}}\left\Vert \mathbf{X}_{t}(\boldsymbol{\theta}^{\ast}) 1\{y_{t} \leq \mathbf{X}_{t}(\boldsymbol{\theta}^{\ast})^{\prime}\boldsymbol{\beta}^{\ast}\} - \mathbf{X}_{t}(\boldsymbol{\theta}) 1\{y_{t} \leq \mathbf{X}_{t}(\boldsymbol{\theta}^{\ast})^{\prime}\boldsymbol{\beta}^{\ast}\}  \right\Vert^2 \right]\\
	\leq &C \sup_{\boldsymbol{\theta}^{\ast}} \Vert \boldsymbol{\theta}^{\ast}-\boldsymbol{\theta}\Vert^2 \mathrm{E}\left[ B(\mathbf{X}_{t})^2 \right]\\
	\leq &\widetilde{C}r_{1}^2
\end{align*}
for some generic constants $C$ and $\widetilde{C}$. For the second term on the other hand, we have that:
\begin{align*}
	&2\mathrm{E}\left[ \sup_{\boldsymbol{\theta}^{\ast},\boldsymbol{\beta}^{\ast}}\left\Vert \mathbf{X}_{t}(\boldsymbol{\theta}) 1\{y_{t} \leq \mathbf{X}_{t}(\boldsymbol{\theta}^{\ast})^{\prime}\boldsymbol{\beta}^{\ast}\} - \mathbf{X}_{t}(\boldsymbol{\theta}) 1\{y_{t} \leq \mathbf{X}_{t}(\boldsymbol{\theta})^{\prime}\boldsymbol{\beta}\}  \right\Vert^2 \right]\\
	\leq &2\left(\sup_{\boldsymbol{\theta}}\mathrm{E}\left[\left\Vert \mathbf{X}_{t}(\boldsymbol{\theta})  \right\Vert^4 \right]\right)^{\frac{1}{2}}\left(\mathrm{E}\left[\sup_{\boldsymbol{\theta}^{\ast},\boldsymbol{\beta}^{\ast}}\left\vert 1\{y_{t} \leq \mathbf{X}_{t}(\boldsymbol{\theta}^{\ast})^{\prime}\boldsymbol{\beta}^{\ast}\} - 1\{y_{t} \leq \mathbf{X}_{t}(\boldsymbol{\theta})^{\prime}\boldsymbol{\beta}\}  \right\vert \right]\right)^{\frac{1}{2}}\\
	\leq &2 C \left(\mathrm{E}\left[\sup_{\boldsymbol{\theta}^{\ast},\boldsymbol{\beta}^{\ast}}\left\vert 1\{\left( y_{t} -\mathbf{X}_{t}(\boldsymbol{\theta})^{\prime}\boldsymbol{\beta}\right) \leq \left(\mathbf{X}_{t}(\boldsymbol{\theta}^{\ast})^{\prime}\boldsymbol{\beta}^{\ast}-\mathbf{X}_{t}(\boldsymbol{\theta})^{\prime}\boldsymbol{\beta}\right)\} - 1\{\left( y_{t} - \mathbf{X}_{t}(\boldsymbol{\theta})^{\prime}\boldsymbol{\beta}\right) \leq 0\}  \right\vert \right]\right)^{\frac{1}{2}}\\
	=& 2 C\left(\mathrm{E}\left[\sup_{\boldsymbol{\theta}^{\ast},\boldsymbol{\beta}^{\ast}} 1\{\left\vert y_{t} -\mathbf{X}_{t}(\boldsymbol{\theta})^{\prime}\boldsymbol{\beta}\right\vert \leq \left\vert \mathbf{X}_{t}(\boldsymbol{\theta}^{\ast})^{\prime}\boldsymbol{\beta}^{\ast}-\mathbf{X}_{t}(\boldsymbol{\theta})^{\prime}\boldsymbol{\beta}\right\vert\}  \right]\right)^{\frac{1}{2}}\\
	\leq & 2 C\left(\mathrm{E}\left[\sup_{\boldsymbol{\theta}^{\ast},\boldsymbol{\beta}^{\ast}} 1\{\left\vert y_{t} -\mathbf{X}_{t}(\boldsymbol{\theta})^{\prime}\boldsymbol{\beta}\right\vert \leq \left\Vert  \mathbf{X}_{t}(\boldsymbol{\theta}^{\ast})-\mathbf{X}_{t}(\boldsymbol{\theta})\right\Vert \Vert \boldsymbol{\beta}^{\ast}\Vert + \left\Vert \mathbf{X}_{t}(\boldsymbol{\theta})\right\Vert \left\Vert \boldsymbol{\beta}^{\ast}- \boldsymbol{\beta}\right\Vert\}  \right]\right)^{\frac{1}{2}},
\end{align*}
where the second inequality uses A2, while the third inequality uses the triangle inequality. Now, by A4 and A6, we have that the last line is bounded by:
\begin{align*}
	&2 C\left(\mathrm{E}\left[\sup_{\boldsymbol{\theta}^{\ast},\boldsymbol{\beta}^{\ast}} 1\{\left\vert y_{t} -\mathbf{X}_{t}(\boldsymbol{\theta})^{\prime}\boldsymbol{\beta}\right\vert \leq \left\vert B(\mathbf{X}_{t,\tau,h})\right\vert r_{1} + \left\Vert \mathbf{X}_{t}(\boldsymbol{\theta})\right\Vert r_{2} \}  \right]\right)^{\frac{1}{2}}\\
	\leq & \widetilde{C} \left(\sqrt{2}\left( \mathrm{E}\left[ B(\mathbf{X}_{t,\tau,h})\right] \vee \mathrm{E}\left[\left\Vert \mathbf{X}_{t}(\boldsymbol{\theta})\right\Vert \right]\right) \left(r_{1}^{2}+ r_{2}^{2}\right)^{\frac{1}{2}}\right)^{\frac{1}{2}}\\
	\leq & \widetilde{\widetilde{C}} \overline{r}^{\frac{1}{2}}
\end{align*}
Thus, using similar arguments for the RHS of Equation (\ref{EQL2CONT}), we obtain that the Left Hand Side of that equation can be bounded by:
\begin{equation*}
	\widetilde{\widetilde{C}} \overline{r}^{\frac{1}{2}} + \widetilde{C}r_{1}^{2} + C r_{3}^{2}
	\leq \widetilde{\widetilde{C}} \overline{r}^{\frac{1}{2}} +  \overline{C}\widetilde{r}
	\leq  \sqrt{2}\left(\widetilde{\widetilde{C}}\vee\overline{C}\right)\sqrt{\overline{r}+\widetilde{r}^2}
	\leq \overline{\overline{C}}r,
\end{equation*}
where all constants are again generic and may differ from previous displays. As noted in \citet[][p. 121]{AP1994}, the $L_{2}$ continuity implies that the bracketing numbers satisfy:
\[
N(\varepsilon,\mathcal{F}) \leq \overline{\overline{C}} \left(\frac{1}{\varepsilon}\right)^{d_{\boldsymbol{\beta}}+d_{\boldsymbol{\theta}}+1},
\]
where $d_{\boldsymbol{\beta}}$ and $d_{\boldsymbol{\theta}}$ denote the dimensions of the corresponding parameter vectors. Thus, the integrability condition for the bracketing numbers in \citet{DMR1995} holds. Finally, since the $\beta$ mixing condition in A1 satisfies the mixing condition of case (2.15), the result holds by Theorem 1 of \citet{DMR1995}.\medskip

\noindent \textbf{(iii)} By Equation (\ref{EQEST}) and A2  \citep[see e.g.][]{GLN2018}, it holds that: 
\begin{eqnarray}
	&&\left\Vert \frac{1}{\sqrt{P}}\sum_{j=R+1}^{T}\mathbf{X}_{j}(\widehat{\boldsymbol{\theta}}_{\tau,t,h})\left( 1\left\{
	y_{j}\leq \mathbf{X}_{j}(\widehat{\boldsymbol{\theta}}_{\tau,t,h})^{\prime }\widehat{\boldsymbol{\beta} }_{P}\left( \tau \right) \right\}
	-\tau \right) \right\Vert \label{EQFIRSTORDER}\\
	&\leq &d \max_{R+1\leq j\leq T}\left\Vert \mathbf{X}_{j}(\boldsymbol{\theta}_{\tau,h}^{\dag})\right\Vert \frac{1}{\sqrt{P}}%
	\sum_{j=R+1}^{T}1\left\{ y_{j}=\mathbf{X}_{j}(\widehat{\boldsymbol{\theta}}_{\tau,t,h})^{\prime }\widehat{\boldsymbol{\beta} }_{P}\left(
	\tau \right) \right\} +o_{\Pr}(1)=o_{\Pr}(1),\notag
\end{eqnarray}%
where the inequality follows from A6, and the last equality follows again from the fact that for every $\tau\in\mathcal{T}$ and $h\in\mathcal{H}$:
\[
\max_{R+1\leq j\leq T}\left\Vert \mathbf{X}_{j}(\boldsymbol{\theta}_{\tau,h}^{\dag})\right\Vert=o_{\Pr}(P^{\frac{1}{2}}).
\] 
Moreover, we have that:\small
\begin{align*}
	&\frac{1}{\sqrt{P}}\sum_{j=R+1}^{T} \mathbf{X}_{j}(\widehat{\boldsymbol{\theta}}_{\tau,t,h})\left( 1\left\{ y_{j}\leq
	\mathbf{X}_{j}(\widehat{\boldsymbol{\theta}}_{\tau,t,h})^{\prime }\widehat{\boldsymbol{\beta} }_{P}\left( \tau \right) \right\} -\tau
	\right)\\
	=&\frac{1}{\sqrt{P}}\sum_{j=R+1}^{T} \mathbf{X}_{j}(\boldsymbol{\theta}_{\tau,h}^{\dag})\left( 1\left\{ y_{j}\leq
	\mathbf{X}_{j}(\boldsymbol{\theta}_{\tau,h}^{\dag})^{\prime }\widehat{\boldsymbol{\beta} }_{P}\left( \tau \right) \right\} -\tau
	\right)+\\
	&\left(\frac{1}{\sqrt{P}}\sum_{j=R+1}^{T}\left( \mathbf{X}_{j}(\widehat{\boldsymbol{\theta}}_{\tau,t,h})\left( 1\left\{ y_{j}\leq
	\mathbf{X}_{j}(\widehat{\boldsymbol{\theta}}_{\tau,t,h})^{\prime }\widehat{\boldsymbol{\beta} }_{P}\left( \tau \right) \right\} -\tau
	\right)\right.\right.\\
	&\left.\left.-\mathrm{E}\left[ \mathbf{X}_{j}(\widehat{\boldsymbol{\theta}}_{\tau,t,h})\left( 1\left\{ y_{j}\leq \mathbf{X}_{j}(\widehat{\boldsymbol{\theta}}_{\tau,t,h})^{\prime }%
	\widehat{\boldsymbol{\beta} }_{P}\left( \tau \right) \right\} -\tau \right) \right]
	\right) \right.\\
	&\left.-\frac{1}{\sqrt{P}}\sum_{j=R+1}^{T}\left( \mathbf{X}_{j}(\boldsymbol{\theta}_{\tau,h}^{\dag})\left( 1\left\{ y_{j}\leq
	\mathbf{X}_{j}(\boldsymbol{\theta}_{\tau,h}^{\dag})^{\prime }\widehat{\boldsymbol{\beta} }_{P}\left( \tau \right) \right\} -\tau
	\right)\right.\right.\\
	&\left.\left.-\mathrm{E}\left[ \mathbf{X}_{j}(\boldsymbol{\theta}_{\tau,h}^{\dag})\left( 1\left\{ y_{j}\leq \mathbf{X}_{j}(\boldsymbol{\theta}_{\tau,h}^{\dag})^{\prime }%
	\widehat{\boldsymbol{\beta} }_{P}\left( \tau \right) \right\} -\tau \right) \right]
	\right) \right)\\
	&+\sqrt{P}\mathrm{E}\left[ \mathbf{X}_{j}(\widehat{\boldsymbol{\theta}}_{\tau,t,h})\left( 1\left\{ y_{j}\leq \mathbf{X}_{j}(\widehat{\boldsymbol{\theta}}_{\tau,t,h})^{\prime }%
	\widehat{\boldsymbol{\beta} }_{P}\left( \tau \right) \right\} -\tau \right)-\mathbf{X}_{j}(\boldsymbol{\theta}_{\tau,h}^{\dag})\left( 1\left\{ y_{j}\leq \mathbf{X}_{j}(\boldsymbol{\theta}_{\tau,h}^{\dag})^{\prime }%
	\widehat{\boldsymbol{\beta} }_{P}\left( \tau \right) \right\} -\tau \right) \right]\\
	=&I_{1,P}+I_{2,P}+I_{3,P},
\end{align*}\normalsize
where the expectation is taken conditional on the realized values $\widehat{\boldsymbol{\beta}}_{P}(\tau)$ and $\widehat{\boldsymbol{\theta}}_{\tau,t,h}$, respectively. By part (ii) of this lemma as well as A6 and A7, we have that $\left\vert I_{2,P}\right\vert =o_{\Pr}(1)$. In addition, note that:
\begin{align*}
	\vert I_{3,P}\vert &\leq \sup_{\boldsymbol{\beta}\in\boldsymbol{\mathcal{B}}}\left\vert \sqrt{P} \mathrm{E}\left[\left(\mathbf{X}_{j}(\widehat{\boldsymbol{\theta}}_{\tau,t,h})-\mathbf{X}_{j}(\boldsymbol{\theta}_{\tau,h}^{\dag})\right)(1\{y_{j}\leq \mathbf{X}_{j}(\boldsymbol{\theta}_{\tau,h}^{\dag})^{\prime }\boldsymbol{\beta}\}-\tau)\right]\right\vert\\
	&+\sup_{\boldsymbol{\beta}\in\boldsymbol{\mathcal{B}}} \left\vert \sqrt{P}\mathrm{E}\left[\mathbf{X}_{j}(\boldsymbol{\theta}_{\tau,h}^{\dag})(1\{y_{j}\leq \mathbf{X}_{j}(\widehat{\boldsymbol{\theta}}_{\tau,t,h})^{\prime }\boldsymbol{\beta}\}- 1\{y_{j}\leq \mathbf{X}_{j}(\boldsymbol{\theta}_{\tau,h}^{\dag})^{\prime }\boldsymbol{\beta}\})\right]\right\vert\\
	&+\sup_{\boldsymbol{\beta}\in\boldsymbol{\mathcal{B}}}\left\vert\sqrt{P} \mathrm{E}\left[\left(\mathbf{X}_{j}(\widehat{\boldsymbol{\theta}}_{\tau,t,h})-\mathbf{X}_{j}(\boldsymbol{\theta}_{\tau,h}^{\dag})\right)(1\{y_{j}\leq \mathbf{X}_{j}(\widehat{\boldsymbol{\theta}}_{\tau,t,h})^{\prime }\boldsymbol{\beta}\}- 1\{y_{j}\leq \mathbf{X}_{j}(\boldsymbol{\theta}_{\tau,h}^{\dag})^{\prime }\boldsymbol{\beta}\})\right]\right\vert\\
	=&\sup_{\boldsymbol{\beta}\in\boldsymbol{\mathcal{B}}}\left\vert A_{1,P}(\tau,\boldsymbol{\beta})\right\vert+\sup_{\boldsymbol{\beta}\in\boldsymbol{\mathcal{B}}}\left\vert A_{2,P}(\tau,\boldsymbol{\beta})\right\vert+\sup_{\boldsymbol{\beta}\in\boldsymbol{\mathcal{B}}}\left\vert A_{3,P}(\tau,\boldsymbol{\beta})\right\vert
\end{align*}
Defining $\boldsymbol{\delta}_{\boldsymbol{\beta}}=\boldsymbol{\beta}-\boldsymbol{\beta}_{h}^{\dag}(\tau)$ and $\varepsilon_{j} = y_{j}-\mathbf{X}_{j}(\boldsymbol{\theta}_{\tau,h}^{\dag})^{\prime }\boldsymbol{\beta}_{h}^{\dag}(\tau)$, note that by iterated expectations, Jensen's inequality, A3, A6,  for some constants $\widetilde{C},\widetilde{\widetilde{C}}>0$:
\begin{align*}
	\sup_{\boldsymbol{\beta}\in\boldsymbol{\mathcal{B}}}\left\vert A_{1,P}(\tau,\boldsymbol{\beta})\right\vert&\leq \widetilde{C}\sqrt{\frac{P}{R}} \int   B(\mathbf{X}_{j}) \sup_{\boldsymbol{\beta}\in\boldsymbol{\mathcal{B}}}\left\vert F_{\varepsilon_{j} |X} (\mathbf{X}_{j}(\boldsymbol{\theta}_{\tau,h}^{\dag})^{\prime }\boldsymbol{\delta}_{\boldsymbol{\beta}}|\mathbf{X}_{j})-\tau \right\vert f_{X}(\mathbf{X}_{j})\mathrm{d}\mathbf{X}_{j}\\
	&\leq \widetilde{\widetilde{C}}\sqrt{\frac{P}{R}}\mathrm{E}\left[B(\mathbf{X}_{j})\right]=o(1).
\end{align*} 
Turning to $A_{2,P}(\tau,\boldsymbol{\beta})$, observe that:
\begin{align*}
	&\sup_{\boldsymbol{\beta}\in\boldsymbol{\mathcal{B}}}\left\vert A_{2,P}(\tau,\boldsymbol{\beta})\right\vert \\
	\leq &\sup_{\boldsymbol{\beta}\in\boldsymbol{\mathcal{B}}}\left\vert \sqrt{P}\mathrm{E}\left[\mathbf{X}_{j}(\boldsymbol{\theta}_{\tau,h}^{\dag})1\{  \mathbf{X}_{j}(\widehat{\boldsymbol{\theta}}_{\tau,t,h})^{\prime }\boldsymbol{\beta}\leq y_{j}\leq \mathbf{X}_{j}(\boldsymbol{\theta}_{\tau,h}^{\dag})^{\prime }\boldsymbol{\beta}\}\right]\right\vert\\
	&+\sup_{\boldsymbol{\beta}\in\boldsymbol{\mathcal{B}}}\left\vert \sqrt{P}\mathrm{E}\left[\mathbf{X}_{j}(\boldsymbol{\theta}_{\tau,h}^{\dag})1\{  \mathbf{X}_{j}(\boldsymbol{\theta}_{\tau,h}^{\dag})^{\prime }\boldsymbol{\beta}\leq y_{j}\leq \mathbf{X}_{j}(\widehat{\boldsymbol{\theta}}_{\tau,t,h})^{\prime }\boldsymbol{\beta}\}\right]\right\vert
\end{align*}
We only examine the second term, the first follows by similar arguments. On the event $\{  \mathbf{X}_{j}(\boldsymbol{\theta}_{\tau,h}^{\dag})^{\prime }\boldsymbol{\beta}\leq y_{j}\leq \mathbf{X}_{j}(\widehat{\boldsymbol{\theta}}_{\tau,t,h})^{\prime }\boldsymbol{\beta}\}$, recalling the definitions of $\boldsymbol{\delta}_{\boldsymbol{\beta}}=\boldsymbol{\beta}-\boldsymbol{\beta}_{h}^{\dag}(\tau)$ and $\varepsilon_{j}$,  we have by iterated expectations that:
\begin{align*}
	&\sqrt{P}\sup_{\boldsymbol{\beta}\in\boldsymbol{\mathcal{B}}}\left\vert \int \mathbf{X}_{j}(\boldsymbol{\theta}_{\tau,h}^{\dag})\left(  F_{\varepsilon_{j} |X} (\mathbf{X}_{j}(\boldsymbol{\theta}_{\tau,h}^{\dag})^{\prime }\boldsymbol{\delta}_{\boldsymbol{\beta}}+V_{j}(\mathbf{X}_{j};\widehat{\boldsymbol{\theta}}_{\tau,t,h})|\mathbf{X}_{j})\right.\right.\\
	&\left.\left. - F_{\varepsilon_{j} |X} (\mathbf{X}_{j}(\boldsymbol{\theta}_{\tau,h}^{\dag})^{\prime }\boldsymbol{\delta}_{\boldsymbol{\beta}}|\mathbf{X}_{j})\right) f_{X}(\mathbf{X}_{j})\mathrm{d}\mathbf{X}_{j}\right\vert
\end{align*}
with:
\[
V_{j}(\mathbf{X}_{j};\widehat{\boldsymbol{\theta}}_{\tau,t,h})=(\mathbf{X}_{j}(\widehat{\boldsymbol{\theta}}_{\tau,t,h})-\mathbf{X}_{j}(\boldsymbol{\theta}_{\tau,h}^{\dag}))^{\prime }\boldsymbol{\beta}_{h}^{\dag}(\tau)+(\mathbf{X}_{j}(\widehat{\boldsymbol{\theta}}_{\tau,t,h})-\mathbf{X}_{j}(\boldsymbol{\theta}_{\tau,h}^{\dag}))^{\prime }\boldsymbol{\delta}_{\boldsymbol{\beta}}
\]
Using A2, A3, A4, A6 together with a mean value expansion around $\mathbf{X}_{j}(\boldsymbol{\theta}_{\tau,h}^{\dag})^{\prime }\boldsymbol{\delta}_{\boldsymbol{\beta}}$, $\sup_{\boldsymbol{\beta}\in\boldsymbol{\mathcal{B}}}\left\vert A_{2,P}(\tau,\boldsymbol{\beta})\right\vert$ can therefore be bounded by:
\[
\sqrt{\frac{P}{R}}C \left(\mathrm{E}\left[ B(\mathbf{X}_{\tau,j,h})^2 \right]\right)^\frac{1}{2} \text{diam}(\boldsymbol{\mathcal{B}})=o(1)
\]
for some generic constant $C$, where $\text{diam}(\boldsymbol{\mathcal{B}})$ denotes the diameter of $\boldsymbol{\mathcal{B}}$, and the equality in the last display follows from A7. 

Finally, the cross-product term $A_{3,P}(\tau,\boldsymbol{\beta})$ can be shown to be $o(1)$ uniformly in $\boldsymbol{\beta}\in\boldsymbol{\mathcal{B}}$ by Cauchy-Schwarz and similar arguments to before, which establishes that for the second term on the RHS of (\ref{EQConsistency}) it holds that:
\[
\left\vert I_{3,P}\right\vert =O\left(\frac{\sqrt{P}}{\sqrt{R}}\right)=o(1).
\]
Now, using parts (i) and (ii), stochastic equicontinuity again yields that:
\begin{align*}
	&\frac{1}{\sqrt{P}}\sum_{j=R+1}^{T}\left( \mathbf{X}_{j}(\boldsymbol{\theta}_{\tau,h}^{\dag})\left( 1\left\{ y_{j}\leq
	\mathbf{X}_{j}(\boldsymbol{\theta}_{\tau,h}^{\dag})^{\prime }\widehat{\boldsymbol{\beta} }_{P}\left( \tau \right) \right\} -\tau
	\right)\right.\\
	&\left.-\mathrm{E}\left[ \mathbf{X}_{j}(\boldsymbol{\theta}_{\tau,h}^{\dag})\left( 1\left\{ y_{j}\leq \mathbf{X}_{j}(\boldsymbol{\theta}_{\tau,h}^{\dag})^{\prime }%
	\widehat{\boldsymbol{\beta} }_{P}\left( \tau \right) \right\} -\tau \right) \right]
	\right) \\
	=&\frac{1}{\sqrt{P}}\sum_{j=R+1}^{T}\left( \mathbf{X}_{j}(\boldsymbol{\theta}_{\tau,h}^{\dag})\left( 1\left\{ y_{j}\leq
	\mathbf{X}_{j}(\boldsymbol{\theta}_{\tau,h}^{\dag})^{\prime }\boldsymbol{\beta} ^{\dag }\left( \tau \right) \right\} -\tau \right)\right.\\
	&\left.-%
	\underbrace{\mathrm{E}\left[ \mathbf{X}_{j}(\boldsymbol{\theta}_{\tau,h}^{\dag})\left( 1\left\{ y_{j}\leq \mathbf{X}_{j}(\boldsymbol{\theta}_{\tau,h}^{\dag})^{\prime
		}\boldsymbol{\beta} ^{\dag }\left( \tau \right) \right\} -\tau \right] \right) }%
	_{=0}\right) \\
	&+o_{\Pr}(1).
\end{align*}%
Combining this result with (\ref{EQFIRSTORDER}), we obtain:
\begin{align*}
	&\sqrt{P}\mathrm{E}\left[ \mathbf{X}_{j}(\boldsymbol{\theta}_{\tau,h}^{\dag})\left( 1\left\{ y_{j}\leq
	\mathbf{X}_{j}(\boldsymbol{\theta}_{\tau,h}^{\dag})^{\prime }\widehat{\boldsymbol{\beta} }_{P}\left( \tau \right) \right\} -\tau
	\right) \right] \\
	=&\frac{1}{\sqrt{P}}\sum_{j=R+1}^{T}\mathbf{X}_{j}(\boldsymbol{\theta}_{\tau,h}^{\dag}) \left(
	1\left\{ y_{j}\leq \mathbf{X}_{j}(\boldsymbol{\theta}_{\tau,h}^{\dag})^{\prime }\boldsymbol{\beta} ^{\dag }\left( \tau \right)
	\right\} -\tau \right)  +o_{\Pr}(1).
\end{align*}%
Applying again iterated expectations and a mean value expansion of the left hand side around $\boldsymbol{\beta} ^{\dag
}\left( \tau \right)$, we have by A4 and A5  that:
\begin{align*}
	&\sqrt{P}\mathrm{E}\left[ \mathbf{X}_{j}(\boldsymbol{\theta}_{\tau,h}^{\dag})\left( 1\left\{ y_{j}\leq
	\mathbf{X}_{j}(\boldsymbol{\theta}_{\tau,h}^{\dag})^{\prime }\widehat{\boldsymbol{\beta} }_{P}\left( \tau \right) \right\} -\tau
	\right) \right]\\
	=& \sqrt{P} \left(\int f_{\varepsilon_{j}|X}(\overline{U}_{j}|\mathbf{X}_{j})\mathbf{X}_{j}(\boldsymbol{\theta}_{\tau,h}^{\dag})\mathbf{X}_{j}(\boldsymbol{\theta}_{\tau,h}^{\dag})^{\prime}f_{X}(\mathbf{X}_{j})\mathrm{d}\mathbf{X}_{j}\right)\left( \widehat{\boldsymbol{\beta}}\left( \tau \right)
	-\boldsymbol{\beta} ^{\dag }\left( \tau \right) \right)  ,
\end{align*} 
where the intermediate value $\overline{U}_{j}$ lies between $0$ and $\mathbf{X}_{j}(\boldsymbol{\theta}^{\dag})^{\prime}\left( \widehat{\boldsymbol{\beta}}\left( \tau \right)
-\boldsymbol{\beta} ^{\dag }\left( \tau \right) \right)$.   Therefore:
\begin{equation*}
	\frac{1}{\sqrt{P}}\sum_{j=R+1}^{T}\mathbf{X}_{j}(\boldsymbol{\theta}^{\dag})\left( 1\left\{ y_{j}\leq
	\mathbf{X}_{j}(\boldsymbol{\theta}^{\dag})^{\prime }\boldsymbol{\beta} ^{\dag }\left( \tau \right) \right\} -\tau \right)
	=\mathbf{J}\left( \tau \right) \sqrt{P}\left( \widehat{\boldsymbol{\beta}}\left( \tau \right)
	-\boldsymbol{\beta} ^{\dag }\left( \tau \right) \right) +o_{\Pr}(1)
\end{equation*}%
where $\mathbf{J}\left( \tau \right)=\mathbf{J}_{h}\left( \tau_{k} \right)$. Since $\mathbf{J}\left( \tau \right)$ is positive definite, the Bahadur representation for $\sqrt{P}\left( 
\widehat{\boldsymbol{\beta} }\left( \tau \right) -\boldsymbol{\beta} ^{\dag }\left( \tau \right)
\right) $ follows. \qedsymbol

\section{Horizon Monotonicity Test}\label{sec:ELoss}

In this section, we outline a further possibility to test for quantile forecast optimality, namely across forecast horizons when forecasts are reported at multiple horizons. The test exploits the fact that,  under optimality, expected quantile loss is monotonically non-decreasing in the forecast horizon. More specifically,	let $\widehat{y}^{\ast}_{\tau,t,h}$, $h\in\mathcal{H}=\{1,\ldots,H\}$, be optimal forecasts for the $\tau$-quantile. It then holds that
\begin{equation*}
	\mathrm{E} \left[L_{\tau} \left( y_{t,h_i} - \widehat{y}^{\ast}_{\tau,t,h_i} \right) \right] \leq \mathrm{E} \left[L_{\tau} \left( y_{t,h} - \widehat{y}^{\ast}_{\tau,t,h_j} \right) \right]
\end{equation*}
for all $h_{i},h_{j}\in\mathcal{H}$ with $h_{i}<h_{j}$, which follows from strict stationarity, iterated expectations, and the monotonicity of conditional expectations. The result is akin to  \citet{PT2012} who demonstrated that the mean squared forecast error (MSFE) of optimal multi-horizon conditional mean forecasts does not decrease with the forecast horizon, and similar tests have in fact been constructed for mean forecasts \citep[see][and references therein]{FG17}. Intuitively, forecasting becomes more difficult the longer the forecast horizon gets as the information
set gets smaller, so optimal forecasts should display non-decreasing expected loss. A decrease in expected loss from a certain horizon to the next therefore indicates non-optimal forecasts, meaning that forecasts further into the future are systematically more accurate than the forecasts at the shorter horizon.  This can arise, for instance, if different models are used for short and
long-term forecasting or if different specifications of models are selected and estimated at different
horizons. The latter is often the case in quantile forecasting where the direct scheme is typically
used, as opposed to the iterative scheme, and e.g. relevant variables may be excluded at
shorter horizons, or modelled and processed in a way such that non-monotonicity in the information
content over the horizons arises.

The notation is as in the previous sections, although we will treat estimation error here as implicit since arguably the most relevant applications (unless different models are used at different horizons) are to situations where forecasts are reported without model like in the case of the Survey of Professional Forecasters. We are interested in the following hypothesis:
\begin{equation}\label{EQHMH0}
	H^{\text{MH}}_{0}: \mathrm{E}[L_{\tau_{k}}(y_{t} - \widehat{y}_{\tau_{k},t,h_{i}}) - L_{\tau_{k}}(y_{t} - \widehat{y}_{\tau_{k},t,h_{j}}) ] \geq 0
\end{equation}
for all $h_{i},h_{j}\in\mathcal{H}$ s.t. $h_{i}<h_{j}$ and $\tau_{k}\in\mathcal{T}$ versus:
\begin{equation}\label{EQHMH1}
	H^{\text{MH}}_{1}:   \mathrm{E}[L_{\tau_{k}}(y_{t} - \widehat{y}_{\tau_{k},t,h_{i}}) - L_{\tau_{k}}(y_{t} - \widehat{y}_{\tau_{k},t,h_{j}}) ] < 0
\end{equation}
for at least some $h_{i},h_{j}\in\mathcal{H}$ and $\tau_{k}\in\mathcal{T}$. As before, define the set:
\[
\mathcal{C}^{\text{MH}}=\left\{(h_{i},h_{j},\tau_{k}):\ (h_{i},h_{j})\in \mathcal{H}\ \text{s.t.}\ h_{i}<h_{j},\ \tau_{k}\in\mathcal{T}\right\}.
\]
We use the same test statistic as in \citet{AS2010}. Hereafter, denote 
\[
\mathcal{F}%
_{0}=\left\{ F_{0}:\; H_{0}^{MH}\;\text{holds}\right\} 
\] as the set of null
DGPs such that (\ref{EQHMH0}) and Assumptions C1 to C3
below hold.  Also, denote, with some abuse of notation, $\vert \mathcal{C}^{\text{MH}}\vert=\kappa$  the cardinality of the set $\mathcal{C}^{\text{MH}}$ and define $L_{\tau_{k},t,h_{j}}\equiv L_{\tau_{k}}(y_{t} - \widehat{y}_{\tau_{k},t,h_{j}}) $ as well as $L_{\tau_{k},t,h_{i}}\equiv L_{\tau_{k}}(y_{t} - \widehat{y}_{\tau_{k},t,h_{i}})$, where $L_{s,t}$ stands for the difference $L_{\tau_{k},t,h_{j}}-L_{\tau_{k},t,h_{i}}$ for a specific $\tau_k$, $h_{i}$, and $h_{j}$ combination. That is, under the null hypothesis, for every $s\in\{1,\ldots,\kappa\}$:
\[
\mathrm{E}_{F_{0}}\left[L_{s,t}\right]\equiv \overline{m}_{s}\geq 0 
\] 
We also let (with some abuse of notation):
\[
\widehat{\overline{m}}_{s}=\frac{1}{P}\sum_{t=R+1}^{T} L_{s,t}
\]
denote the empirical moment condition. As mentioned above, we keep implicit the (possible) dependence of $L_{s,t}$ on estimated parameters. The test statistic is then given by:
\begin{equation}\label{EQTS3}
	\widehat{U}_{\text{MH}}=\sum_{s=1}^{\kappa} \left[\frac{\sqrt{P}\widehat{\overline{m}}_{s}}{\widehat{\overline{\sigma}}_{s,s}}\right]_{-}^{2},
\end{equation}
where $[x]_{-}=x1\{x < 0\}$ and $\widehat{\sigma}_{s,s}^2 $ is an estimator of the diagonal element of $\overline{\boldsymbol{\Sigma}}$, the variance-covariance matrix of the moment inequalities. That is, the statistic in (\ref{EQTS3}) gives positive weight only to those empirical moment inequalities which are indeed violated. A corresponding HAC estimator of $\overline{\boldsymbol{\Sigma}}$ may be chosen as:
\[
\widehat{\overline{\boldsymbol{\Sigma}}}_{T}=\frac{1}{P}\sum_{t=R+s_{T}}^{T-s_{T}}\sum_{k=-s_{T}}^{s_{T}}\lambda_{k,t}\left(\mathbf{m}_{t}-\widehat{{\overline{\mathbf{m}}}}_{T}\right)\left(\mathbf{m}_{t+k}- 	\widehat{\overline{\mathbf{m}}}_{T}\right)^{\prime}.
\] 
where:
\[
\mathbf{m}_{t}=\left(\begin{array}{c}L_{1,t}\\\vdots\\L_{\kappa, t}\end{array}\right)\quad\text{and}\quad
\widehat{{\overline{\mathbf{m}}}}_{T}=\left(\begin{array}{c}\frac{1}{P}\sum_{t=R+1}^{T} L_{1,t}\\\vdots\\\frac{1}{P}\sum_{t=R+1}^{T}L_{\kappa,t}\end{array}\right).
\]
Note, however, that the statistic in (\ref{EQTS3})  in principle only requires estimates of the diagonal elements, which may be an advantage when $\mathcal{H}$ and $\mathcal{T}$ are large. The downside, of course, is that the limiting distribution is again non-pivotal as it depends on the (unknown) correlation structure between the different moment (in-)equalities. Thus, as in the Mincer-Zarnowitz tests, we will generate bootstrap critical values using the MBB of \citet{K1989} with the resampling  as explained before. For each bootstrap sample, we obtain the bootstrap equivalent of $L_{s,t}$, say $L^{b}_{s,t}$, to construct the bootstrap statistic as:
\begin{equation}\label{EQBS2}
	\widehat{U}^{b}_{\text{MH}}=\sum_{s=1}^{\kappa} \left[\frac{\sqrt{P}(\widehat{\overline{m}}^{b}_{s}-\widehat{\overline{m}}_{s})}{\widehat{\overline{\sigma}}^{b}_{s,s}}\right]_{-}^{2}1\left\{\frac{\widehat{\overline{m}}_{s}}{\widehat{\overline{\sigma}}_{s,s}}\leq \sqrt{2\ln(\ln(P))/P}\right\},
\end{equation}
where $\widehat{\overline{m}}^{b}_{s}$ is defined as before but using the bootstrap series instead of the original sample, i.e.:
\[
\widehat{\overline{m}}^{b}_{s}=\frac{1}{P}\sum_{t=R+1}^{T}L_{s,t}^{b}.
\]
The term $\left(\widehat{\overline{\sigma}}^{b}_{s,s} \right)^2$, on the other hand, is the bootstrap variance estimator, which requires more careful consideration for first order validity of the bootstrap variance estimator \citep{GK1996,GW2004}. More specifically, we follow \citet{GW2004} and use the following bootstrap variance estimator:
\[
\left(\widehat{\overline{\sigma}}^{b}_{s,s} \right)^2=\frac{1}{K_{b}}\sum_{k=1}^{K_{b}}\frac{1}{l}\left( \sum_{i=1}^{l} \left(L_{s,I_{k}+i}^{b}-\frac{1}{P}\sum_{t=R+1}^{T}L_{s,t}^{b}\right)\right)^{2}
\]
Finally, note that the second term on the right hand side of (\ref{EQBS2}) implements the Generalised Moment Selection (GMS) procedure introduced by \citet{AS2010} that uses information about the slackness of the sample moment conditions to
infer which population moment conditions are most likely to be binding, and thus will enter into the limiting distribution. The critical value will then be based on the $(1-\alpha)$ quantile of the empirical bootstrap distribution of $\widehat{U}^{b}_{\text{MH}}$ over $B$ draws, which we denote by $\overline{c}_{B,P,(1-\alpha)}$. We make the following additional assumptions:

\medskip

\noindent \textbf{C1}: For all $s\in\{1,\ldots,\kappa\}$ and $F_{0}\in\mathcal{F}_{0}$, $L_{s,t}$ is strictly stationary with $\beta$-mixing coefficients satisfying the mixing condition from A1 with $\epsilon=1.1$  and $\mathrm{E}_{F_{0}}[\vert L_{s,t}\vert^{3r}]\leq M<\infty$ with $r=2+\epsilon$.
\medskip

\noindent \textbf{C2}: For all $s\in\{1,\ldots,\kappa\}$, $F_{0}\in\mathcal{F}_{0}$, and some $0< \delta\leq 2$, it holds that: 
\[
\frac{1}{P}\sum_{t=R+1}^{T}\left\vert \mathrm{E}_{F_{0}}\left[L_{s,t}\right]- \frac{1}{P}\sum_{t=R+1}^{P}\mathrm{E}_{F_{0}}\left[ L_{s,t}\right]\right\vert^{2+\delta}=o(l^{-1-\delta/2}).
\]

\medskip

\noindent \textbf{C3}: The variance-covariance matrix $\overline{\boldsymbol{\Sigma}}$ is positive definite for any $F_{0}\in\mathcal{F}_{0}$. Moreover, it holds that:
\[
\widehat{\overline{\mathbf{D}}}_{T}^{-\frac{1}{2}}\widehat{\overline{\boldsymbol{\Sigma}}}_{T}\widehat{\overline{\mathbf{D}}}_{T}^{-\frac{1}{2}}\stackrel{\mathrm{Pr}_{F_{0}}}{\rightarrow}\overline{\mathbf{D}}^{-\frac{1}{2}}\overline{\boldsymbol{\Sigma}}\overline{\mathbf{D}}^{-\frac{1}{2}},
\]
where $\widehat{\overline{\mathbf{D}}}_{T}=\text{diag}(\widehat{\overline{\sigma}}_{1,1},\ldots,\widehat{\overline{\sigma}}_{\kappa,\kappa})$ and $\overline{\mathbf{D}}=\text{diag}(\overline{\sigma}_{1,1},\ldots,\overline{\sigma}_{\kappa,\kappa})$ are $(\kappa\times\kappa)$ diagonal matrices and:
\[
\widehat{\overline{\sigma}}_{s,s}/\overline{\sigma}_{s,s}\stackrel{\mathrm{Pr}_{F_{0}}}{\rightarrow}1
\]
for all $s\in\{1,\ldots,\kappa\}$.   \medskip

Assumption C1 implies Assumption 2.1 in \citet{GW2002} and, together with C2, ensures the first order validity of the block bootstrap procedure in our set-up. It entails, for expositional simplicity, a homogeneity assumption across $s\in\{1,\ldots,\kappa\}$, which is stronger than what is required in  \citet{GW2002}. The latter allows for considerable heterogeneity across the series.  Assumption C2 on the other hand is identical Assumption A.2.2$^{\prime}$ in \citet{GW2004}.  We obtain the following result:\medskip

\begin{theorem}\label{DLossTest} Assume that C1 to C3 hold, and $T\rightarrow \infty$, $B\rightarrow \infty$, $l\rightarrow \infty$, $\frac{l}{\sqrt{T}}\rightarrow 0$. Then, under $H^{\text{MH}}_{0}$: 
	\[
	\lim\sup_{T,B\rightarrow \infty} \sup_{F_{0}\in\mathcal{F}_{0}}\mathrm{Pr}_{F_{0}}\left(\widehat{U}_{\text{MH}}>\overline{c}_{B,P,(1-\alpha)}\right)\leq \alpha.
	\]
\end{theorem}
Theorem \ref{DLossTest} states that the monotonicity test proposed in this section has asymptotic size at most equal $\alpha$. As pointed out in \citet{AS2010}, the test is non-conservative whenever some weak inequality holds with equality. Note also that Theorem \ref{DLossTest} required a tightening of the block length condition for the (first order) validity of the bootstrap variance estimator.

In analogy to the extension of the autocalibration test in Section \ref{sec:testing} to a multivariate set-up, the Horizon Monotonicity test may also be extended to a group of time series. More specifically, we may be interested in testing the following null hypothesis: 
\begin{equation}\label{EQHMH0P}
	H^{\text{MMH}}_{0}: \mathrm{E}[L_{\tau_{k}}(y_{i,t} - \widehat{y}_{i,\tau_{k},t,h_{i}}) - L_{\tau_{k}}(y_{i,t} - \widehat{y}_{i,\tau_{k},t,h_{j}}) ] \geq 0
\end{equation}
for all $h_{l},h_{j}\in\mathcal{H}$ s.t. $h_{l}<h_{j}$, $\tau_{k}\in\mathcal{T}$, $i=1,\ldots,G$, versus:
\begin{equation}\label{EQHMH1P}
	H^{\text{MMH}}_{1}:   \mathrm{E}[L_{\tau_{k}}(y_{i,t} - \widehat{y}_{i,\tau_{k},t,h_{i}}) - L_{\tau_{k}}(y_{i,t} - \widehat{y}_{i,\tau_{k},t,h_{j}}) ] < 0
\end{equation}
for at least some $h_{l},h_{j}\in\mathcal{H}$, $\tau_{k}\in\mathcal{T}$, $i=1,\ldots,G$. As before, define the set:
\[
\mathcal{C}^{\text{MMH}}=\left\{(i,h_{l},h_{j},\tau_{k}):\ i\in\{1,\ldots,G\},\ (h_{l},h_{j})\in \mathcal{H}\ \text{s.t.}\ h_{l}<h_{j},\ \tau_{k}\in\mathcal{T}\right\},
\]
with $\kappa$ denoting the cardinality again. Then, for every difference $L_{i,\tau_{k},t,h_{j}}-L_{i,\tau_{k},t,h_{l}}$ with $L_{i,\tau_{k},t,h_{j}}\equiv L_{i,\tau_{k}}(y_{i,t} - \widehat{y}_{i,\tau_{k},t,h_{j}}) $ and $L_{i,\tau_{k},t,h_{l}}\equiv L_{i,\tau_{k}}(y_{i,t} - \widehat{y}_{i,\tau_{k},t,h_{l}})$, which we denote by $\overline{L}_{s,t}$, $s\in\{1,\ldots,\kappa\}$, we may test:
\[
\mathrm{E}\left[\overline{L}_{s,t}\right]\equiv \overline{m}_{s,G}\geq 0 ,\ s\in\{1,\ldots,\kappa\}.
\] 
This population moment inequality can be replaced by its sample analogue:
\[
\widehat{\overline{m}}_{s,G}=\frac{1}{P}\sum_{t=R+1}^{T} \overline{L}_{s,t}.
\]
The test statistic $\widehat{U}_{\text{MMH}}$ and the bootstrap statistic $\widehat{U}^{b}_{\text{MMH}}$ is then constructed in analogy to before, with the only difference consisting in the fact that for $t=R+1,\ldots,T$, the series to be resampled comes again in array form as in the multivariate extension of Subsection \ref{ssec:AugmentedMZTest}. Under Assumptions C1-C3, the bootstrap distribution of $\widehat{U}^{b}_{\text{MMH}}$, $b=1,\ldots,B$, provides asymptotically valid critical values that yield a test of size at most $\alpha$ by Theorem \ref{DLossTest}.
\bigskip

\noindent\textbf{Proof of Theorem \ref{DLossTest}}:  Firstly, let:
\[
h_{s}=\lim_{P\rightarrow \infty}\sqrt{P}\mathrm{E}_{F_{0}}\left[L_{s,t}\right].
\]
Then, under the null hypothesis and Assumptions C1 and C3, it holds by arguments from the proof of Theorem 1 in \citet{AG2009} that for a given $F_{0}\in\mathcal{F}_{0}$:
\begin{equation}\label{Thm6EQ1}
	\sum_{s=1}^{\kappa} \left[\frac{\sqrt{P}\widehat{\overline{m}}_{s}}{\widehat{\overline{\sigma}}_{s,s}} \right]_{-}^{2}\stackrel{d}{\rightarrow}\sum_{s=1}^{\kappa}\left[\sum_{j=1}^{\kappa}\omega_{s,j} Z_{s}+ h_{s}\right]_{-}^{2},
\end{equation}
where $Z_{s}$ is an element of:
\begin{equation}\label{Thm6EQ2}
	\mathbf{Z}=\left(\begin{array}{c} Z_{1}\\\vdots\\ Z_{\kappa}\end{array}\right)\sim N\left(\mathbf{0},\mathbf{I}\right),
\end{equation} 
and $\omega_{s,j}$ is the square root of a generic element of the correlation matrix:
\[
\boldsymbol{\Omega}=\boldsymbol{D}^{-\frac{1}{2}}\overline{\boldsymbol{\Sigma}}\boldsymbol{D}^{-\frac{1}{2}},
\]
with $\boldsymbol{D}=\text{Diag}(\overline{\boldsymbol{\Sigma}})$ and $\overline{\boldsymbol{\Sigma}}$ denoting the population variance-covariance matrix of the moment inequalities.\footnote{Note that in the case where $h_{s}=-\infty$, it follows by definition that $Z_{s}+ h_{s}=-\infty$.} Next we need to show that the $(1-\alpha)$ percentile of this limiting distribution is accurately approximated by the corresponding $(1-\alpha)$ percentile of the bootstrap (limiting) distribution. To this end, note that for all $s\in\{1,\ldots,\kappa\}$ it holds by Assumptions C1 and the law of iterated logarithms \citep[][Theorem 5]{OY1971} that:
\[
\lim_{P\rightarrow \infty}\sup\left(\frac{P}{2\ln(\ln(P))}\right)^{\frac{1}{2}} \left(\frac{\widehat{\overline{m}}_{s}}{\overline{\sigma}_{s,s}}\right)=1
\]
a.s. under $\mathrm{Pr}_{F_{0}}$ when $\mathrm{E}_{F_{0}}\left[L_{s,t}\right]=0$, while
\[
\lim_{P\rightarrow \infty}\left(\frac{P}{2\ln(\ln(P))}\right)^{\frac{1}{2}} \left(\frac{\widehat{\overline{m}}_{s}}{\overline{\sigma}_{s,s}}\right)> 1
\]
a.s. under $\mathrm{Pr}_{F_{0}}$ when $\mathrm{E}_{F_{0}}\left[L_{s,t}\right]>0$. Moreover, note that by C3 for all $s$ it holds that $\frac{\widehat{\overline{\sigma}}_{s,s}}{\overline{\sigma}_{s,s}}\stackrel{\mathrm{Pr}_{F_{0}}}{\rightarrow}1$, so that by standard arguments:
\[
\lim_{P\rightarrow \infty}\mathrm{Pr}_{F_{0}}\left(\left(\frac{P}{2\ln(\ln(P))}\right)^{\frac{1}{2}} \left(\frac{\widehat{\overline{m}}_{s}}{\widehat{\overline{\sigma}}_{s,s}}\right)>1\right)=0
\]
when $\mathrm{E}_{F_{0}}\left[L_{s,t}\right]=0$, while
\[
\lim_{P\rightarrow \infty}\mathrm{Pr}_{F_{0}}\left(\left(\frac{P}{2\ln(\ln(P))}\right)^{\frac{1}{2}} \left(\frac{\widehat{\overline{m}}_{s}}{\widehat{\overline{\sigma}}_{s,s}}\right)>1\right)=1
\]
when $\mathrm{E}_{F_{0}}\left[L_{s,t}\right]>0$. Thus, for sufficiently large $T$ only moment conditions holding with equality will contribute to the bootstrap limiting distribution, and the probability of eliminating a binding moment equality approaches zero as $T\rightarrow \infty$.

Moreover, from Theorem 2.2 of \citet{GW2002}, it follows by C1 and C2 that for any $\nu>0$ and $F_{0}\in\mathcal{F}_{0}$:
\[
\mathrm{Pr}_{F_{0}}\left(\sup_{x\in\mathbb{R}}\left\vert \mathrm{Pr}_{F_{0}}^{b} \left(\sqrt{T}(\widehat{\overline{m}}^{b}_{s}-\widehat{\overline{m}}_{s})\leq x\right)-\mathrm{Pr}_{F_{0}}\left(\sqrt{T}(\widehat{\overline{m}}_{s}-\overline{m}_{s})\leq x\right)\right\vert >\nu \right)\rightarrow 0
\]
for all $s\in\{1,\ldots,\kappa\}$, where $\mathrm{Pr}_{F_{0}}^{b} $ denotes the probability measure induced by the bootstrap under $F_{0}$. Likewise, for a given $F_{0}\in\mathcal{F}_{0}$, by Lemma B1 in \citet{GW2004} we have that for any $\epsilon>0$:
\[
\mathrm{Pr}_{F_{0}}\left(\mathrm{Pr}_{F_{0}}^{b} \left(\lvert \widehat{\overline{\sigma}}^{b 2}_{s,s}-\widehat{\overline{\sigma}}^{2}_{s,s}\rvert>\nu\right)>\nu\right)\rightarrow 0
\]
This suggests that for any $\nu>0$ and a given $F_{0}$: 
\[
\mathrm{Pr}_{F_{0}}\left(\sup_{x\in\mathbb{R}}\left\vert \mathrm{Pr}_{F_{0}}^{b} \left(\frac{\sqrt{T}(\widehat{\overline{m}}^{b}_{s}-\widehat{\overline{m}}_{s})}{\widehat{\overline{\sigma}}_{s,s}^{b}}\leq x\right)-\mathrm{Pr}_{F_{0}}\left(\frac{\sqrt{T}(\widehat{\overline{m}}_{s}-\overline{m}_{s})}{\widehat{\overline{\sigma}}_{s,s}}\leq x\right)\right\vert >\nu \right)\rightarrow 0
\]
for all $s\in\{1,\ldots,\kappa\}$. 

Now, let $\overline{c}_{B,P,(1-\alpha) }$ be the $(1-\alpha )$ critical value of $%
\widehat{U}_{\text{MH}}^{b}$ based on $B$ bootstrap replications. Also, consider a sequence $\{\boldsymbol{\gamma}_{P}\}_{P=1}^{\infty}$ with $\boldsymbol{\gamma}_{P}=\left( \gamma _{1,P},...,\gamma
_{\kappa,P}\right) $ and each $\boldsymbol{\gamma}_{P}\in\mathcal{F}_{0}$ such that $\sqrt{P}\gamma _{P}\rightarrow \boldsymbol{h}=(h_{1},\ldots,h_{\kappa})^{\prime}$ and $(\ln(\ln(P)))^{-1}\sqrt{%
	P}\gamma _{P}\rightarrow \boldsymbol{\xi}$ where $\boldsymbol{h},\boldsymbol{\xi} \in \mathbb{R}%
_{-,\infty}^{\kappa}$ with $\mathbb{R}_{-}=\{x\in\mathbb{R}: x\leq 0\}$ and $\mathbb{R}%
_{-,\infty}=\mathbb{R}%
_{-}\cup \{-\infty\}$. Then, let $\overline{c}_{P,1-\alpha }$ be the $(1-\alpha )$
critical values of $\widehat{U}_{\text{MH},\gamma}$ defined as:
\[
\widehat{U}_{\text{MH},\gamma}=\sum_{s=1}^{\kappa} \left[\frac{\sqrt{P}(\widehat{\overline{m}}_{s}-\gamma_{s,P})}{\widehat{\overline{\sigma}}_{s,s}}\right]_{-}^{2}.
\]
By Lemma 2(a) in the supplement of \citet{AS2010}, $\overline{c}^{b}_{P,1-\alpha
}\leq \overline{c}_{P,(1-\alpha) }$ almost surely for all $P$ for a sequence such that $\overline{c}^{b}_{P,1-\alpha
}\stackrel{\mathrm{Pr}_{F_{0}}}{\rightarrow}\overline{c}_{1-\alpha
}^{b }=\lim_{B,T\rightarrow \infty}\overline{c}_{B,P,1-\alpha
}^{b }$  noting that the assumptions together with the
HAC estimator $\widehat{\overline{\sigma}}_{s,s}$ satisfy conditions (A.2) and (A.3) of %
\citet{AS2010} for dependent data. Also, under the drifting
sequence $\{\boldsymbol{\gamma}_{P}, P\geq 1\}$, $\lim_{P\rightarrow \infty }\overline{c}_{P,(1-\alpha)
}=c_{1-\alpha }^{\dag }$ which is the $(1-\alpha )$ critical value of the
limiting distribution of $\widehat{U}_{\text{MH}}$ in Theorem \ref{DLossTest}. The result then follows from subsequence arguments
analogous to the ones used in the proof of Theorem 1(i)-(ii) in %
\citet{AS2010}.   \qedsymbol

\section{Additional Results - Empirical Application 1} \label{app:Finance_Application}

\subsection{Related Risk Measures and their Evaluation} \label{subsec:expected_shortfall}

Expected shortfall, which is a coherent risk measure and more sensitive to the shape of the tail distribution beyond the VaR, is just the expectation of this tail distribution and can be written as an integral over all quantiles with levels below $\tau$:
$$ES(\tau)=\frac{1}{\tau}\int_{0}^{\tau} VaR(\alpha)\mathrm{d} \alpha.$$
Median shortfall, a natural alternative to expected shortfall, is the median of the tail distribution and just a quantile itself:
$MS(\tau)= VaR \left(\frac{\tau}{2} \right)$.

The evaluation of expected shortfall forecasts is more difficult than the evaluation of VaR or median shortfall forecasts as they cannot be evaluated on their own due to the non-identifiability and non-elicitability of expected shortfall \citep[e.g.,][]{fissler2015}. \cite{bayer2022} exploit the joint identifiability of expected shortfall and VaR to jointly test their optimality for a given quantile level and horizon, while \cite{kratz2018} propose to test optimality of expected shortfall forecasts implicitly by testing forecasts for multiple quantiles of the tail distribution for a given horizon. The latter is possible naturally in our multi-quantile evaluation framework, and so our tests can give guidance regarding the optimality of forecasts for all three risk measures.

\subsection{Additional Tables and Figures}\label{ssec:AdditionalMaterial}

\begin{figure}[H]
	\centering
	\caption{Quantile Forecasts and Realisations, $h=1$}
	\includegraphics[width=0.7\linewidth]{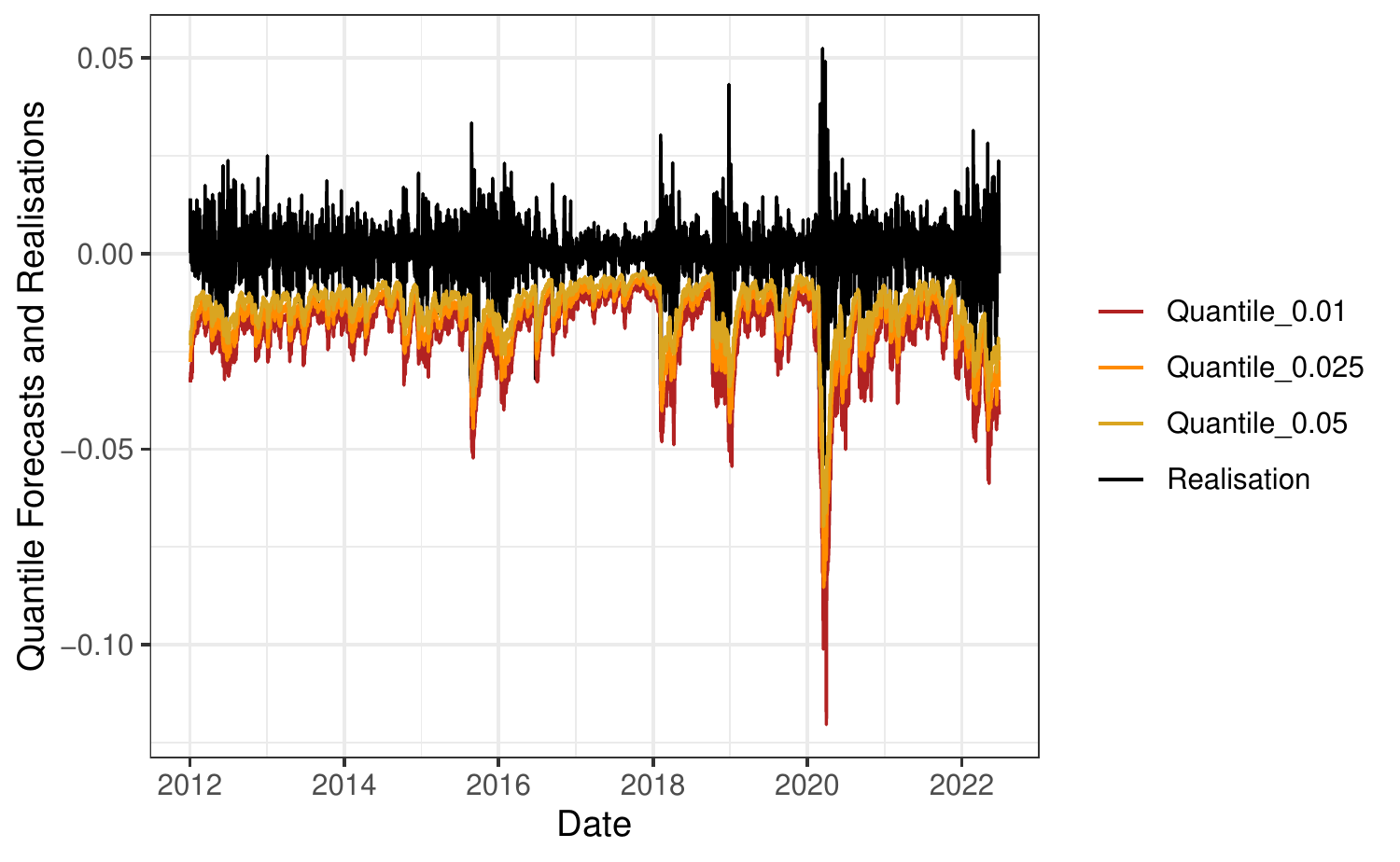}
	\label{fig:forecastploth1}
\end{figure}

\begin{table}[H]
	\centering
	\caption{Individual $p$-values, Mincer-Zarnowitz Test, Finance Application}
	\begin{tabular}{lcccc}
		\toprule
		& $\tau=0.01$ & $\tau=0.025$ & $\tau=0.05$ & all   \\
		\midrule
		$h=1 $ & 0.000        & 0.092     & 0.294    & 0.006 \\
		$h=2 $ & 0.000        & 0.011     & 0.161    & 0.002 \\
		$h=3 $ & 0.008    & 0.011     & 0.131    & 0.001 \\
		$h=4 $ & 0.002    & 0.023     & 0.176    & 0.008 \\
		$h=5 $ & 0.005    & 0.001     & 0.141    & 0.010  \\
		$h=6 $ & 0.010     & 0.032     & 0.236    & 0.012 \\
		$h=7 $ & 0.312    & 0.073     & 0.069    & 0.137 \\
		$h=8 $ & 0.228    & 0.065     & 0.091    & 0.125 \\
		$h=9 $ & 0.030     & 0.113     & 0.029    & 0.028 \\
		$h=10$ & 0.122    & 0.044     & 0.021    & 0.010  \\
		all  & 0.013    & 0.011     & 0.063    & 0.010  \\
		\bottomrule
	\end{tabular}
	\label{tab:MZ_Finance_Application_pValues}
\end{table}

\begin{table}[H]
	\centering
	\caption{Intercepts, Mincer-Zarnowitz Regressions, Finance Application}
	\begin{tabular}{lccc}
		\toprule
		& $\tau=0.01$ & $\tau=0.025$ & $\tau=0.05$ \\
		\midrule
		$h=1 $ & -0.009   & -0.003    & -0.001   \\
		$h=2 $ & -0.009   & -0.004    & -0.002   \\
		$h=3 $ & -0.011   & -0.006    & -0.003   \\
		$h=4 $ & -0.011   & -0.006    & -0.002   \\
		$h=5 $ & -0.011   & -0.007    & -0.003   \\
		$h=6 $ & -0.011   & -0.007    & -0.003   \\
		$h=7 $ & -0.006   & -0.006    & -0.004   \\
		$h=8 $ & -0.008   & -0.007    & -0.004   \\
		$h=9 $ & -0.012   & -0.007    & -0.005   \\
		$h=10$ & -0.011   & -0.009    & -0.005  \\
		\bottomrule
	\end{tabular}
	\label{tab:MZ_Finance_Application_Intercepts}
\end{table}

\begin{table}[H]
	\centering
	\caption{Slopes, Mincer-Zarnowitz Regressions, Finance Application}
	\begin{tabular}{lccc}
		\toprule
		& $\tau=0.01$ & $\tau=0.025$ & $\tau=0.05$ \\
		\midrule
		$h=1 $ & 0.597    & 0.824     & 0.878    \\
		$h=2 $ & 0.591    & 0.727     & 0.862    \\
		$h=3 $ & 0.494    & 0.682     & 0.78     \\
		$h=4 $ & 0.525    & 0.682     & 0.805    \\
		$h=5 $ & 0.543    & 0.595     & 0.768    \\
		$h=6 $ & 0.541    & 0.594     & 0.791    \\
		$h=7 $ & 0.761    & 0.667     & 0.704    \\
		$h=8 $ & 0.686    & 0.663     & 0.708    \\
		$h=9 $ & 0.538    & 0.607     & 0.609    \\
		$h=10$ & 0.565    & 0.537     & 0.575   \\
		\bottomrule
	\end{tabular}
	\label{tab:MZ_Finance_Application_Slopes}
\end{table}

\subsection{Robustness Checks and Additional Tests} \label{subsec:robustness_rolling}

We compare the results of our main specification, where we use a recursive estimation window with initial window size 3000, to specifications using rolling estimation windows of size 1500 and 750 (everything else unchanged) making sure that the forecasts are generated for the same evaluation sample (by discarding the first 1500 and 2250 observations, respectively). The results are robust against the choice of the estimation scheme and the window size as the $p$-values barely change: from 0.01 in the original specification to 0.00 in the two rolling window specifications (see Table \ref{tab:MZ_Finance_Application_Rolling} below). The same patterns in terms of individual contributions to the test statistic and slopes (all smaller than 1) and intercepts (all negative) of the MZ regression lines arise (we do not report those additional tables here). These findings are in line with \citet{Hassler2023}, who report a systematic comparison of different rolling window sizes and an expanding window for different forecasting approaches for daily return volatility and find that from a certain minimum window size on the forecasting performance is virtually unchanged when the estimation window size changes. 

\begin{table}[H]
	\centering
	\caption{Mincer-Zarnowitz Test Results, Finance Application, Rolling Estimation Windows}
	\begin{tabular}{lcccccc}
		\toprule
		& Window and Size & Stat     & 90\%      & 95\%      & 99\%      & $p$-value \\
		\midrule
		& Expanding, 3000 & 9834.131 & 5009.153 & 6569.754 & 9821.539 & 0.01  \\
		& Rolling, 1500 & 13396.985 & 3647.580 & 4736.266 &6993.295 & 0.00  \\
		& Rolling, 750 & 11583.056 & 4541.195 & 5946.943 & 8025.501 & 0.00  \\
		\bottomrule
	\end{tabular}
	\label{tab:MZ_Finance_Application_Rolling}
\end{table}

Similar to the (more systematic) analysis of \citet{GLLS11}, we also compare  single horizon, single quantile level Mincer-Zarnowitz tests with the autocalibration test of \citet{engle2004caviar}, using the respective quantile forecast as regressor, and with the test of \citet{christoffersen1998}. The \texttt{R} package \texttt{GAS} \citep{Ardia2019} was used to execute both tests. The results are contained in Table \ref{tab:Alternative_Tests_Finance_Application} below.

\begin{table}[H] 
	\centering
	\caption{$p$-values, Mincer-Zarnowitz, Engle-Manganelli and Christoffersen test for $h=1$}
	\begin{tabular}{lccc}
		\toprule
		& $\tau=0.01$ & $\tau=0.025$ & $\tau=0.05$ \\
		\midrule
		Mincer-Zarnowitz  & 0.000   & 0.092    & 0.294   \\
		Engle-Manganelli  & 0.000   & 0.070    & 0.455   \\
		Christoffersen  & 0.039   & 0.609    & 0.651   \\
		
		\bottomrule
	\end{tabular}
	\label{tab:Alternative_Tests_Finance_Application}
\end{table}


\section{Additional Results - Empirical Application 2}\label{app:Macro_Application}

\subsection{Figures: Out-of-sample Predictions for $h=1$}\label{secappoos}

\begin{center}
	\includegraphics[width=0.7\linewidth]{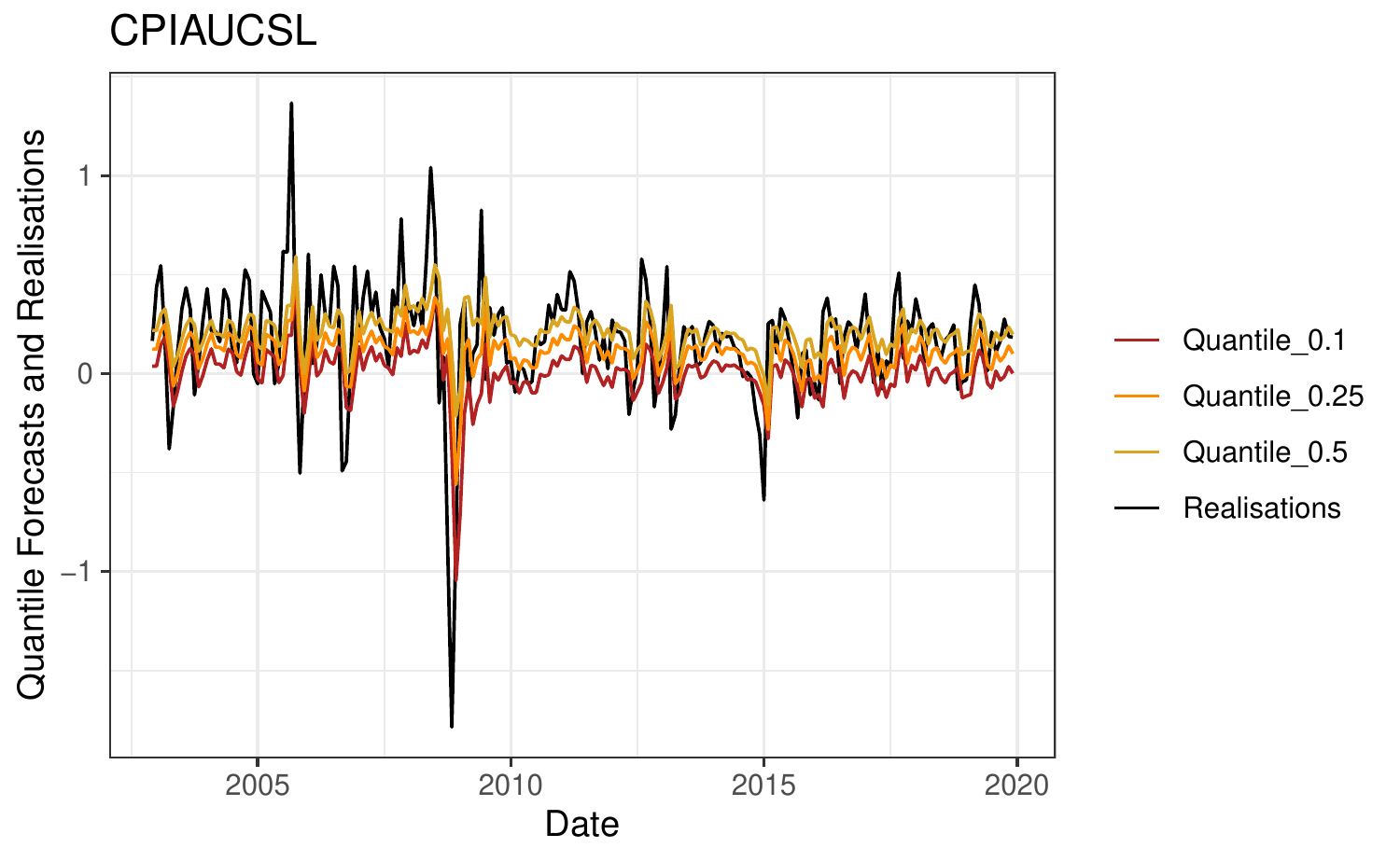}
\end{center}
\begin{center}
	\includegraphics[width=0.7\linewidth]{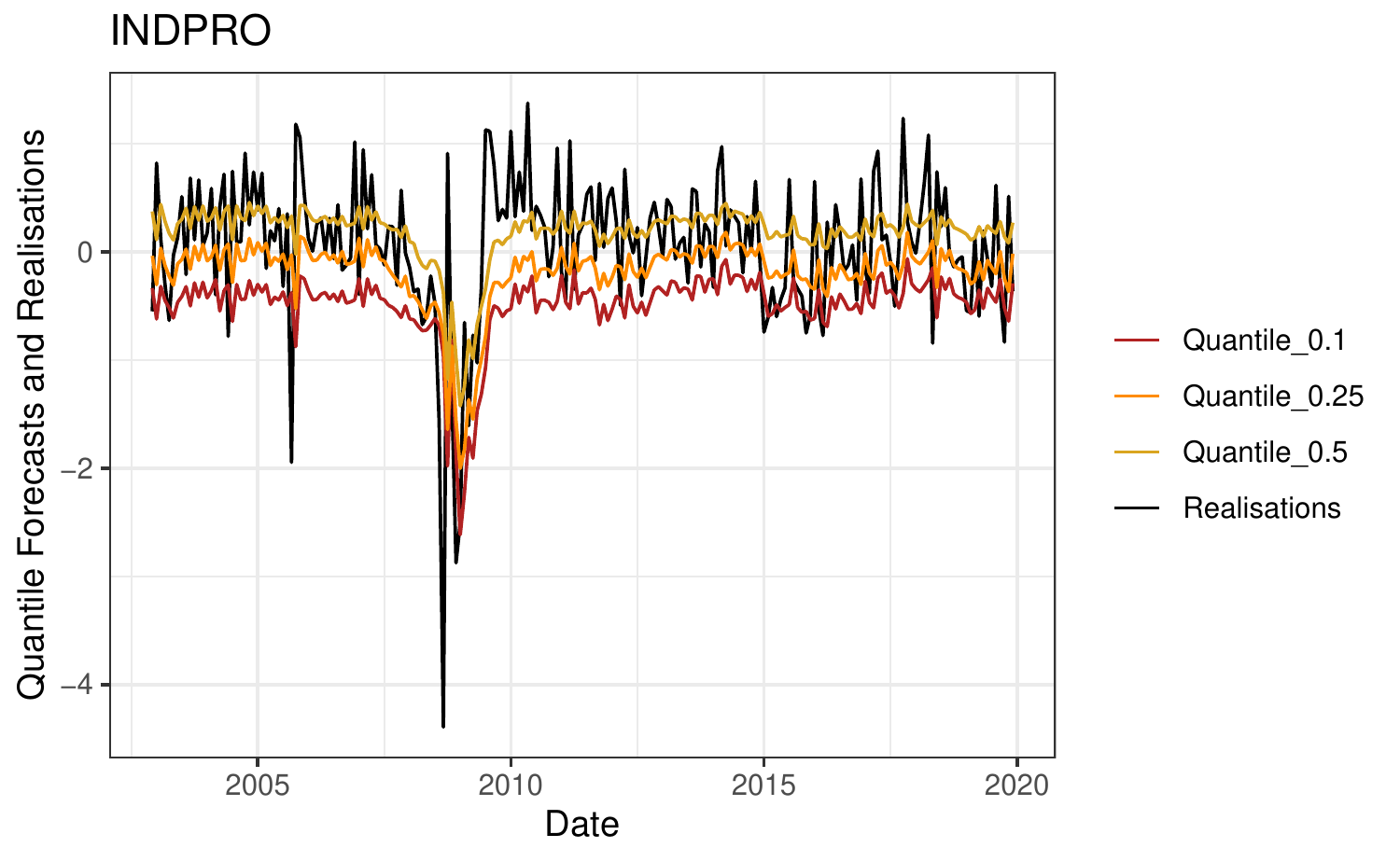}
\end{center}
\begin{center}
	\includegraphics[width=0.7\linewidth]{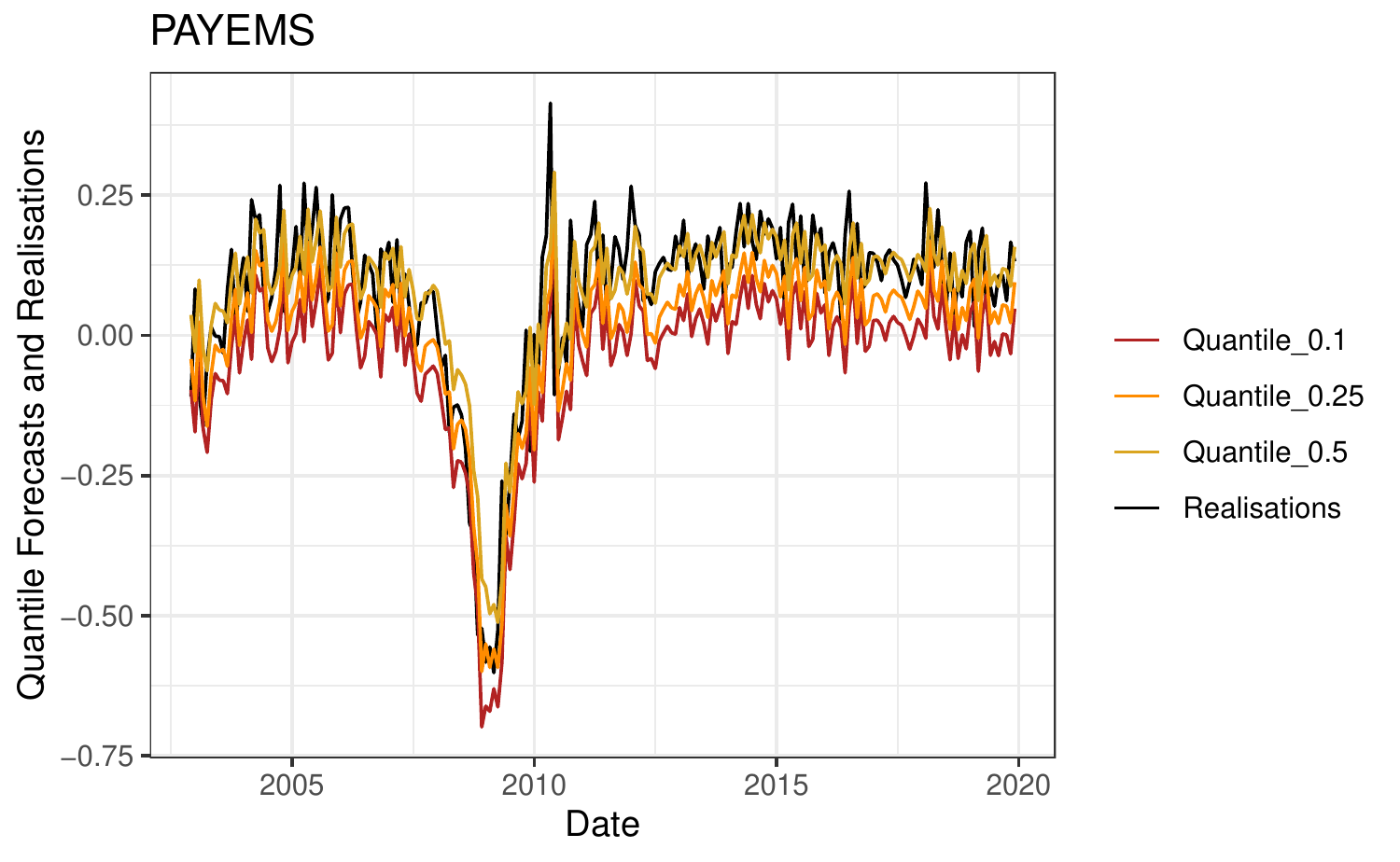}
\end{center}
\begin{center}
	\includegraphics[width=0.7\linewidth]{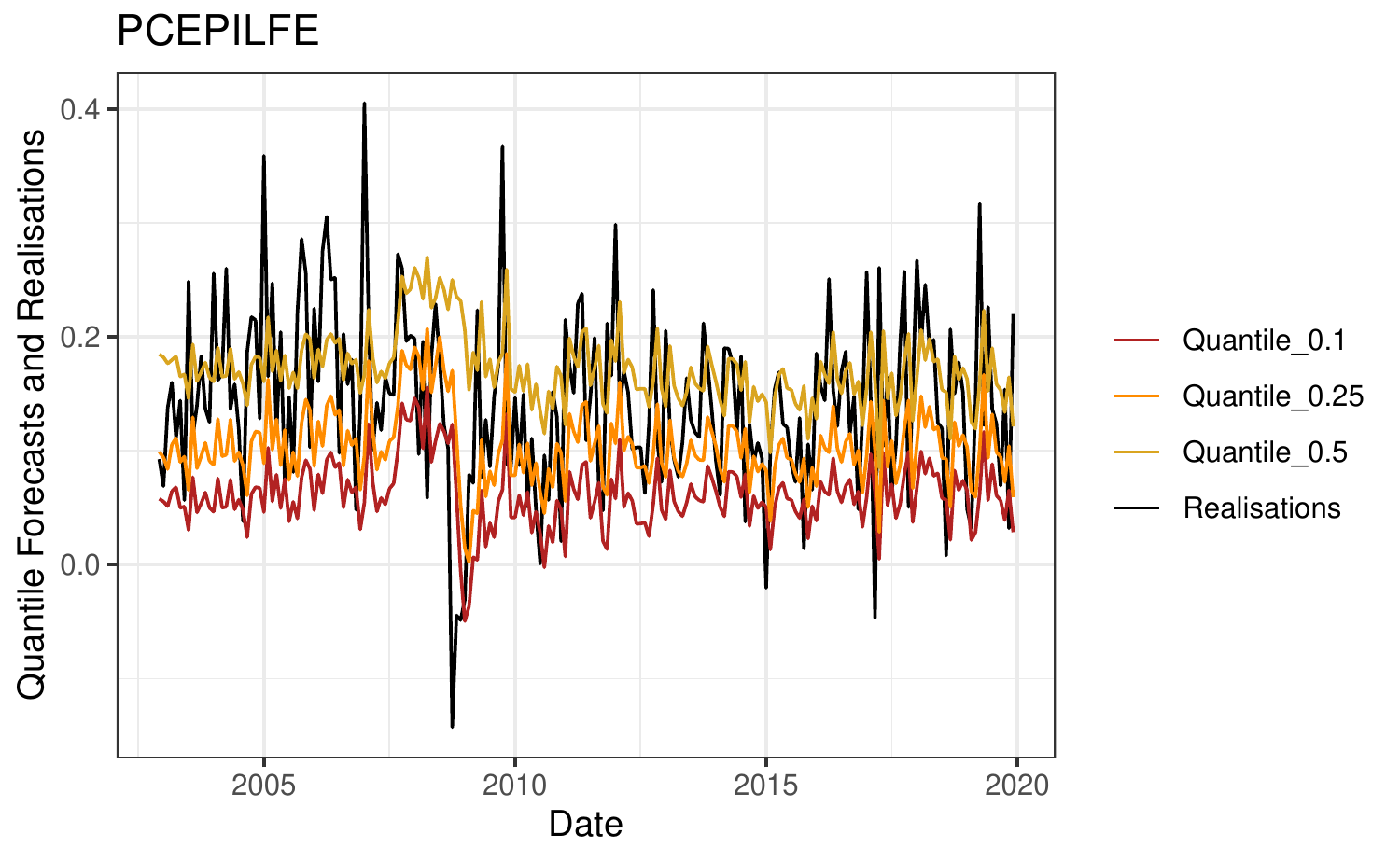}
\end{center}

\subsection{Figures: MZ Regression Lines for $h=1$ and $\tau=0.1$}\label{sec:appMZ}

\begin{center}
	\includegraphics[width=0.7\linewidth]{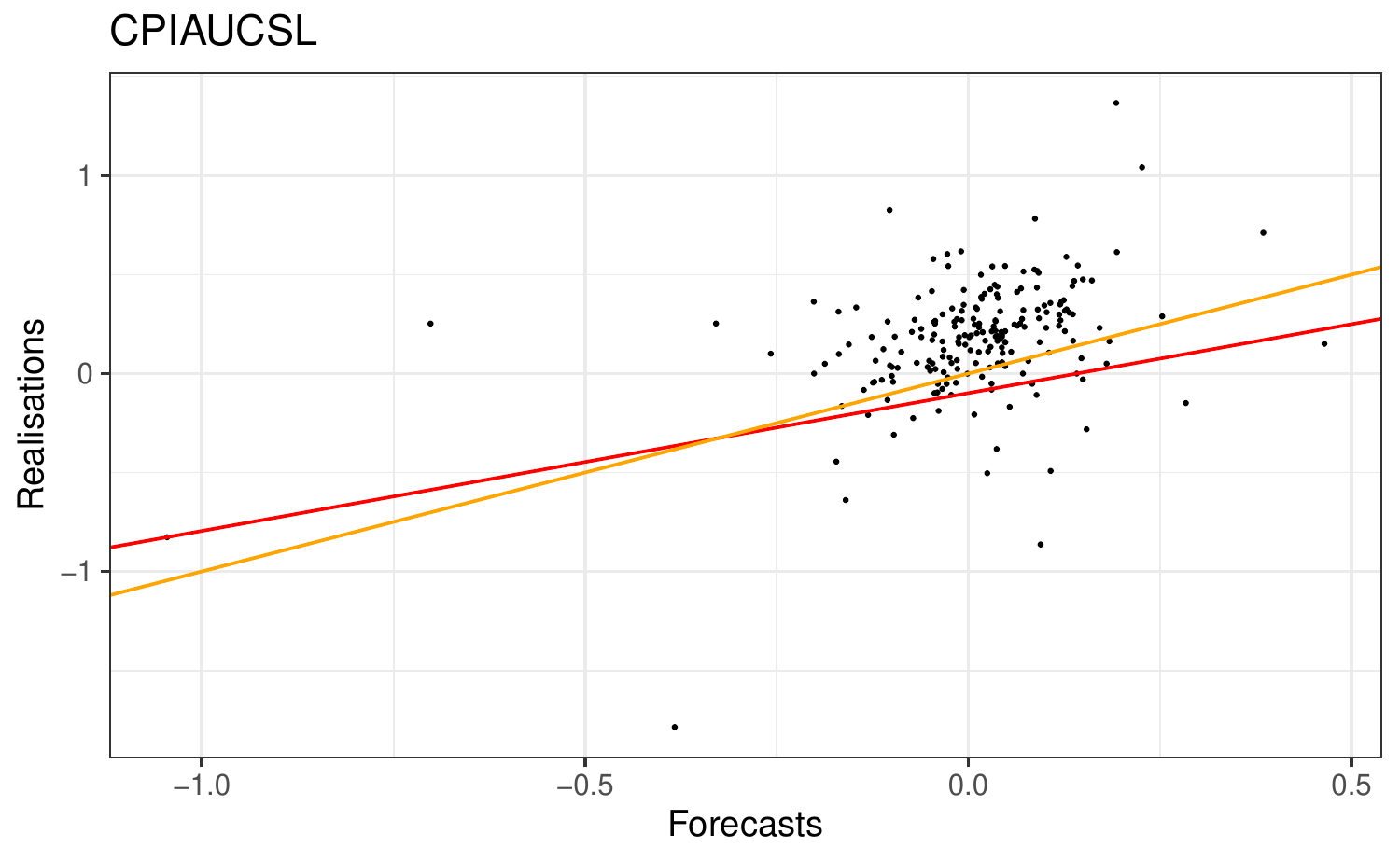}
\end{center}
\begin{center}
	\includegraphics[width=0.7\linewidth]{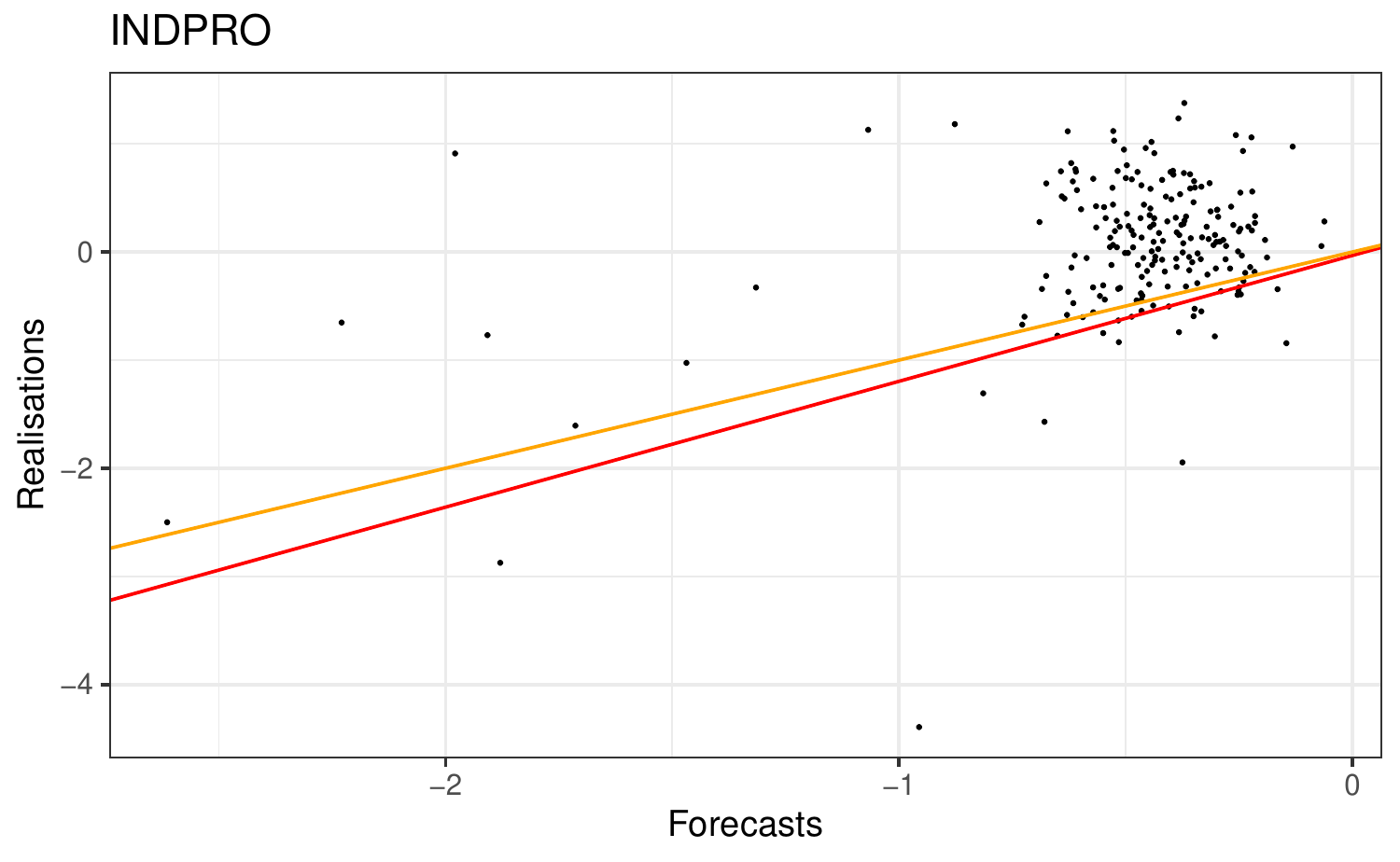}
\end{center}
\begin{center}
	\includegraphics[width=0.7\linewidth]{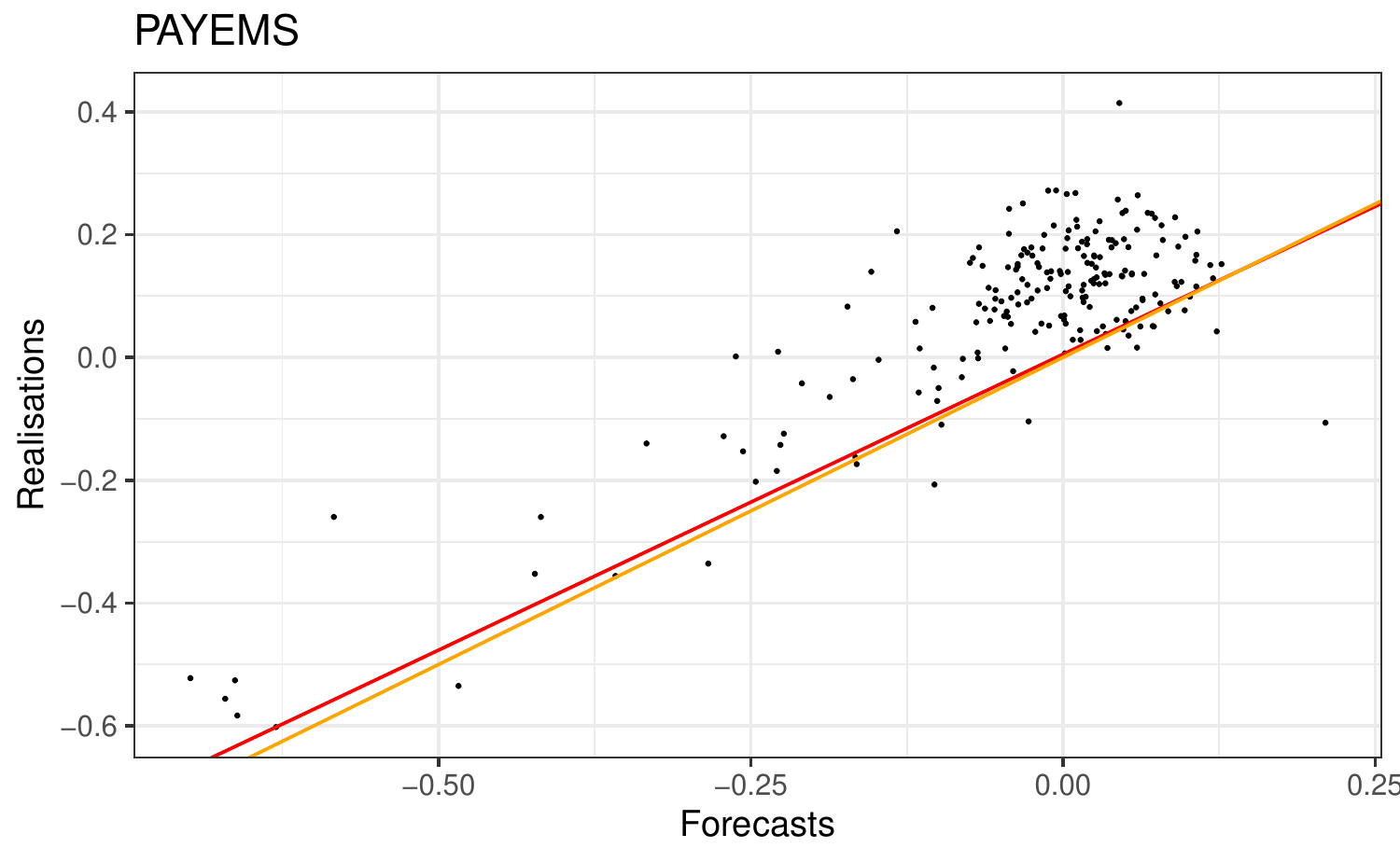}
\end{center}
\begin{center}
	\includegraphics[width=0.7\linewidth]{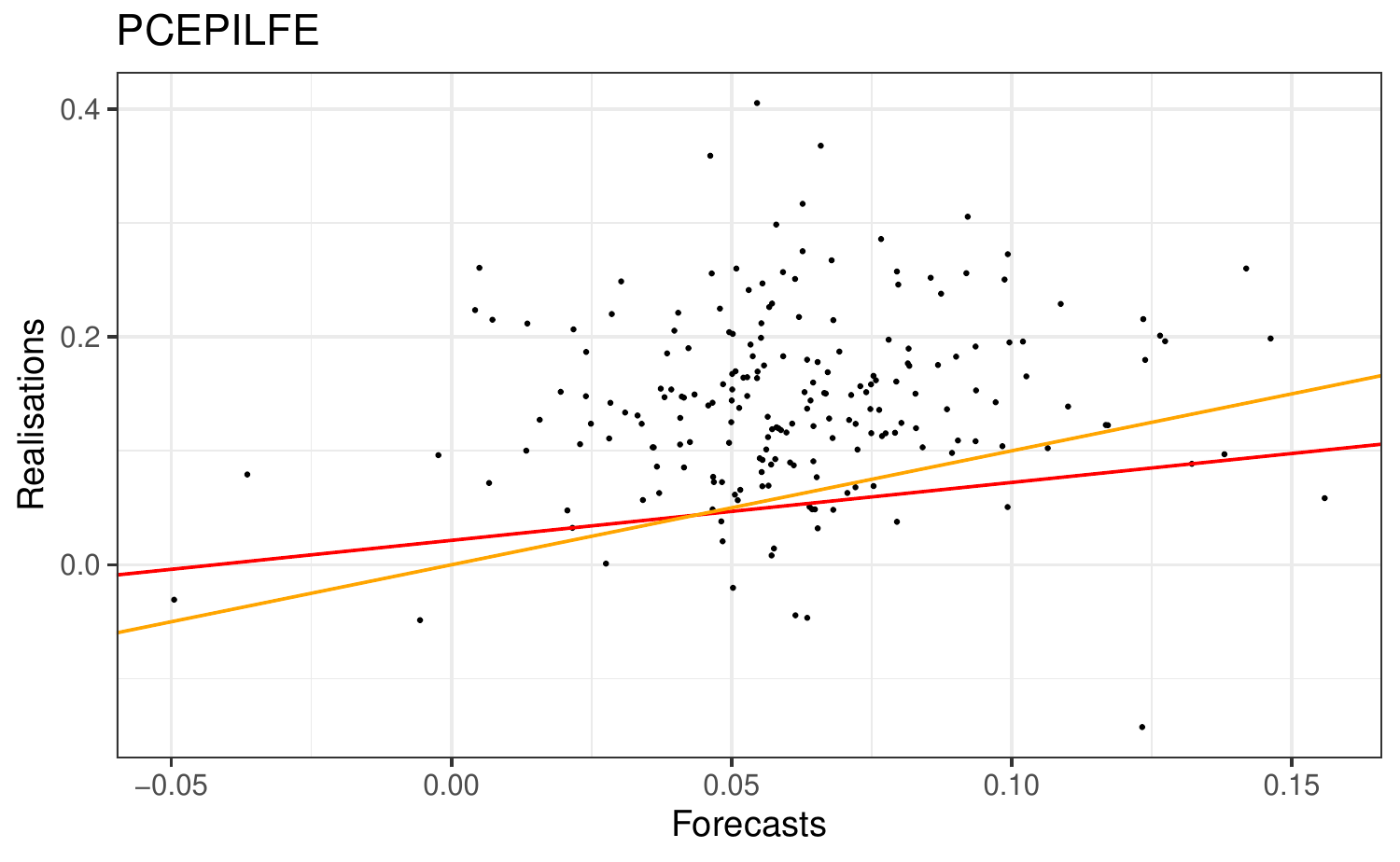}
\end{center}

\newpage

\subsection{Table: Horizon and Quantile Contributions}

\begin{table}[!ht]
	\caption{Contributions to the Test Statistic by Horizon and Quantile Level}
	\centering
	\begin{tabular}{lcccllccc}\toprule
		
		&  & CPIAUCSL &  &  &  &  & INDPRO &  \\ \midrule
		& $\tau=0.1$ & $\tau=0.25$ & $\tau=0.5$ &  &  & $\tau=0.1$ & $\tau=0.25$ & $\tau=0.5$ \\ \midrule
		$h=1$ & 20.848 & 0.441 & 0.778 &  & $h=1$ & 5.720 & 4.196 & 18.968 \\ 
		$h=2$ & 647.848 & 900.753 & 168.374 &  & $h=2$ & 1.340 & 6.251 & 4.954 \\ 
		$h=3$ & 603.497 & 544.596 & 356.875 &  & $h=3$ & 6.800 & 1.778 & 9.367 \\ 
		$h=4$ & 743.235 & 768.847 & 93.374 &  & $h=4$ & 3.014 & 1.558 & 21.279 \\ 
		$h=5$ & 386.648 & 544.058 & 264.158 &  & $h=5$ & 7.245 & 52.469 & 31.338 \\ 
		$h=6$ & 781.774 & 720.688 & 314.995 &  & $h=6$ & 43.711 & 15.228 & 96.149 \\ 
		$h=7$ & 498.539 & 1070.123 & 325.264 &  & $h=7$ & 9.559 & 46.382 & 284.909 \\ 
		$h=8$ & 1133.387 & 1758.585 & 721.687 &  & $h=8$ & 99.920 & 41.575 & 288.573 \\ 
		$h=9$ & 1349.857 & 857.559 & 510.451 &  & $h=9$ & 10.912 & 215.386 & 288.707 \\ 
		$h=10$ & 217.077 & 174.061 & 226.312 &  & $h=10$ & 108.957 & 330.662 & 310.437 \\ 
		$h=11$ & 361.884 & 230.551 & 252.698 &  & $h=11$ & 221.439 & 404.278 & 318.836 \\ 
		$h=12$ & 174.499 & 232.623 & 313.021 &  & $h=12$ & 226.706 & 406.279 & 313.196 \\ \midrule
		&  &  &  &  &  &  &  &  \\ \midrule
		&  & PAYEMS &  &  &  &  & PCEPILFE &  \\ \midrule
		& $\tau=0.1$ & $\tau=0.25$ & $\tau=0.5$ &  &  & $\tau=0.1$ & $\tau=0.25$ & $\tau=0.5$ \\ \midrule
		$h=1$ & 0.280 & 0.316 & 0.046 &  & $h=1$ & 49.667 & 83.078 & 124.573 \\ 
		$h=2$ & 0.297 & 0.483 & 0.067 &  & $h=2$ & 442.858 & 127.057 & 279.434 \\ 
		$h=3$ & 0.836 & 4.267 & 9.897 &  & $h=3$ & 622.015 & 529.844 & 245.981 \\ 
		$h=4$ & 0.022 & 1.628 & 20.176 &  & $h=4$ & 350.324 & 281.311 & 281.12 \\ 
		$h=5$ & 0.170 & 11.645 & 26.387 &  & $h=5$ & 755.340 & 483.032 & 275.909 \\ 
		$h=6$ & 0.007 & 9.931 & 38.356 &  & $h=6$ & 311.629 & 312.997 & 299.373 \\ 
		$h=7$ & 29.247 & 53.612 & 31.944 &  & $h=7$ & 1226.937 & 687.718 & 306.836 \\ 
		$h=8$ & 4.018 & 40.909 & 48.416 &  & $h=8$ & 711.826 & 431.873 & 313.763 \\ 
		$h=9$ & 19.371 & 52.964 & 53.312 &  & $h=9$ & 787.800 & 326.569 & 279.552 \\ 
		$h=10$ & 27.079 & 51.491 & 34.155 &  & $h=10$ & 388.266 & 337.273 & 365.069 \\ 
		$h=11$ & 12.010 & 80.984 & 36.712 &  & $h=11$ & 885.383 & 300.752 & 321.414 \\ 
		$h=12$ & 6.047 & 92.947 & 71.674 &  & $h=12$ & 659.990 & 337.922 & 340.049 \\ \bottomrule
	\end{tabular}
	\label{tab:Cont}
\end{table}

\end{document}